\documentclass[preprint,12pt,authoryear]{elsarticle}

\usepackage{amssymb}
\usepackage{amsmath}
\usepackage{gensymb}

\journal{Planetary and Space Science}

\begin{document}

\begin{frontmatter}



\title{Dynamical regimes of small bodies perturbed by an eccentric giant planet} 


\author{Tabar\'{e} Gallardo and Rodrigo Cabral} 

\affiliation{organization={Facultad de Ciencias, Udelar},
	addressline={Igua 4225}, 
	postcode={11400}, 
	state={Montevideo},
	country={Uruguay}}

\begin{abstract}

The dynamics of small bodies perturbed by an eccentric planet was done mostly under the assumption of well separated orbits using analytical approximations appropriate for the hierarchical case. In this work
we study the dynamics of small bodies in a wide range of eccentricities and inclinations perturbed by a giant planet with $e_p=0.4$, in the non-hierarchical case. We consider small bodies  both interior and exterior to the planet.  We apply semi-analytical models for the study of resonances and the properties of the secular disturbing function. We perform a frequency analysis of numerical integration of the exact equations of motion  to obtain the proper frequencies and corresponding dynamical secular paths. We study the dependence of proper frequencies with the initial mutual inclination and we find a critical inclination around 30  degrees for which the pericenter proper frequency vanishes giving rise to the increase of small bodies eccentricities followed by unstable dynamics. This happens for both interior and exterior small bodies and constitutes a stability barrier in the inclination. For greater inclinations the ZLK mechanism dominates both populations.
By means of numerical integration of thousands of small bodies we reproduce the well known pericenter shepherding, but for the exterior populations we also find concentrations of the longitude of the ascending node in the direction of the planetary line of apsides.

\end{abstract}


%
%
%
%
%
%
%
%
%
%
%
%
%


\end{frontmatter}




\section{Introduction}\label{intro}

Small body dynamics has been studied in depth and extensively in the context of the Solar System, with planets in low eccentricity $(e)$ and low inclination  $(i)$ orbits. The particular case of small bodies with high $(e,i)$ orbits perturbed by circular and coplanar planets was studied since the works of \cite{LIDOV1962719} and \cite{1962AJ.....67..591K}, and the reader should see the exhaustive historical revision by \cite{Ito2019}. The particular dynamics found in that case was called the Lidov-Kozai mechanism, or LK for short. It consists on the existence of some minimum orbital inclination for which oscillations of the argument of the pericenter $(\omega)$ around $90\degree$ or $270\degree$ take place associated with correlated oscillations in $(e,i)$ that can make strong orbital changes. The oscillations of $\omega$ constitute also a protection mechanism for asteroids that grow in eccentricity because when they reach their perihelia and the terrestrial planetary zone they are out of the ecliptic, avoiding collisions with the planets.
This kind of evolution, which preserves the z component of the particle's orbital angular momentum, has been called LK-resonance or LK-mechanism indistinctly in the literature.
 It is clear that it is a secular long term dynamical mechanism, but whether is a resonance or not depends on the existence of some separatrix and if the period for a complete cycle close to it tends to infinity or not, see discussion in \cite{Sidorenko2018}. 
After a comprehensive description of the dynamical evolution of particles perturbed by circular and coplanar planets, several authors have addressed
the problem of the secular evolution of an interior particle under the perturbation of an eccentric planet (the \textit{eccentric} LK or ELK mechanism as it was called) using progressively higher-order (quadrupolar or octupolar) expansions of the disturbing function in terms of the ratio of semi-major axes, $\alpha=a_1/a_2$, which are appropriate for hierarchical systems, i.e., bodies well separated in space, or $\alpha < < 1$. For example,
\cite{Katz2011} and \cite{Lithwick2011} studied the case of a massless particle perturbed by an eccentric exterior perturber, while \cite{Libert2012}, \cite{Li2014} and \cite{weldon2024analytical} studied the planetary case (both are massive bodies), but derived some results for the limit of a test particle with vanishing mass.
Some of these works focused on the interesting property of flipping orbits, oscillations of the inclination around $90\degree$ \citep{Naoz2011} 
which was a forbidden property in the standard LK mechanism and that could explain the origin of some retrograde objects by secular evolution in the Solar System. Numerical explorations of a particle in circular orbit perturbed by a giant exterior and eccentric planet were done by \cite{2011A&A...526A..98F} for different $\alpha$ for non-hierarchical cases and different planetary masses and eccentricities finding a region of stable motion for mutual $i\lesssim 35\degree$ that is more or less independent of planetary mass and $\alpha$, provided that $\alpha$ is smaller than some value that depends on the perturber eccentricity.

\cite{Naoz2016} provides a review of the LK mechanism in the context of two planets in eccentric hierarchical orbits and in the cases of a particle perturbed by a
circular and also by an eccentric exterior perturber in the hierarchical three-body problem employing quadrupolar and octupolar levels of the approximation of the secular disturbing function. It was shown that for high-inclination interior particles ($39.2^\circ<i<140.7^\circ$), the ELK mechanism can cause large-amplitude oscillations in inclination and eccentricity, which can lead to orbits going from prograde to retrograde and objects reaching almost linear trajectories ($e\sim 1$). These particles can exhibit oscillations, circulations, or chaotic behavior in 
the dynamical evolution of their $\omega$ and longitude of the ascending node, $\Omega$. More recently \citet{Lei2022} made a systematic study about orbital flips for an inner test particle in the context of the ELK mechanism.

\cite{2010MNRAS.401.1189F} was one of the first works that studied the case of a far away exterior particle perturbed by an eccentric inner planet, with some results for the planetary case also. Some years later \cite{2017A&A...605A..64Z}
studied this problem without analytical approximations using numerical simulations, followed by
\cite{Naoz2017}, \cite{Vinson2018} and \cite{DeElia2019} that  used the octupolar or hexadecapolar level of the approximation instead. 
The case of the exterior particle was called
the \textit{inverse} ELK mechanism and these authors  found a dependency between the maximum and minimum inclination a particle achieves and the perturber's eccentricity, finding that flipping also occurs for the inverse ELK. 
\cite{Vinson2018} studied the ELK mechanism, finding not only cases where the argument of pericenter oscillates but cases where a combination of angular variables oscillates and generates large eccentricity variations.
Other works considered the modification due to General Relativity \citep{zan2018,Zanardi2023,Coronel2024}.

Several works have addressed this problem  in the context of planets evolving in  binary stars, see for example \cite{verrier, Saleh2009, 2010MNRAS.401.1189F} and subsequent references.
In recent years this mechanism has also been known as von Zeipel-Lidov-Kozai mechanism (ZLK for short) since \cite{Ito2019} pointed out the role of von Zeipel in the formulation and study of the problem.

In this paper, we will address the problem of interior and exterior particles perturbed by an eccentric giant planet from a general configuration without distinctions between the different types of the ZLK mechanism. As hierarchical cases were very well studied we focus mainly in non-hierarchical ones, that is, semimajor axes not very different, typical of situations commonly found in populations of small bodies perturbed by a planet.  Our study is based on the analysis of the output of the numerical integrations of the exact equations of motion and the use of semi-analytical models based on the numerical calculation of the disturbing function and its derivatives. Specifically, we will determine the proper frequencies for the particles and analyze their role in the general behavior of the secular dynamical regimes of small bodies perturbed by an eccentric planet. To achieve this, firstly
in Section 2 we characterize the mean motion resonances (MMRs) that the giant planet generates for particles with arbitrary $(e,i)$ to separate resonant and secular regions. Then, in Section 3, we study the secular evolution focusing on the determination of the proper frequencies and their role in the different dynamical regimes. In Section 4 we summarize our results.  A previous recent paper \citep{Gallardo2025} deals with the more complex case of the dynamics of two planets.
The main finding of that work is that for some critical mutual inclination between 30 and 40 degrees a pericenter secular resonance takes place dividing the secular dynamics in two different regimes.
In the general case addressed by \citep{Gallardo2025}, the conservation of the system's angular momentum constrains the dynamics of both planets. However, in our restricted scenario, neither the angular momentum of the particle nor its z-component is conserved, thus permitting more intricate dynamics for the particle, as we will demonstrate. We will follow approximately the same scheme of work by  \citep{Gallardo2025}.

\section{Resonances generated by an eccentric planet}

The star-planet system we consider all along in this work is summarized in Table \ref{tab:system}. We chose a system analogous to the Sun-Jupiter system, but with Jupiter on a highly eccentric orbit, to understand the potential effects of such an eccentricity within our own Solar System. Without losing generality, we choose the plane of the planet's orbit as the reference plane and its pericenter direction as the origin for the direction of the particle's ascending node, $\Omega$. 

\begin{figure*}[t]
	\centering
	\begin{tabular}{c c}
		\includegraphics[width=0.45\textwidth]{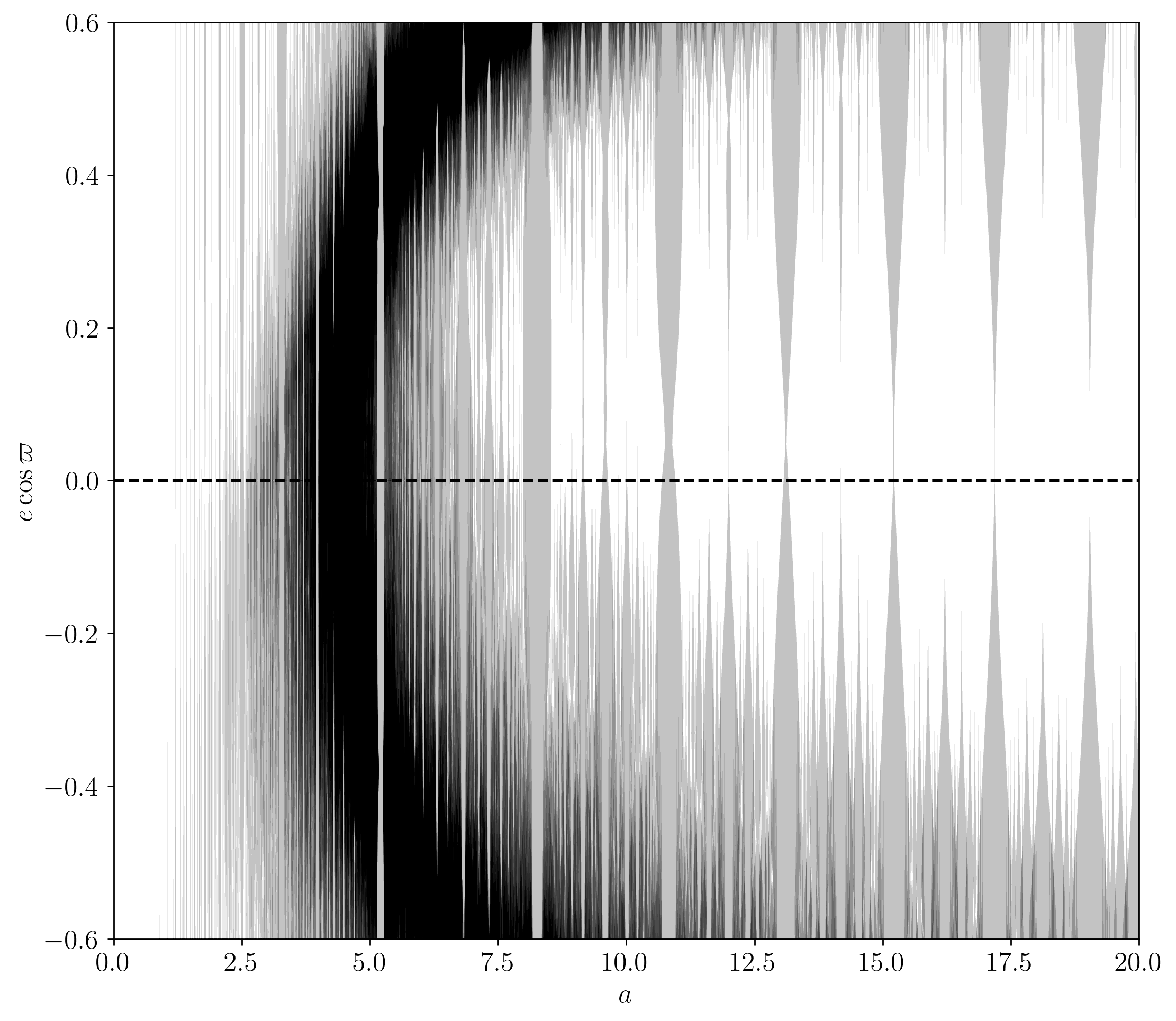}  & \includegraphics[width=0.57\textwidth]{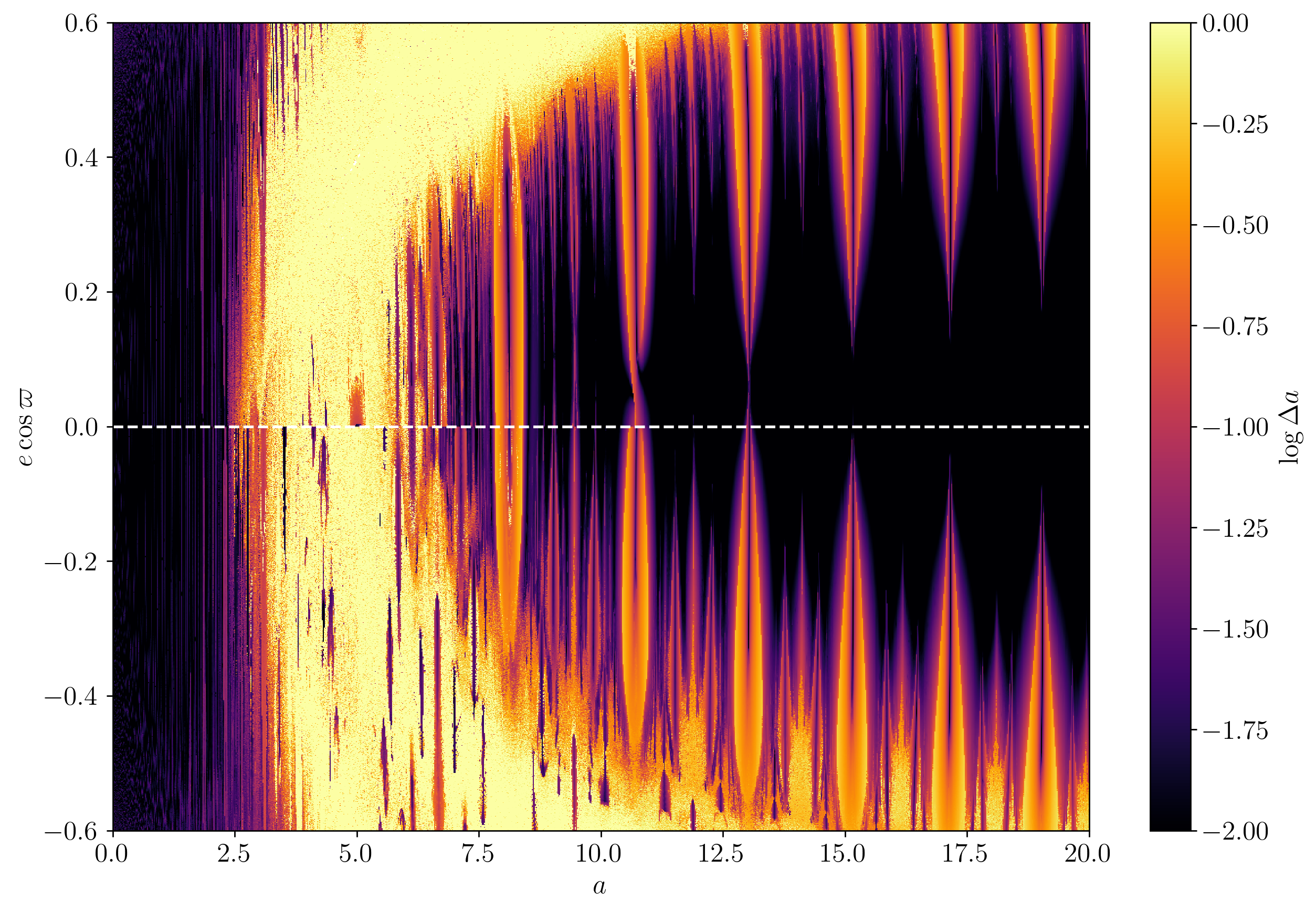}  \\ 
	\end{tabular}
	\caption{Left: Atlas of 1446 resonances $N$:$N_p$ with $N,N_p\leq 50$ generated by the model for particles with $i=30\degree$, $\Omega=270\degree$  and two values of $\varpi=0\degree, 180\degree$. Right: Dynamical map generated integrating numerically the exact equations of motion for $10^4$ years showing $\log(\Delta a)$  in color scale for the same initial conditions and initial mean anomalies $M=M_p=0\degree$. Parameters for the star and giant planet are given at Table \ref{tab:system}.}
	\label{resonances}
\end{figure*}

The Jupiter-like planet generates a web of resonances given by the integers $(N,N_p)$ verifying $Nn \simeq N_p n_p$ being $n,n_p$ the mean motions of the asteroid and planet respectively.
\begin{table}
	\centering
	\begin{tabular}{cc}
		$M_{\star}$ &  $1 M_{\odot}$\\
		$m_p$  & 0.001\\
		$a_p$  & 5.2 au\\
		$e_p$ & 0.4 \\
		$i_p$  & $0\degree$\\
		$\varpi_p$  & $0\degree$\\
	\end{tabular}
	\caption{Properties of the system star-planet adopted in all along this work. From top to bottom: stellar mass, planetary mass, semi-major axis, eccentricity, inclination and longitude of the pericenter.}
	\label{tab:system}
\end{table}
Using the model by \cite{Gallardo2020}, based on the numerical calculation of the resonant disturbing function without any limitations in orbital elements, 
and the codes for an eccentric planet available at the site\footnote{https://sites.google.com/view/mmresonances/home/2basteroidal} 
we calculated the atlas of resonances shown in Fig. \ref{resonances} left panel. The atlas is composed of all resonances defined by  $N$ and $N_p$ between 1 and 50 and was calculated for a nominal small body with $i=30\degree$, $\Omega=270\degree$ and two configurations for $\Delta \varpi = \varpi - \varpi_p$: $0\degree$ and $180\degree$. Each resonance is colored in gray, so overlap or resonances, generally associated with chaos, appear as darker gray or black. On the other hand, white regions correspond to absence of resonances, indicating regions where purely secular evolutions can take place. In the right panel of Fig. \ref{resonances} we present a dynamical map generated integrating numerically the exact equations of motion with the same initial conditions in a grid of 
$(a,e)$ showing the changes in semimajor axis, $\Delta a$, obtained in the integrations, which is a measure of the dynamical effect of the resonances and also unstable evolutions. There is an excellent match between the resonant model and the numerical results of the map which give us confidence in the use of the atlas for predicting resonant regions.  The main structures exterior to the planet are the $1$:$N$ resonances and between them the absence of strong resonances allows the secular dynamics to be present. Close to the planet strong chaos is evident due to close encounters and resonance overlap. Note that the model (left panel of the figure) predicts there is a narrow gray region at $a\simeq a_p$ where only one resonance exists, the coorbitals, without other MMRs overlapping. It was not detected as stable in the dynamical map of the right panel because the initial conditions were not chosen adequately to obtain the characteristic librations of the resonant coorbital motion.

The plots of Fig. \ref{resonances} correspond to test particles with $i=30\degree$, for lower inclinations the MMRs overlap and the chaotic region grows while for higher inclinations diminishes. Planets in circular orbits generate atlases with the same pattern for $\varpi=0\degree$ as for $\varpi=180\degree$ but, as can be deduced from the plots, for eccentric planetary orbits there is some asymmetry of the resonance structures with respect to the horizontal line $e=0$, so that there is more space free from resonances for $\varpi=0\degree$ than for $\varpi=180\degree$.
The plots show that for an eccentric Jupiter the evolution could be secular in the range $a \lesssim 2.5$ au and $a\gtrsim 9$ au, otherwise resonances and overlap of resonances dominate the evolution, resulting in large variations of semi-major axis (see right panel of Fig. \ref{resonances}). 

\begin{figure}
	\centering
	\includegraphics[width=0.6\textwidth]{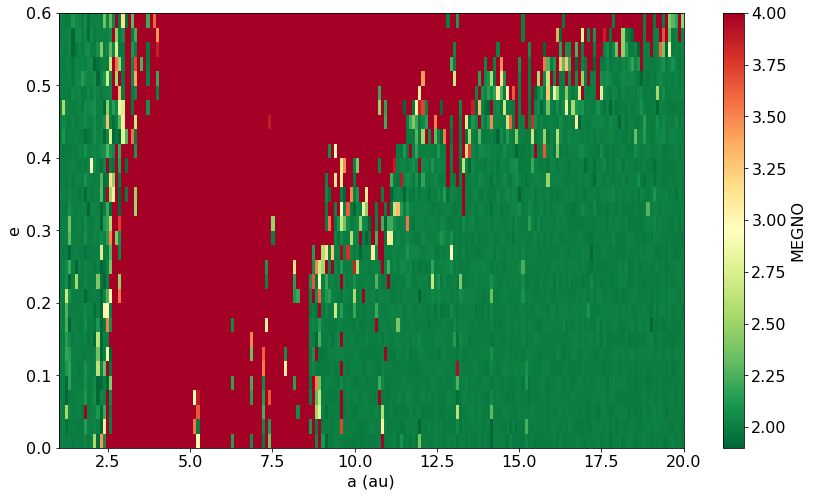}
	\includegraphics[width=0.6\textwidth]{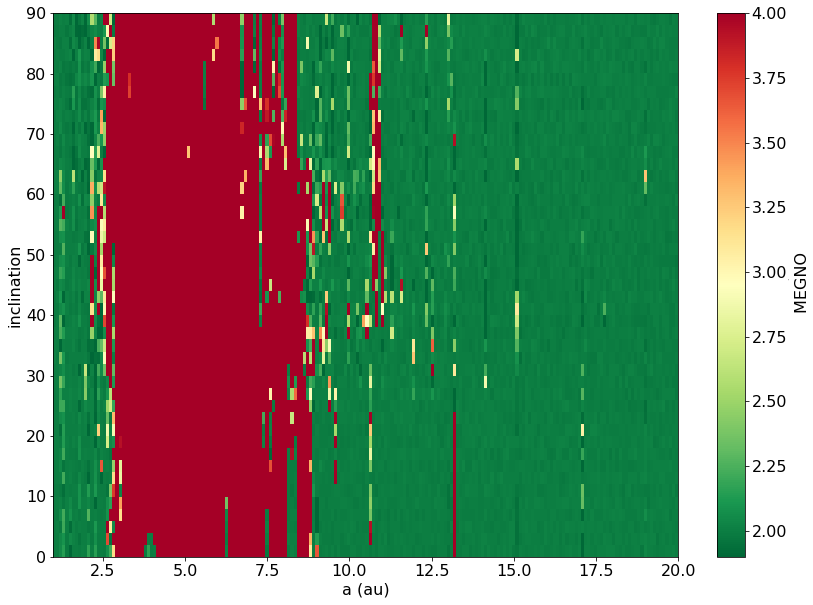}    
	\caption{Upper panel: MEGNO map in the plane $(a,e)$ constructed with test particles with  initial conditions corresponding to the positive part of y-axis of Fig. \ref{resonances} ($i=30\degree$, $\varpi=0\degree$). Lower panel: MEGNO map in the plane $(a,i)$ 
		constructed with test particles with  initial conditions
		$\Omega=270\degree$, $\omega=90\degree$ and $e=0.2$. Integration time of 5000 orbital revolutions of the test particle for both panels. In both map it was taken  initial mean anomalies  $M=M_p=0\degree$.}
	\label{megno}
\end{figure}

In Fig. \ref{megno} we present the MEGNO chaos indicator \citep{2000A&AS..147..205C}
in the planes $(a,e)$ for initial $i=30\degree$ in upper panel
and  $(a,i)$ for initial $e=0.2$ in the low panel calculated with REBOUND, integrating the test particles for 5000 orbital revolutions with ias15 algorithm \citep{2015MNRAS.446.1424R}. Green color means regular dynamics while red means chaotic.
Note that for particles with $e=0.2$ (lower panel) the region $2.5 \lesssim  a 
\lesssim  9$ au is chaotic independently of the orbital inclination. Outside this region dynamics is more or less regular, which is a basic condition when looking for secular dynamics. 
It is important to note that regular regions do not necessarily mean stable regions, as it is possible that, due to a regular secular evolution (on a much longer timescale than computed here), the eccentricity could increase and the particle could eventually collide with the planet or star. Thus, in longer numerical integrations, we can find unstable evolutions within the region of initial regular motion.

\section{Secular dynamics of small bodies}

Obtaining the equations describing the secular evolution is a basic procedure that we include here for completeness.
In terms of canonical Poincar\'{e} variables 
$(\lambda,\gamma,z,\Lambda,\Gamma,Z)=(\lambda,-\varpi,-\Omega,L,L-G,G-H)$, being $L=\sqrt{\mu a}$, $G=L\sqrt{1-e^2}$ and $H=G\cos{i}$, the Hamiltonian of an asteroid evolving under the perturbation of a unique eccentric planet  is
\begin{equation}\label{ham}
	\mathcal{H}=-\frac{\mu^2}{2\Lambda^2}  -R(\Lambda,\Gamma,Z,\lambda,\gamma,z,\lambda_p)
\end{equation}
where 
$\mu=k^2M_{\star}$, $k$ is the Gaussian gravitational constant, $R$ is the disturbing function proportional to $m_p$, $\lambda$ is the mean longitude of the particle  and
$\lambda_p$ is the mean longitude of the planet in its fixed orbit.
Assuming there is no mean motion commensurability between asteroid and planet and excluding close encounter situations we can perform a method of averaging, eliminating the fast variables $\lambda, \lambda_p$ to obtain the new time-independent  Hamiltonian
\begin{equation}
	\mathcal{H}=-\frac{\mu^2}{2\Lambda^2} -R_s(\Lambda,\Gamma,Z,\gamma,z)
\end{equation}
where $R_s$ is the secular disturbing function obtained by a  numerical double averaging in $\lambda$ and $\lambda_p$. For simplicity we use the same nomenclature for the new variables and Hamiltonian but, after averaging, the new variables are mean variables in the sense that now they have no short period variations and strictly they are not the same shown in Eq. \ref{ham}.
Since $\mathcal{H}$ now does  not depend on $\lambda$ we have that $\Lambda$ is constant, that means $a$ is constant. Then, the conservation of  $\mathcal{H}$ implies $R_s$ is constant and the canonical equations for the secular motion of the asteroid become:

\begin{equation}
	\frac{d\Gamma}{dt}=\frac{\partial R_s}{\partial \gamma},  \hspace{1cm} \frac{d\gamma}{dt}=-\frac{\partial R_s}{\partial \Gamma}   
\end{equation}
\begin{equation}
	\frac{dZ}{dt}=\frac{\partial R_s}{\partial z},  \hspace{1cm} \frac{dz}{dt}=-\frac{\partial R_s}{\partial Z}   
\end{equation}
This is a system with 2 degrees of freedom, which, if it admits a regular solution, must contain 2 constant frequencies, which in small body dynamics are called proper frequencies. These frequencies and some of their combinations will appear in the time evolution of the small bodies orbital elements $(e,i,\varpi,\Omega)$. Considering the  perturbing planet and the particle both with low $(e,i)$ orbits, the Lagrange-Laplace secular linear  theory predicts that these two frequencies 
have the same absolute value with opposite signs
and the time evolution of the orbital elements in the plane $(k,h)=e(\cos\varpi,\sin\varpi)$ and $(q,p)=i(\cos\Omega,\sin\Omega)$ can be modeled by a vector rotating with constant rate, counterclockwise for $(k,h)$ and clockwise for $(q,p)$ \citep{1999ssd..book.....M}. For the more general case where large $(e,i)$ are involved the linear theory and simple vector model cannot represent the orbital evolution and more elaborate theories and methods are necessary, see \cite{2019CeMDA.131...27K} for the state of art.
The basic and more general property of the secular dynamics is that particles must follow curves of constant $R_s$ in the space $(e,i,\varpi,\Omega)$ or $(k,h,q,p)$. We will explore the properties of $R_s$ with the aim of obtaining some properties of the dynamics of particles interior and exterior to the perturbing planet.

\subsection{Properties of $R_s$ in the subspace $(\omega,\Omega)$}

Considering the success of the multipole models to the understanding  of the dynamics of hierarchical systems, as we have already stated, in this work we are interested in the study of the more general case without restrictions in $\alpha$ but avoiding the chaotic region around $0.5 \lesssim a/a_p \lesssim 1.8$. To accomplish this objective we will start our study based on the numerical calculation of $R_s$. This methodology eliminates the restrictions in $(\alpha,e,i)$ but, anyway, there is a limit for the use of the secular model imposed by falling in MMRs or in close encounters situations. Given $R$ in rectangular coordinates the numerical mean, $R_s$, is computed as
\begin{equation}
	R_s=\frac{1}{4\pi^2}\int_0^{2\pi}\int_0^{2\pi} R d\lambda d\lambda_p
	\label{rsecu}
\end{equation}
fixing all other parameters, so assuming two fixed ellipses. We can also calculate the partial derivatives with respect to a generic element $e$ as:
\begin{equation}
	\frac{\partial R_s}{\partial e} \simeq \frac{\Delta R_s}{\Delta e}
	\label{rsecuder}
\end{equation}
where $\Delta R_s = R_s(e+\Delta e)-R_s(e)$ is also numerically calculated. The canonical equations or Lagrange-Laplace system of planetary equations can also be computed this way.

The study of the dynamics in terms of $R_s$ is analogous to the study of the dynamics of a solid  ellipse (small body) perturbed by another solid and fixed ellipse (planet), see for example \cite{1999ssd..book.....M}. One can conjecture that when both ellipses  are placed in space with some symmetry there should be an equilibrium configuration with angles $(\omega,\Omega)$ trying to maintain that configuration. Or, on the contrary, if there is no symmetry in the positions of the two ellipses the imbalance of forces will destroy the configuration. Taking the planetary orbit as reference plane and remembering that $\varpi_p=0\degree$,  there are two groups of symmetric configurations for arbitrary $(e,i)$. One group is composed of those obtained for $(\omega,\Omega)$ taking the values
$(90\degree,90\degree)$, $(90\degree,270\degree)$, $(270\degree,90\degree)$ and $(270\degree,270\degree)$, where there is a plane of symmetry containing both line of apses.  The other group of configurations is the one composed by $(0\degree,0\degree)$, $(0\degree,180\degree)$,
$(180\degree,0\degree)$ and $(180\degree,180\degree)$ where there is an axial symmetry along the common line of apsides. 

\begin{figure*}
	\centering
	\begin{tabular}{cc}
		\includegraphics[width=0.45\linewidth]{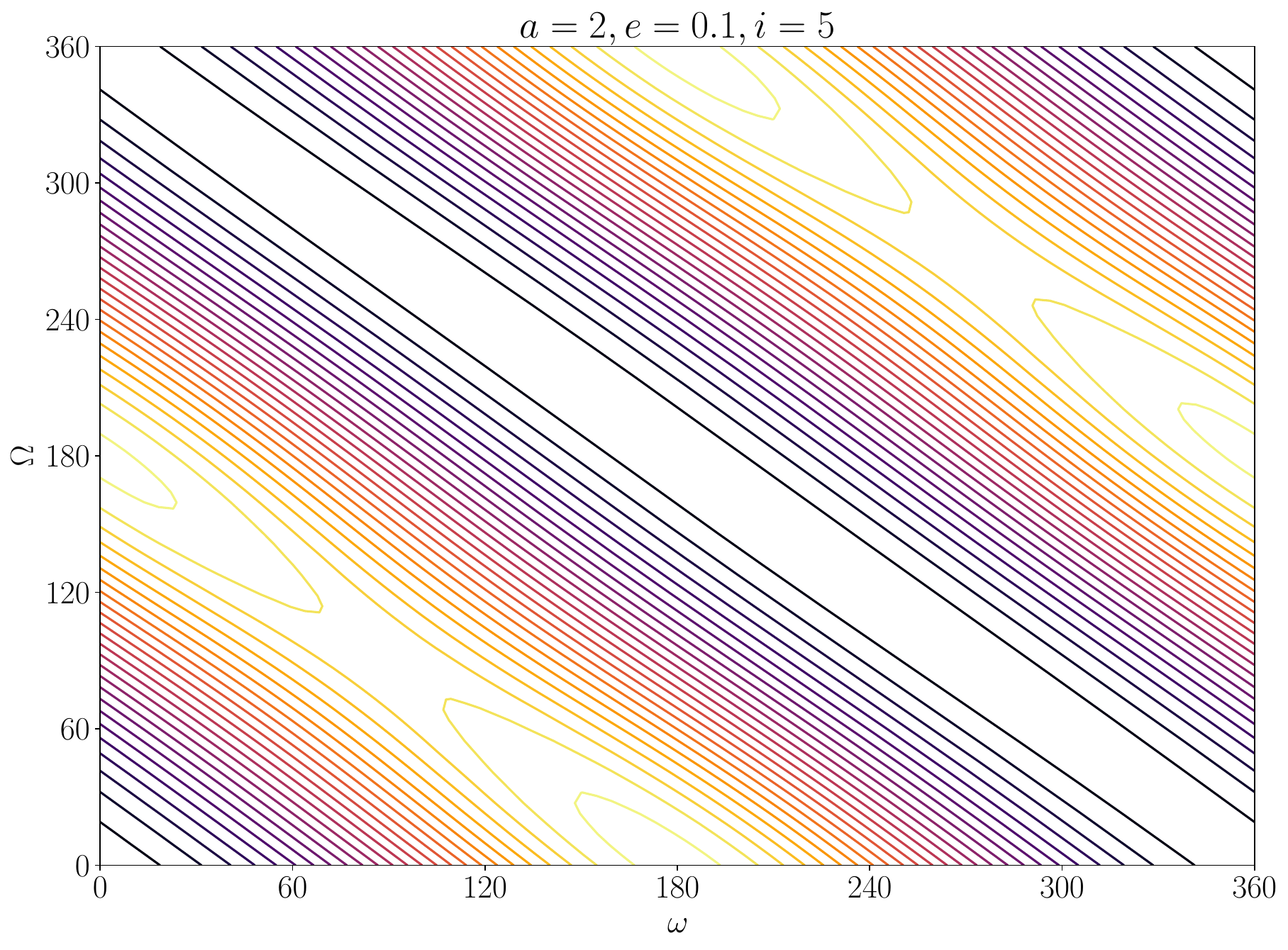} & \includegraphics[width=0.45\linewidth]{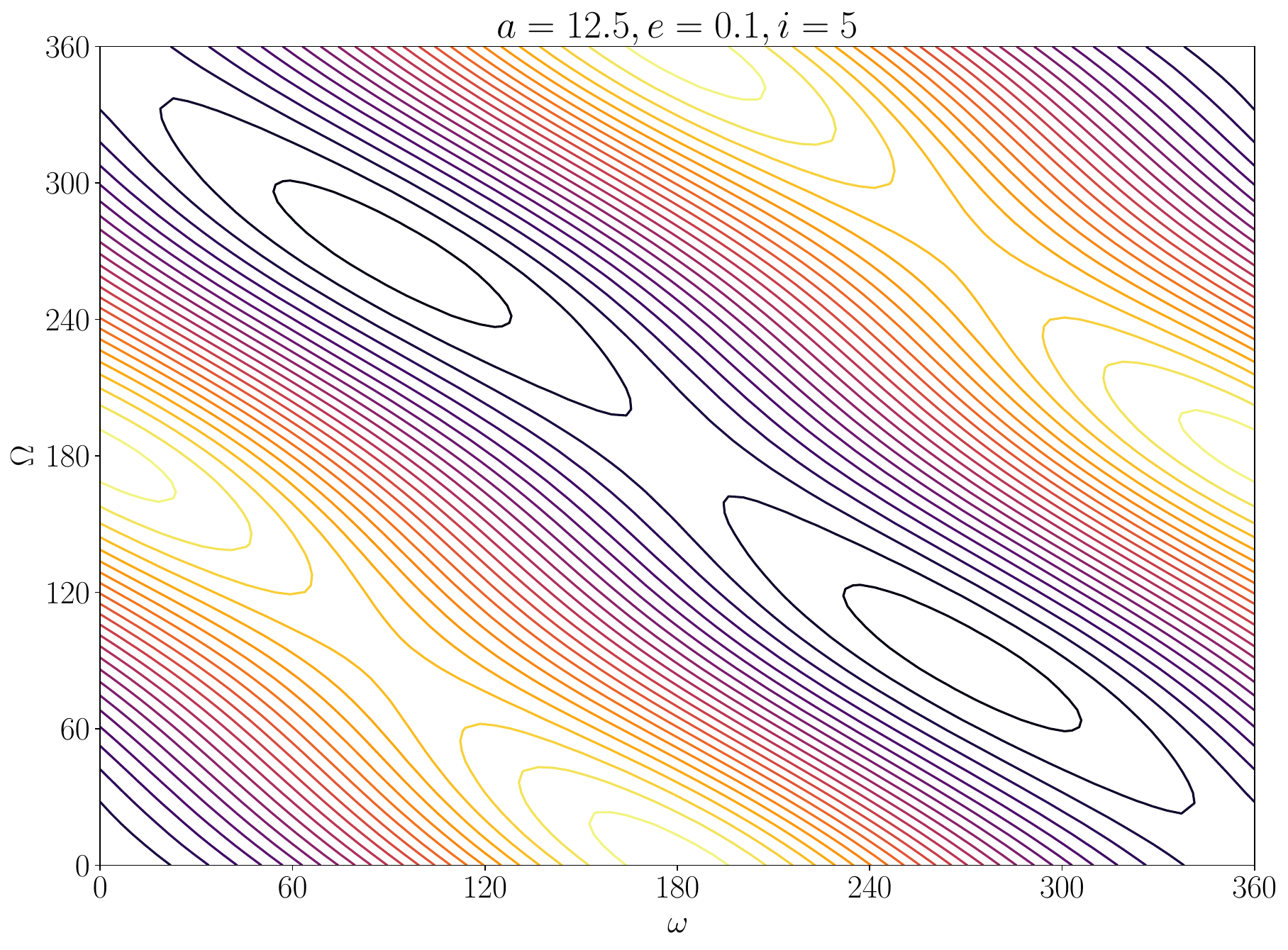}  \\
		\includegraphics[width=0.45\linewidth]{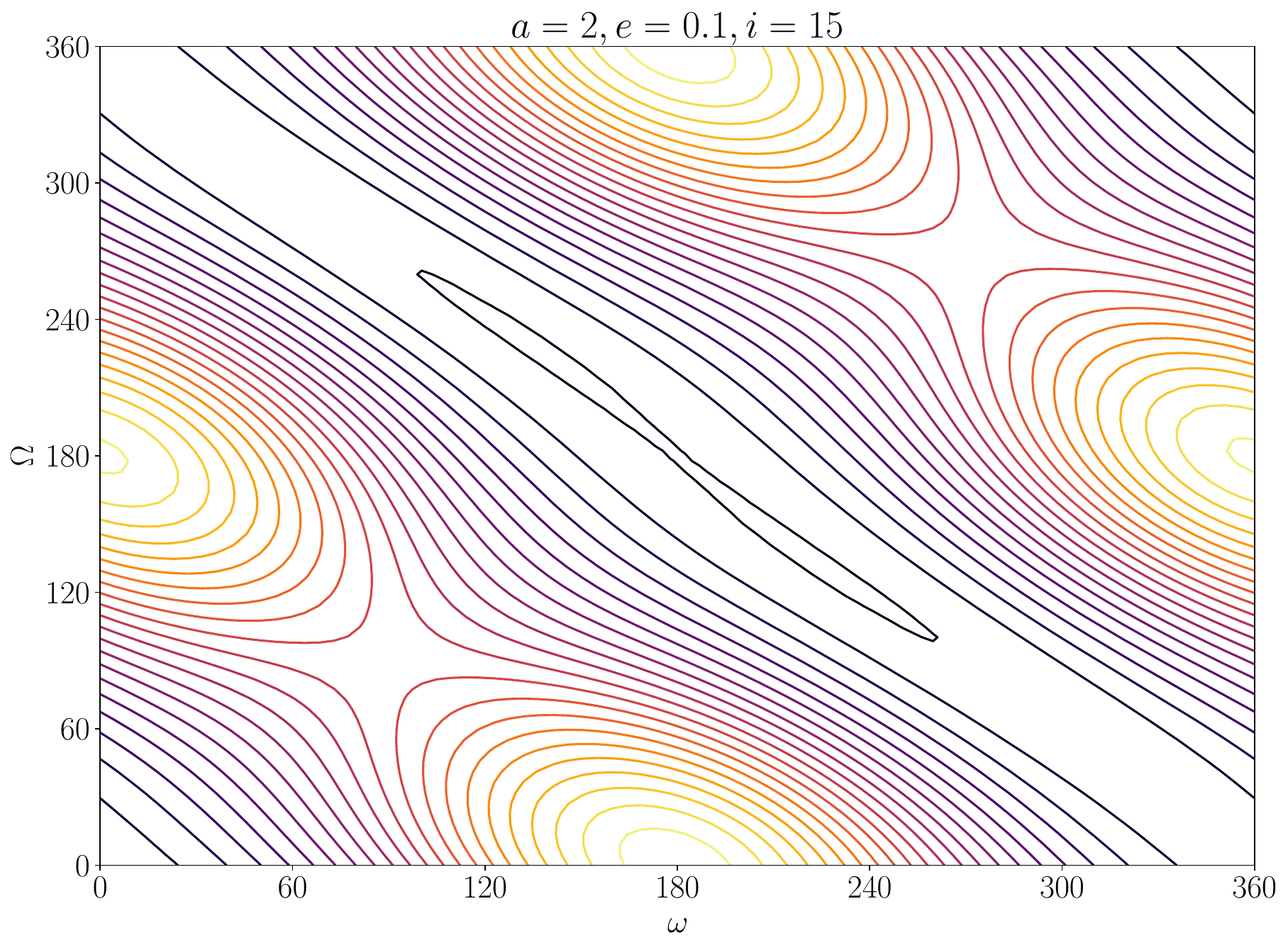} & \includegraphics[width=0.45\linewidth]{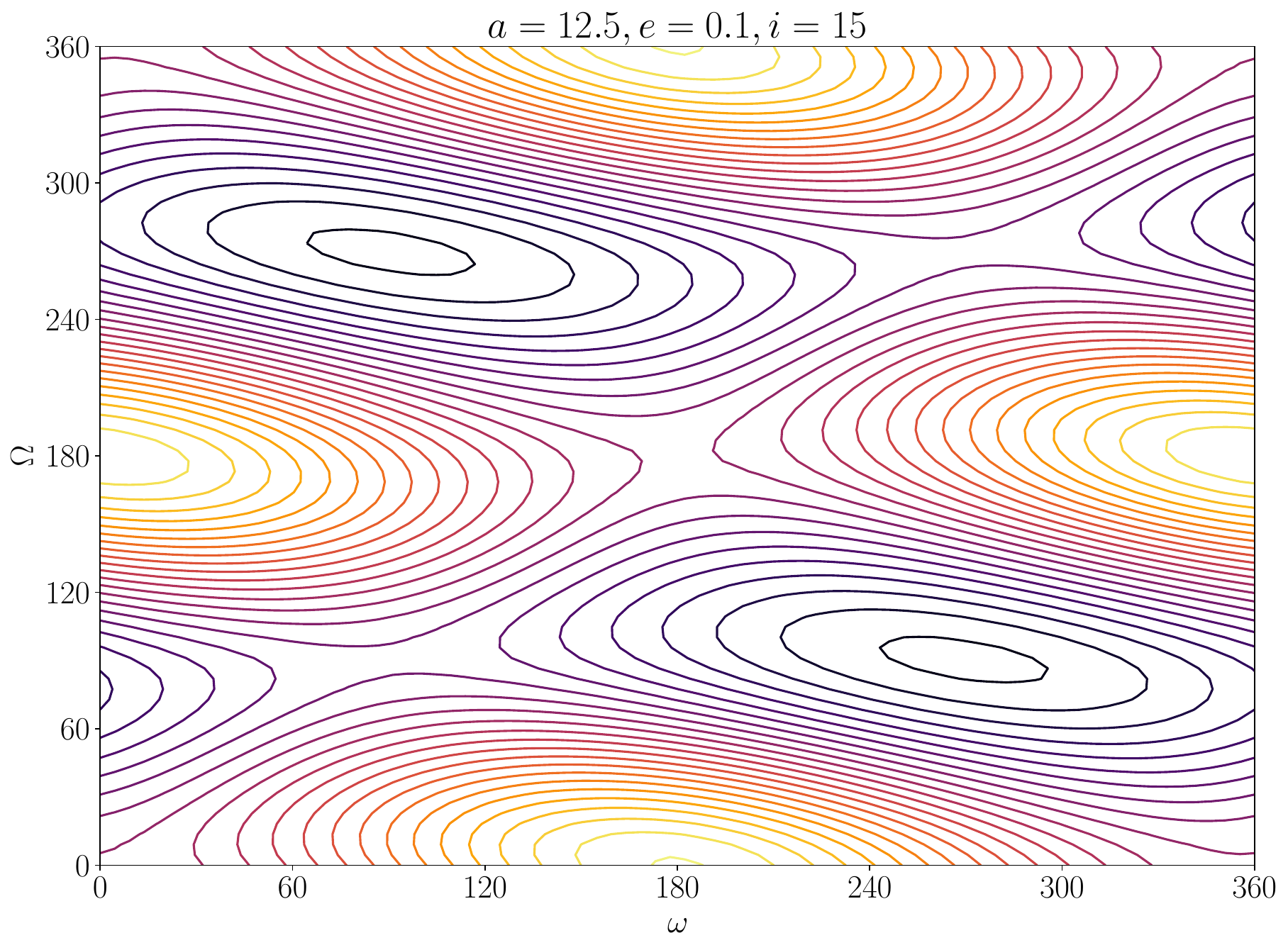} \\
		\includegraphics[width=0.45\linewidth]{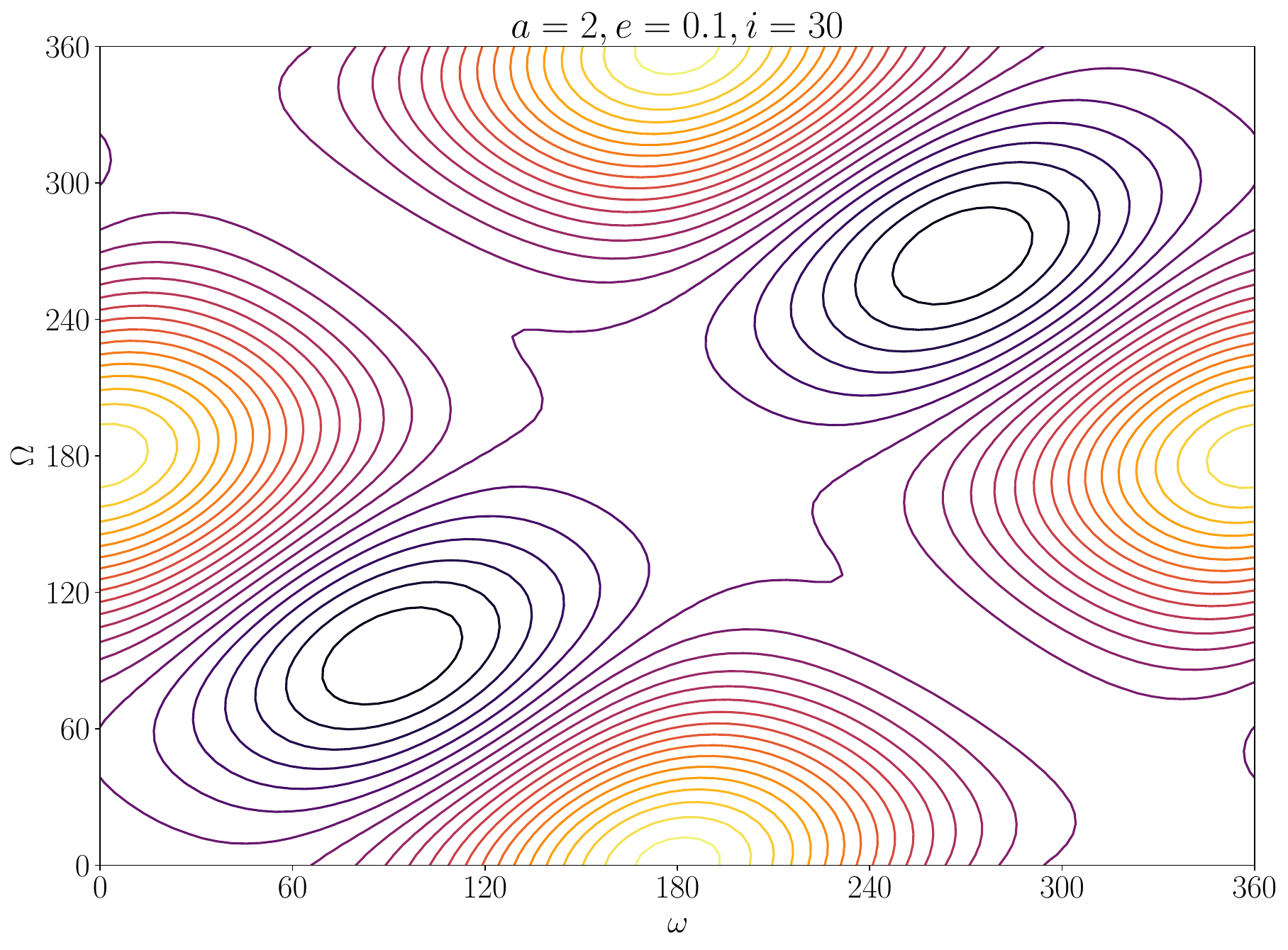} & \includegraphics[width=0.45\linewidth]{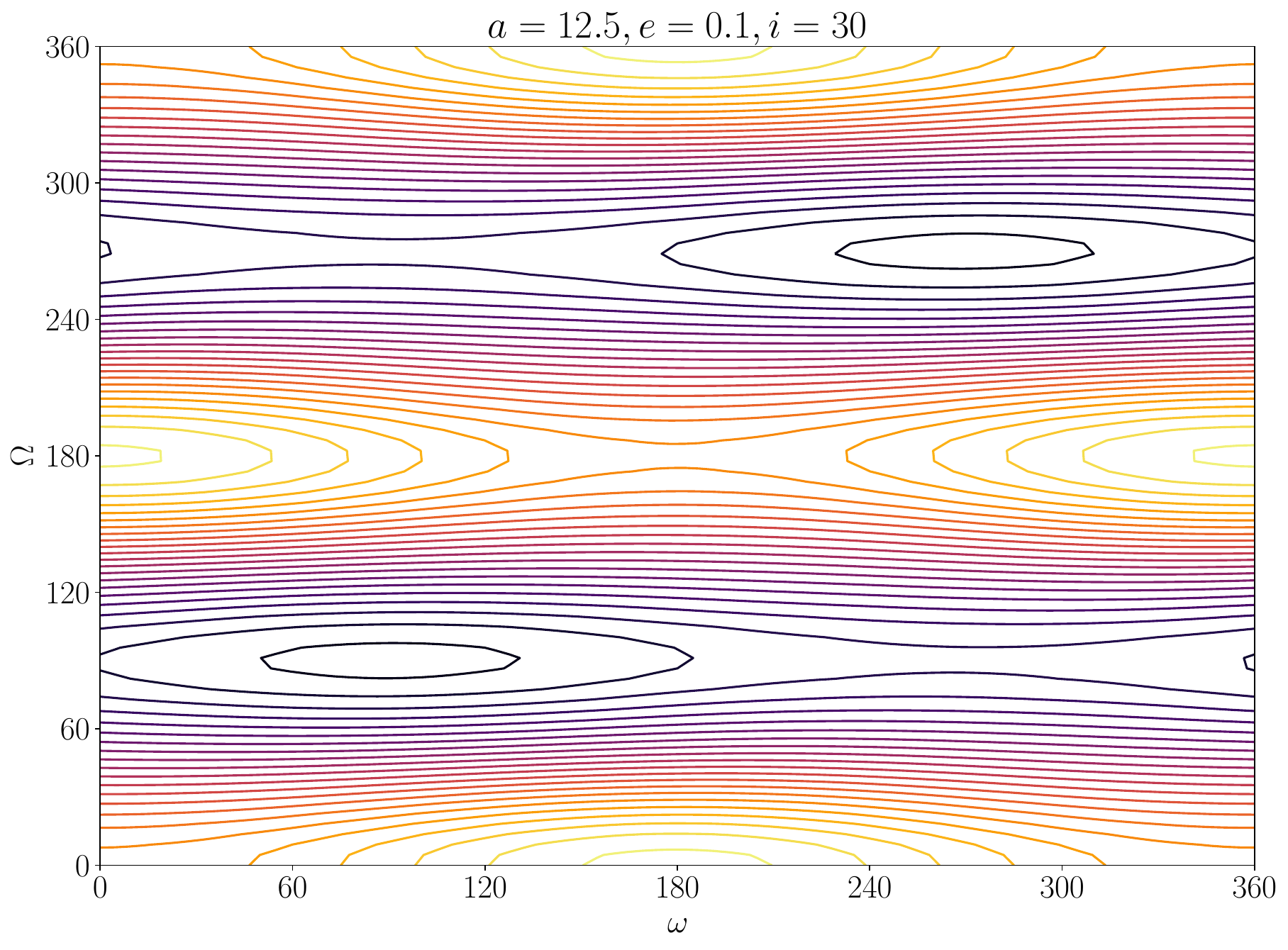} \\ \includegraphics[width=0.45\linewidth]{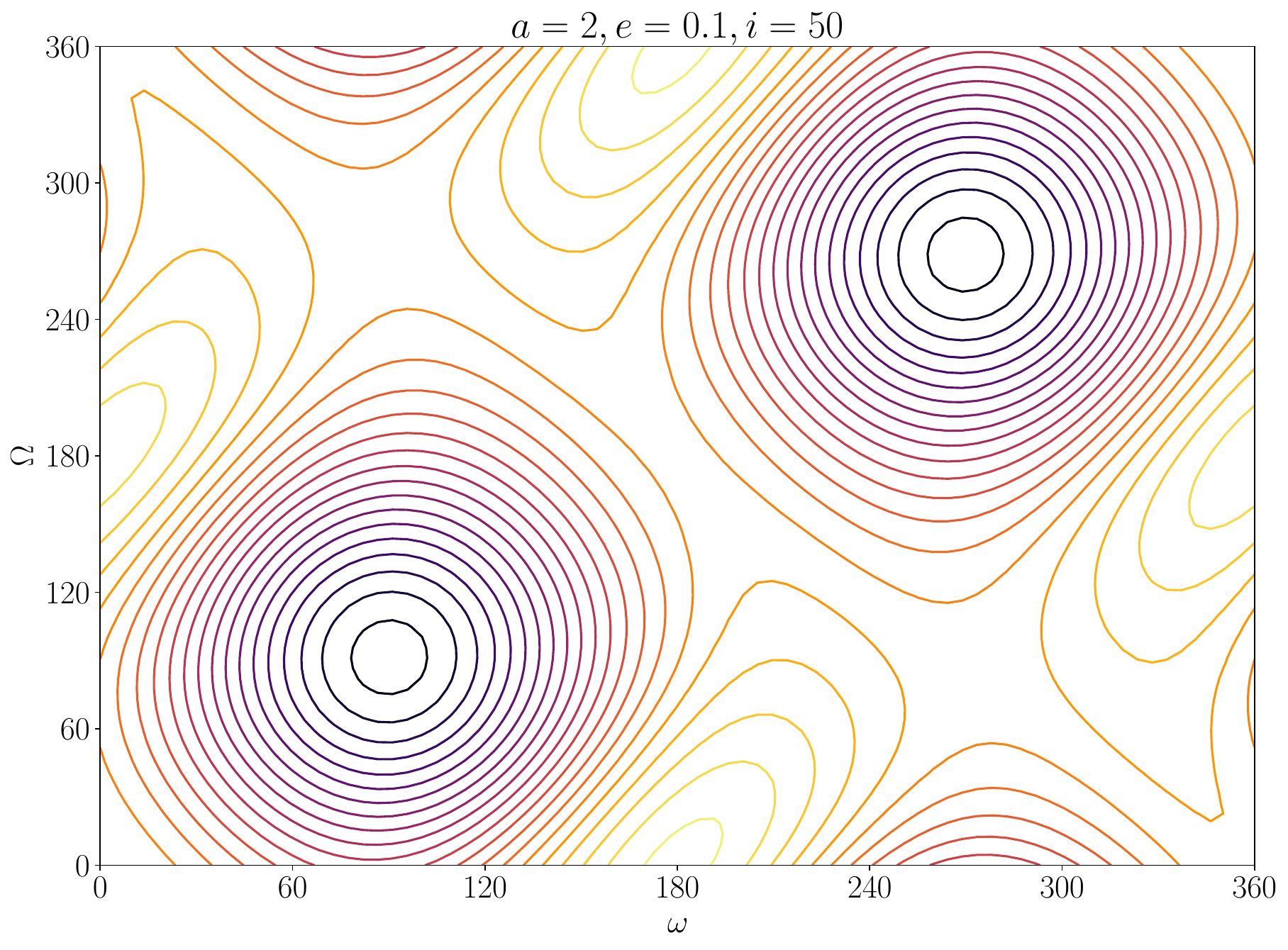} & \includegraphics[width=0.45\linewidth]{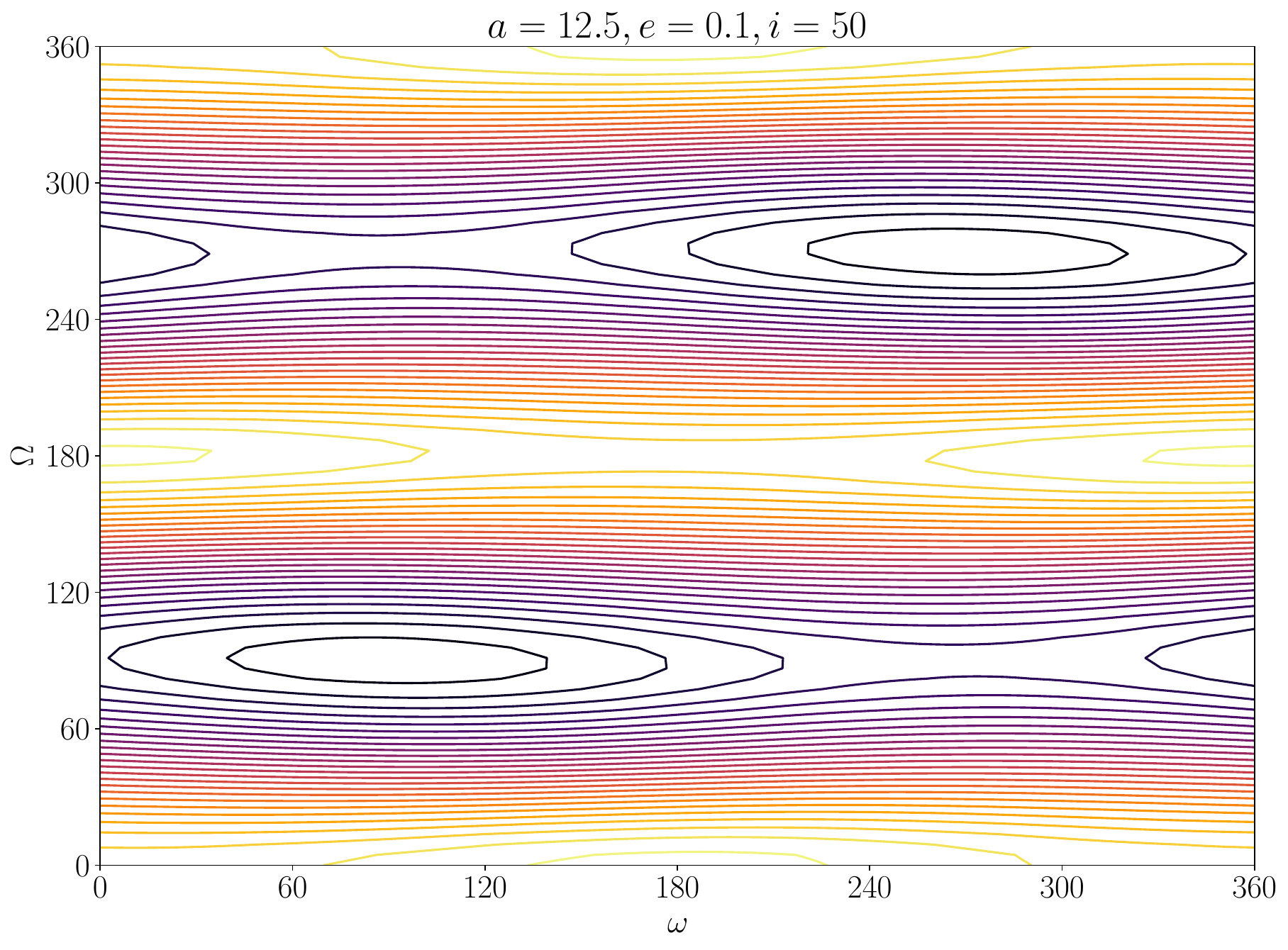}
		
	\end{tabular}
	\caption{Contour plots of $R_s(\omega,\Omega)$ for particles with $a=2$ au (left) and $a=12.5$ au (right) and for increasing inclination from top to bottom. Minima of $R_s$ are shown with dark colored lines and maxima with light colors.	In all cases it was adopted $e=0.1$.}
	\label{composedlc}
\end{figure*}

\begin{figure}
	\centering
	\includegraphics[width=0.5\textwidth]{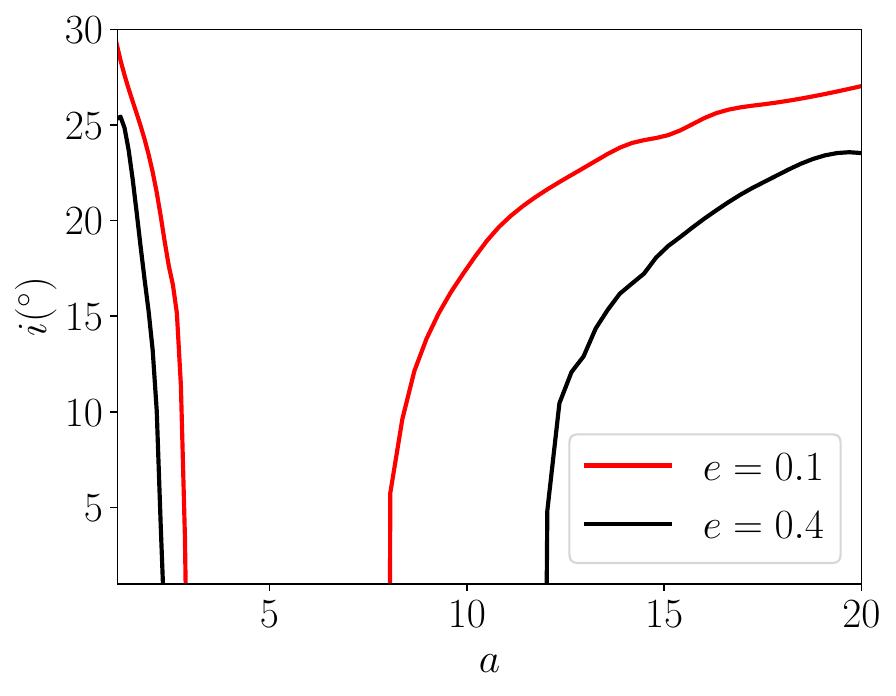}
	\caption{Inclination where minima of the $R_s$ switch from $\varpi=0\degree$ to $(\omega,\Omega) = (90\degree,90\degree)$ as a function of the particle's semi-major axis and for two values of the particle's eccentricity. Results for semi-major axes near the planet were not plotted due to the non-secular nature of the evolutions.}
	\label{fig:2fideos}
\end{figure}

The secular evolution of the asteroid is confined to hypersurfaces of constant $R_s(e,i,\omega,\Omega)$ and the representations of trajectories in that space are impossible. 
We can explore some properties fixing two variables.
For example, 
in Fig. \ref{composedlc}  we show level curves of $R_s(\omega,\Omega)$ for interior (left columns) and exterior (right column) small bodies with increasing inclination from top to bottom.  Although these plots correspond to defined values of $(a,e,i)$, more or less similar results are obtained for a wide range of values of these orbital elements.
If the asteroid's $(e,i)$ remains constant, which in general is not the case, then trajectories in the plane $(\omega,\Omega)$ will follow oscillations around the extrema of $R_s$.  Note that for low inclination orbits (first row) the minima of $R_s$ verify $\varpi\sim 0\degree$ while for high inclinations orbits ($i\geq 30\degree$ two lower rows) minima are specifically located at $(90\degree,90\degree)$ and $(270\degree,270\degree)$. There is also a transition region that we show for $i= 15\degree$ at the second row. There is a difference between interior and exterior case: while for a high inclination  interior asteroid the minima are well defined at $(90\degree,90\degree)$ and $(270\degree,270\degree)$, for the high inclination exterior asteroid $\Omega= 90\degree, 270\degree$ is a stronger condition than the particular value of $\omega$.
From these figures we cannot assure these extrema are the equilibrium points in all space because  we have assumed constant $(e,i)$ for the asteroid which, in general,  is not the case. Anyway, the figures show the relevance of the symmetric configurations for the location of minima and maxima of $R_s$, which are associated with the equilibrium configurations. The switch of extrema from $\varpi\sim 0\degree$ to defined values of $(\omega,\Omega)$ occurs for different inclinations according to $(a,e)$ as it is shown in Fig. \ref{fig:2fideos}. 
In this figure, we numerically calculated the minima of $R_s$ and register the minimum inclination for which the minima of $R_s$ sets at $(90\degree,90\degree)$, as function of the asteroid semimajor axis and for two values of the asteroid's eccentricity.
In all situations when $i\gtrsim 30\degree$ the extrema are at $(90\degree,90\degree)$ and $(270\degree,270\degree)$.
As it was explained in \cite{Gallardo2025} properties of $R_s$ regarding extrema locations are independent of planetary mass because it just appears as a multiplicative factor.

\begin{figure*}
	\centering
	\begin{tabular}{ccc}
		\includegraphics[width=0.33\linewidth]{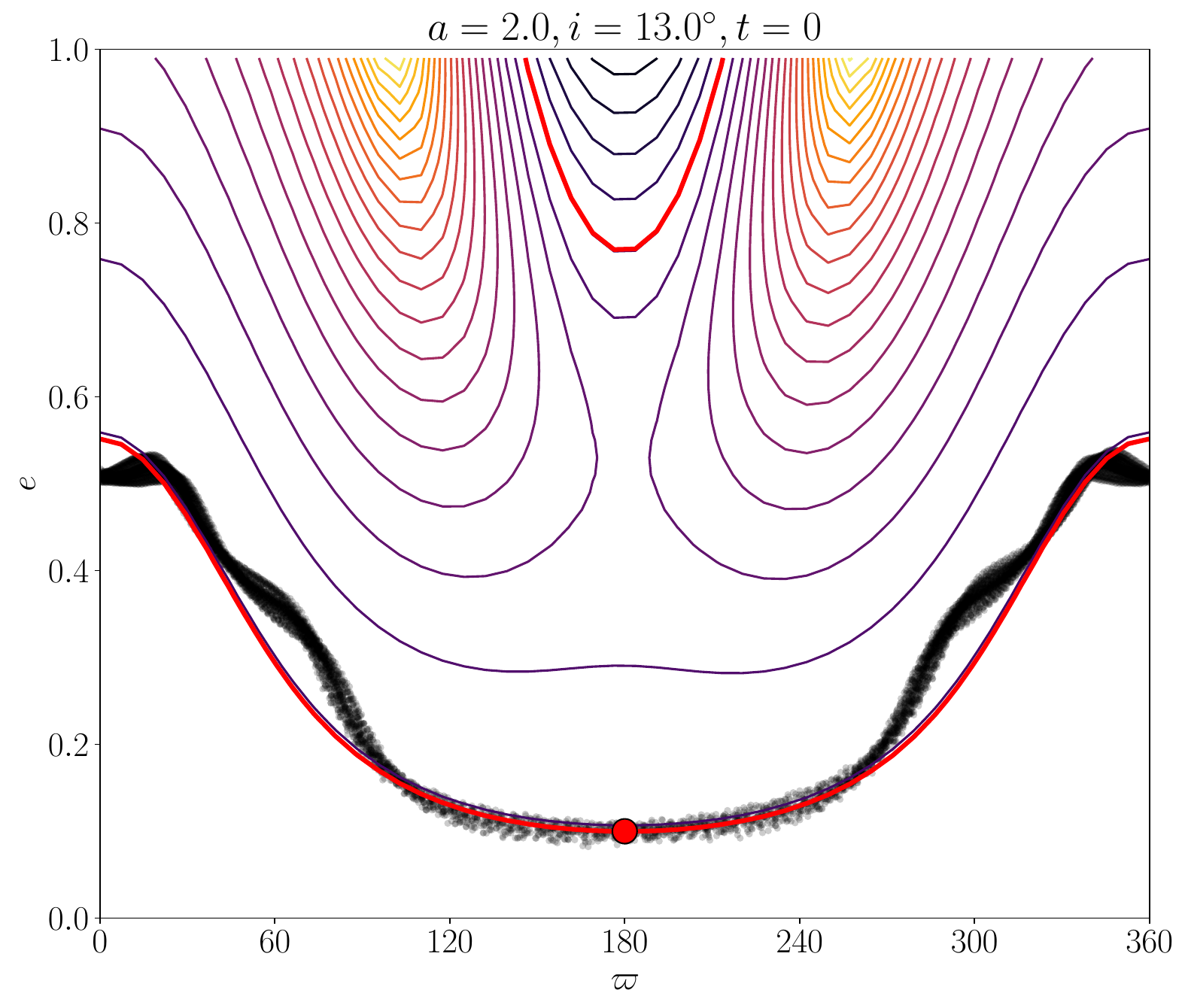} & \includegraphics[width=0.33\linewidth]{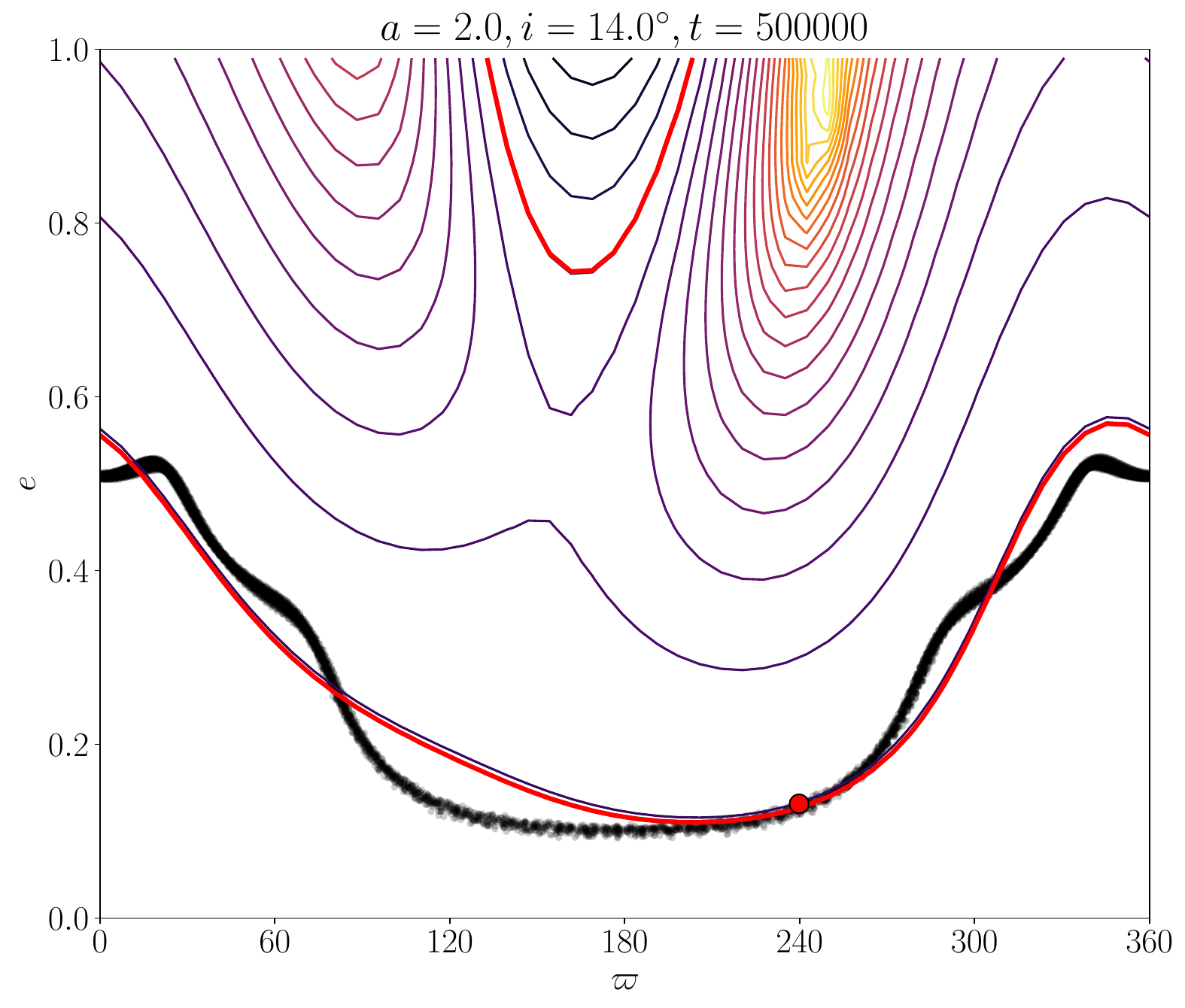}&
		\includegraphics[width=0.33\linewidth]{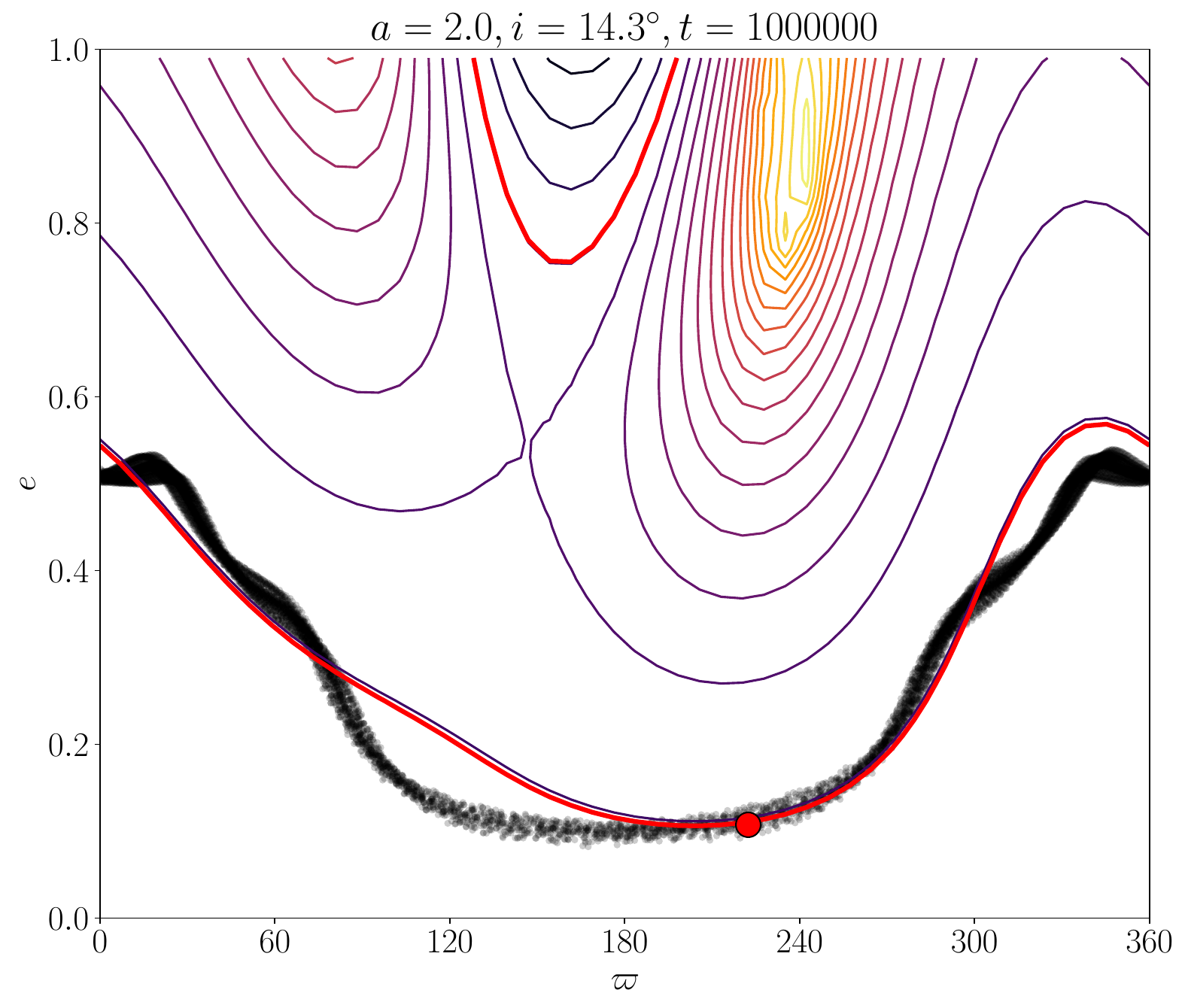} \\
		\includegraphics[width=0.33\linewidth]{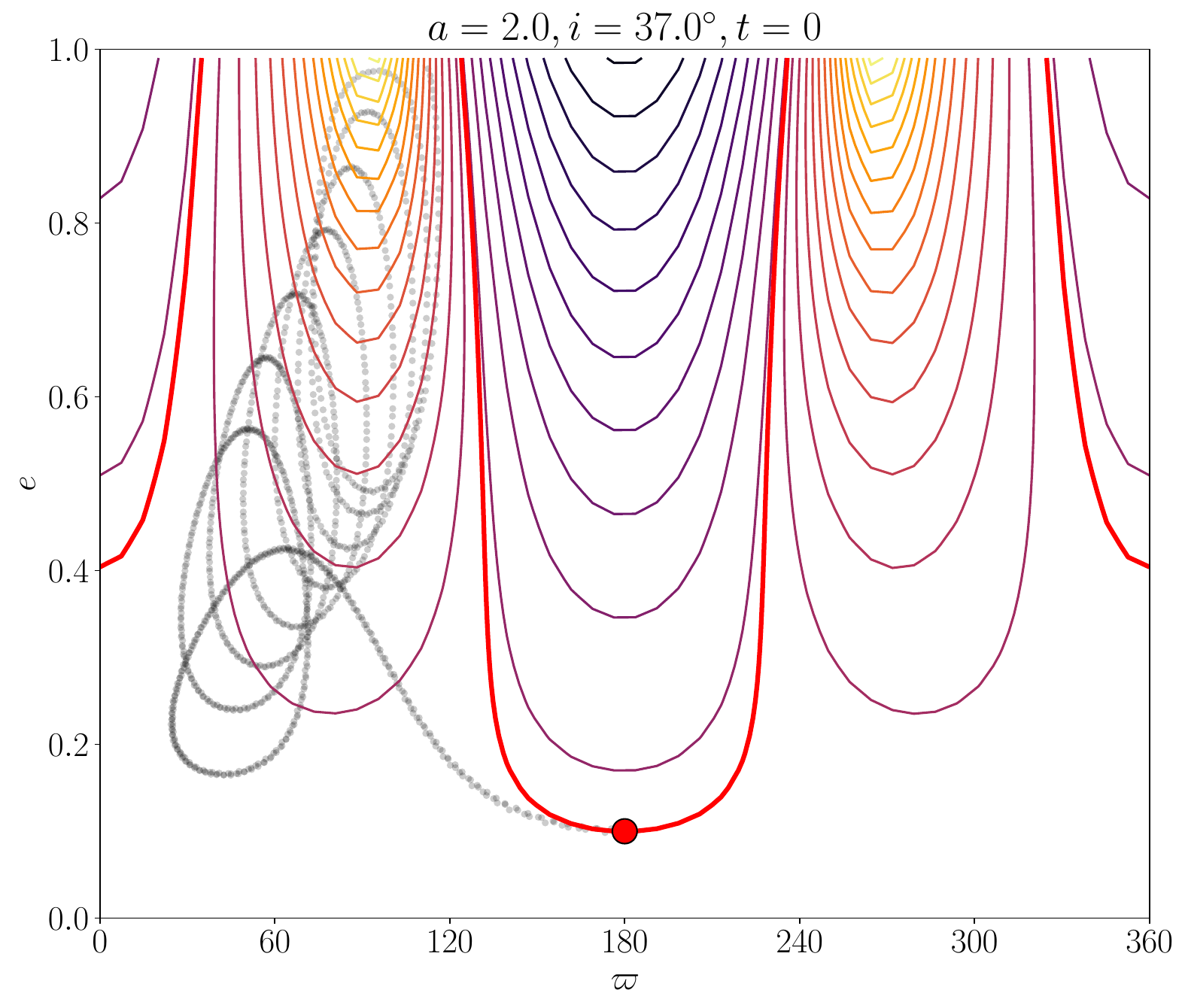} & \includegraphics[width=0.33\linewidth]{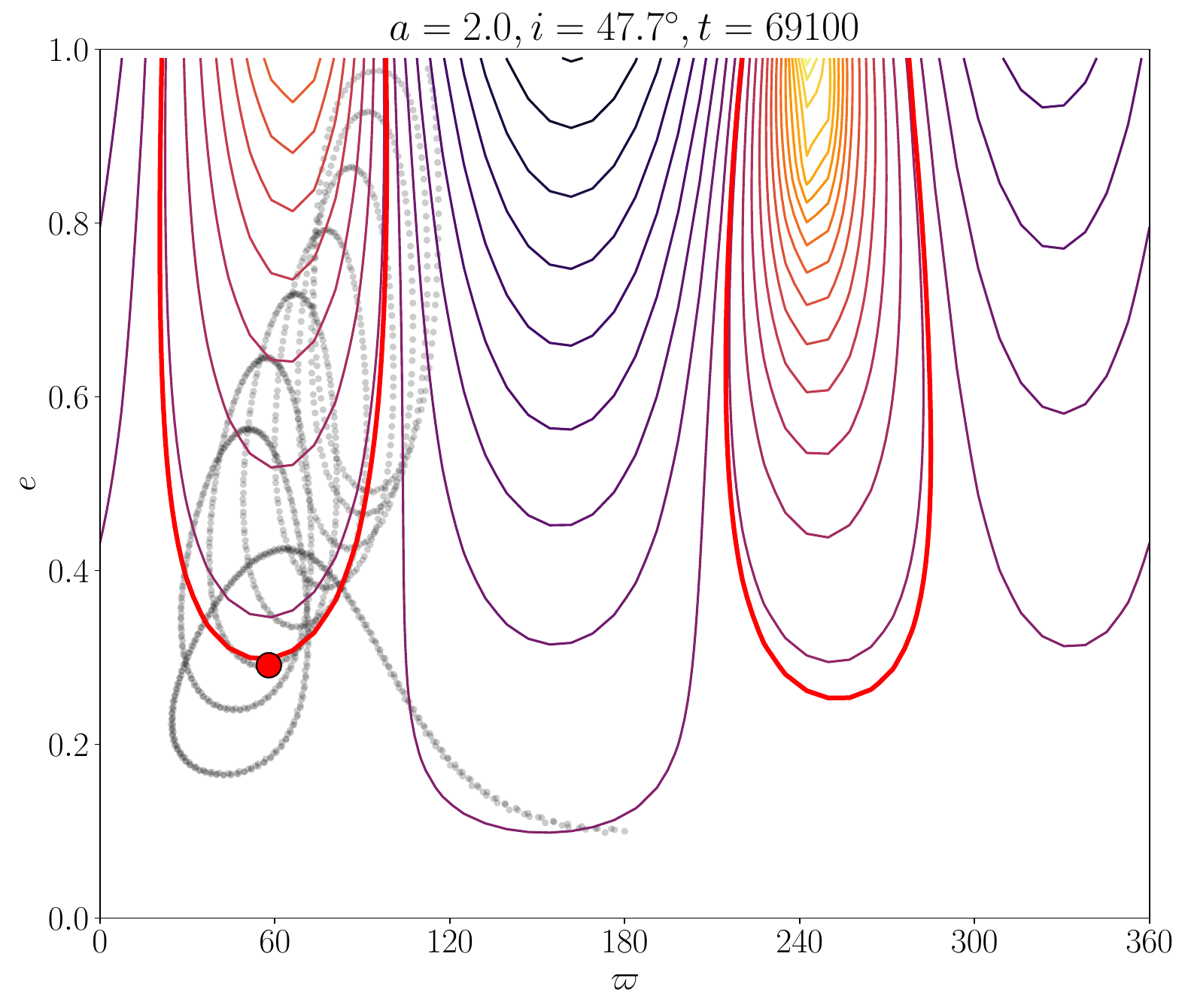} & \includegraphics[width=0.33\linewidth]{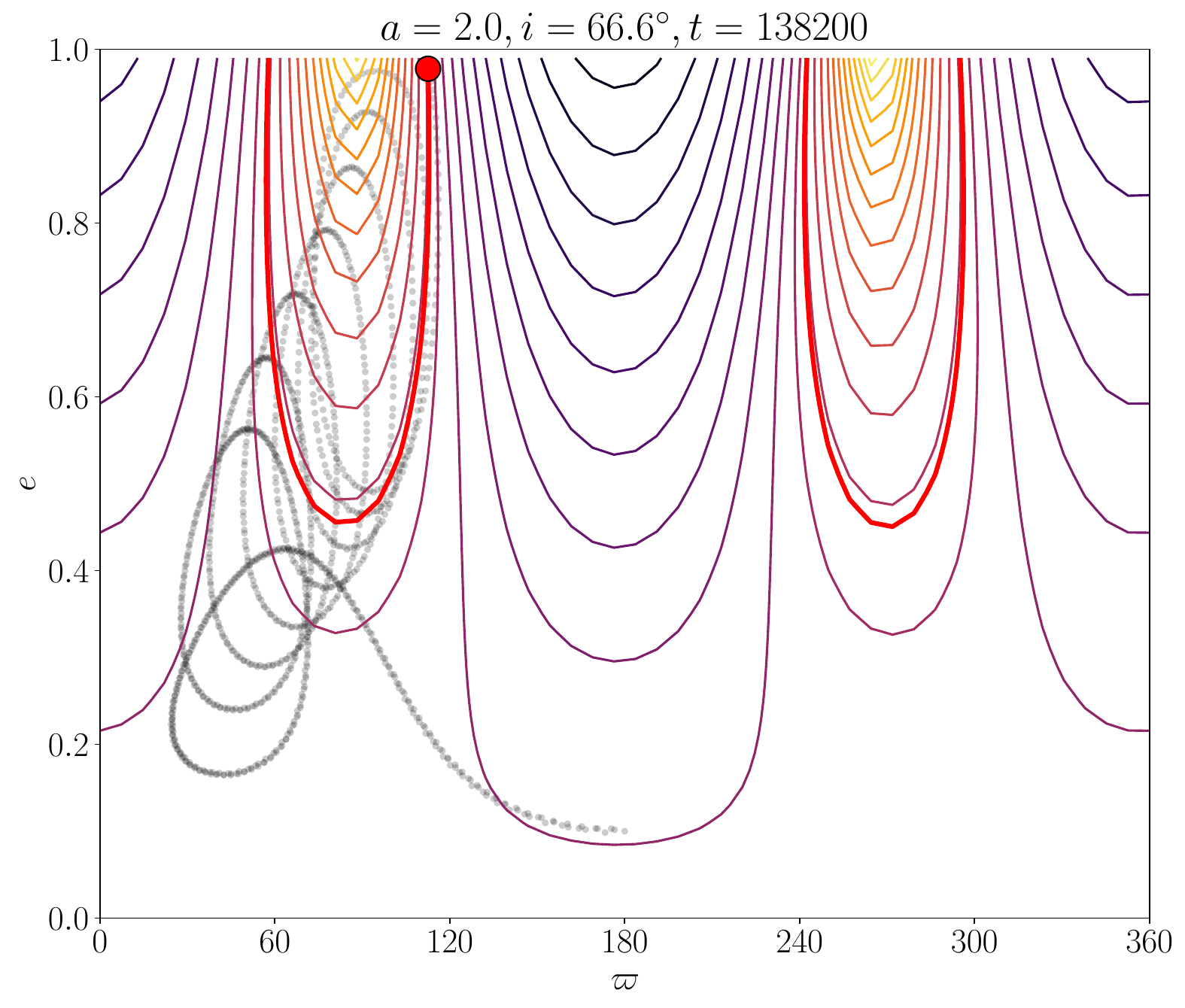}
	\end{tabular}
	\caption{Three instants of the secular evolution of two interior small bodies with $a=2$ au, initial $e=0.1$, $\Omega=\omega = 270\degree$ and initial $i=13\degree$ (upper panels)
		and initial $i=37\degree$ (bottom).
		Changing contour plots of $R_s(\varpi,e)$ are shown in all panels. The red contour correspond to the initial value of $R_s$ for each particle (indicated with a red point at each instant). Snapshots of the location of the particle taken each 100 yrs are shown as tiny black dots.}
	\label{2paths}
\end{figure*}

The differences in dynamical behavior between both configurations of low and high inclination orbits can be seen if we compare level curves of $R_s(\varpi,e)$
in the two situations.
Figure \ref{2paths} shows two cases showing the time evolution of two interior particles  in the plane $(\varpi,e)$ computed by the numerical integration of the full equations of motion planet-particle. They are accompanied by level curves of $R_s$ showing in red the actual $R_s$ level for the particle at each instant. The particles remain stuck to each level curve while they change with time. 
The value of $R_s$ for each particle is constant over time but the level curves change according to the variable orbital state.
We show three different moments in their dynamical evolution. In the low inclination case (top panels), the contour lines show no connections between low and high eccentricities after 1 million years of dynamical evolution, while in the high inclination case (bottom panels), they do, and on a shorter time scale. This suggests that a particle with a small initial inclination will evolve along changing $R_s$ contour lines that keep the particle with a low $(e,i)$. In contrast, for initial high $i$ orbits, the particles evolve along level curves of $R_s$ that connect to very high $(e,i)$ orbits generating instability. Note that the high inclination particle goes to large eccentricities with oscillations of $\varpi$ around some point approximately located at $\varpi\sim 90\degree$, we will discuss this point in the next subsection.
Note also that the low-inclination particle evolves over a level curve with a center at $\varpi=0\degree$ in agreement with the level curves shown in Fig. \ref{composedlc}, while the high-inclination particle does not follow that behavior.
This reinforces the idea that for low inclination orbits the proximity between the line of apsides $(\Delta \varpi \sim 0\degree)$ is determinant for the equilibrium, while for high inclination this rule is broken. As we have shown in Fig. \ref{fig:2fideos}, this change occurs for some $i<30\degree$ which its exact value depends on $(a,e)$. This behavior is very similar to the one obtained for the planetary case in \cite{Gallardo2025}.

\subsection{Proper frequencies and dynamical paths}\label{pfadp}

Using numerical integrations of the full equations of motion of particles perturbed by the giant planet using codes EVORB \citep{2002Icar..159..358F} and REBOUND \citep{rebound,2015MNRAS.446.1424R,reboundwhfast,wh} we analyzed the time evolution of variables $(k,h)$ and $(q,p)$ in order to obtain the main frequencies present and identify the proper frequencies.
We obtained the frequencies as function of $(e,i)$ for some values of semimajor axes considering interior and exterior orbits respect to the planet avoiding chaotic regions and MMRs. To obtain the frequencies we used the Lomb-Scargle method \citep{lomb1976,scargle1982,vanderplas2018} keeping just the one with the greatest power in each pair of variables $(k,h)$ and $(q,p)$. For low-inclination orbits, these are typically the proper frequencies, that we will call $g$ and $f$. However, for high-inclination orbits, the main frequencies are often identified as a harmonic or a combination of proper frequencies.

\begin{figure*}[t]
	\centering
	\begin{tabular}{c c}
		\includegraphics[width=0.49\textwidth]{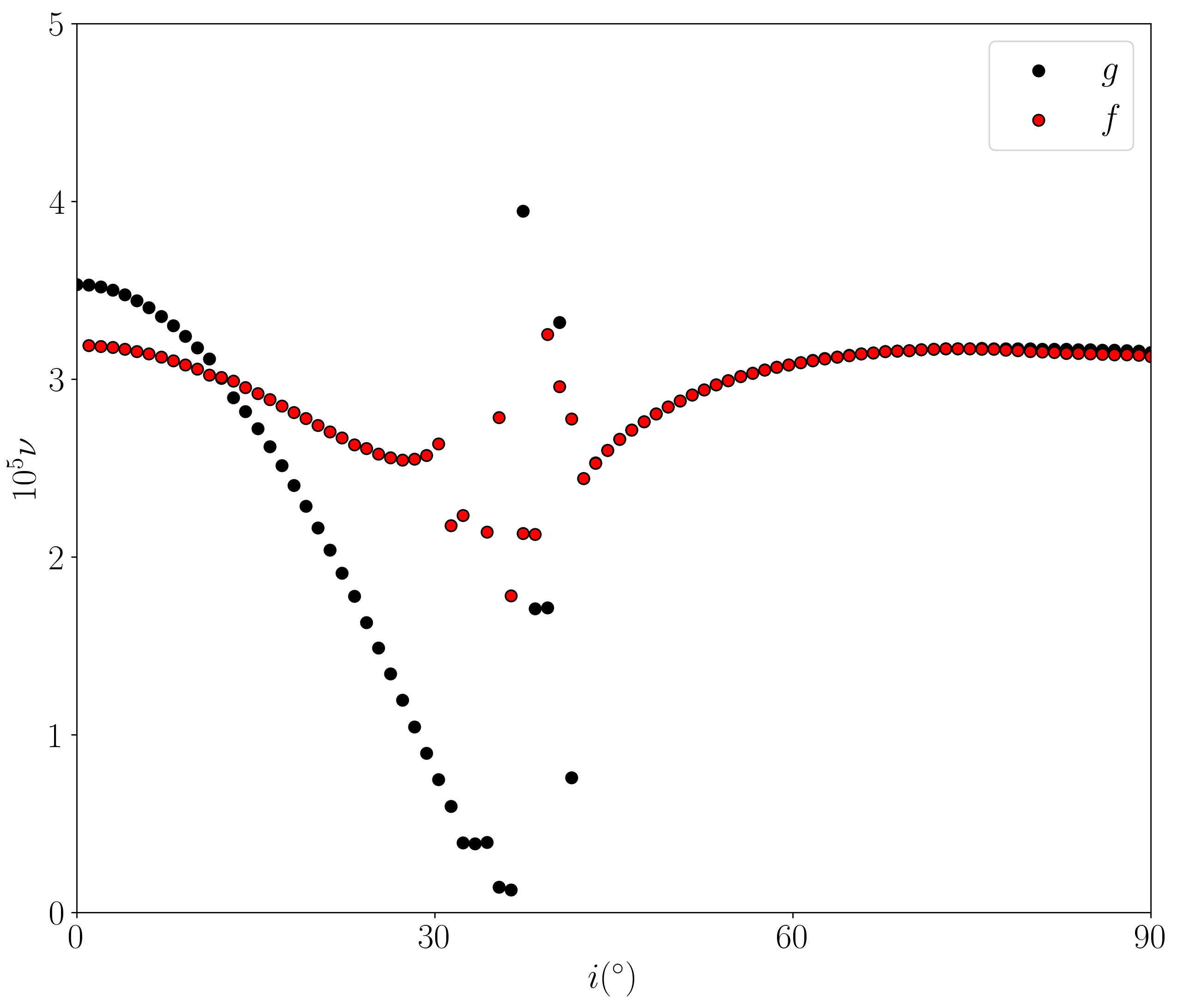}  & \includegraphics[width=0.5\textwidth]{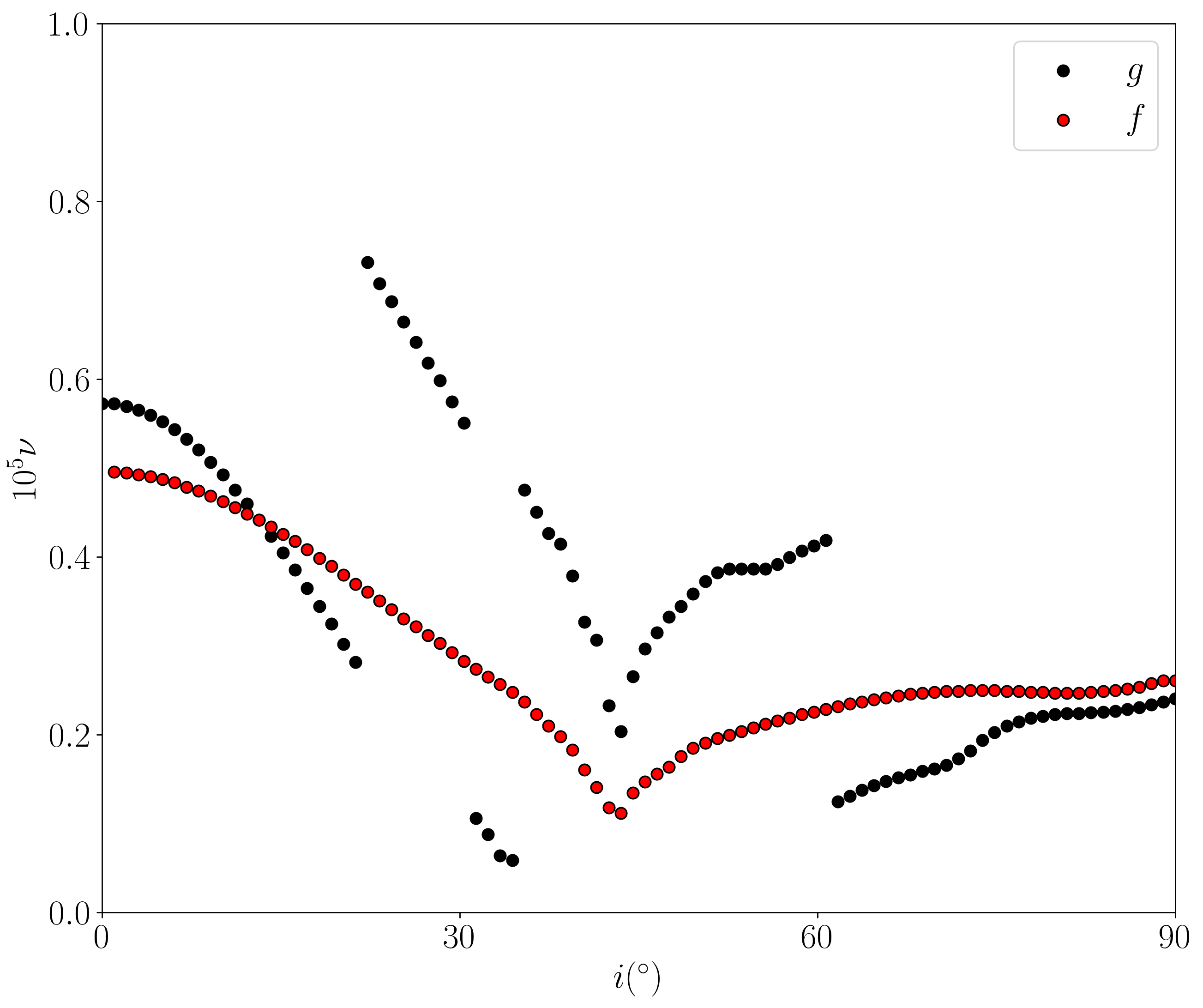}  \\ 
	\end{tabular}
	\caption{Main frequencies identified as proper frequencies in units of $1/yr$ associated with $(k,h)$ in black and $(q,p)$ in red for different initial inclinations for interior particle with $a=2$ au and $e=0.4$ (left) and exterior particle with $a=12.5$ au and $e=0.1$ (right). In all cases initial $\Omega=270\degree$ and $\omega=90\degree$. }
	\label{frecs}
\end{figure*}

The evolution of orbital elements is highly dependent on initial conditions, and can be described in terms of the frequencies associated with the evolution of the $(k,h)$ and $(q,p)$ pairs. Following \cite{Gallardo2025} we focused in  how frequency varies with initial inclination. For example, taking initial $\omega=90^\circ$ and $\Omega = 270^\circ$ we get the general behavior shown in Fig. \ref{frecs}, left panel for an interior particle and right panel for exterior one.
We have two well behaved frequencies for low inclination orbits, until at $i\sim 34\degree$ the frequency associated with the $(k,h)$ evolution ($g$) approaches zero, instabilities arise and the particle is ejected. For higher initial inclinations ($i>40\degree$) the evolution is different depending on whether it is an interior or exterior particle. For the interior particle the most evident phenomenon is that the main frequencies present in  $(k,h)$ and $(q,p)$ are similar giving rise to the ZLK mechanism as we will see below.
For the exterior particle we note that $f$ substantially decreases and then it grows again. In this case the main frequency for $(k,h)$ verifies $g\simeq f$ but also harmonics $g\simeq 2f$ or $g\simeq f/2$ are present. We will see below that the main dynamical property in this region is oscillation of $\Omega$ associated with orbital flips. 

We want to stress that when $g\sim 0$ we have unstable evolutions with large changes in $e$. It is interesting that we have a hint to explain this dynamics looking at the simplified Lagrange planetary equation for low eccentricity orbits:
\begin{equation}
	\frac{de}{dt}\simeq C e_p m_p \sin \varpi + ...
	\label{dedt}
\end{equation}
where $C$ is constant \citep{1999ssd..book.....M}. Integrating we obtain that the eccentricity oscillates with amplitude $\Delta e \sim |C e_p m_p/\dot{\varpi}|$. So, if $\varpi$ is constant or varying very slowly  around a value different from $0\degree$ and $180\degree$ (where $de/dt=0$) the eccentricity will have large amplitude oscillations making the evolution unstable. In other words, the asteroid's elliptical orbit is not stationary except in a symmetric  equilibrium configuration ($\varpi=0\degree,180\degree$), otherwise its eccentricity will change enough to make its evolution unstable. Then, $g\sim 0$ means unstable evolution unless $\Delta\varpi=0\degree, 180\degree$. In Fig. \ref{2paths} bottom panels there is an example where $\varpi$ is oscillating around $\sim 90\degree$ generating an increase in eccentricity up to 1.
We will call \textit{critical inclination}, $i_c$, the inclination for which $g=0$, in analogy with the planetary case studied by \cite{Gallardo2025}. This situation is also analogous to the well known pericenter secular resonance that happens in the Solar System but with the peculiarity that in our case the planetary orbit is fixed. This critical inclination is different from the one defined in the context of the ZLK mechanism which refers to the minimum inclination for the existence of equilibrium points where $\dot{\omega}=0$.

\begin{figure}
	\centering
	\includegraphics[width=0.6\textwidth]{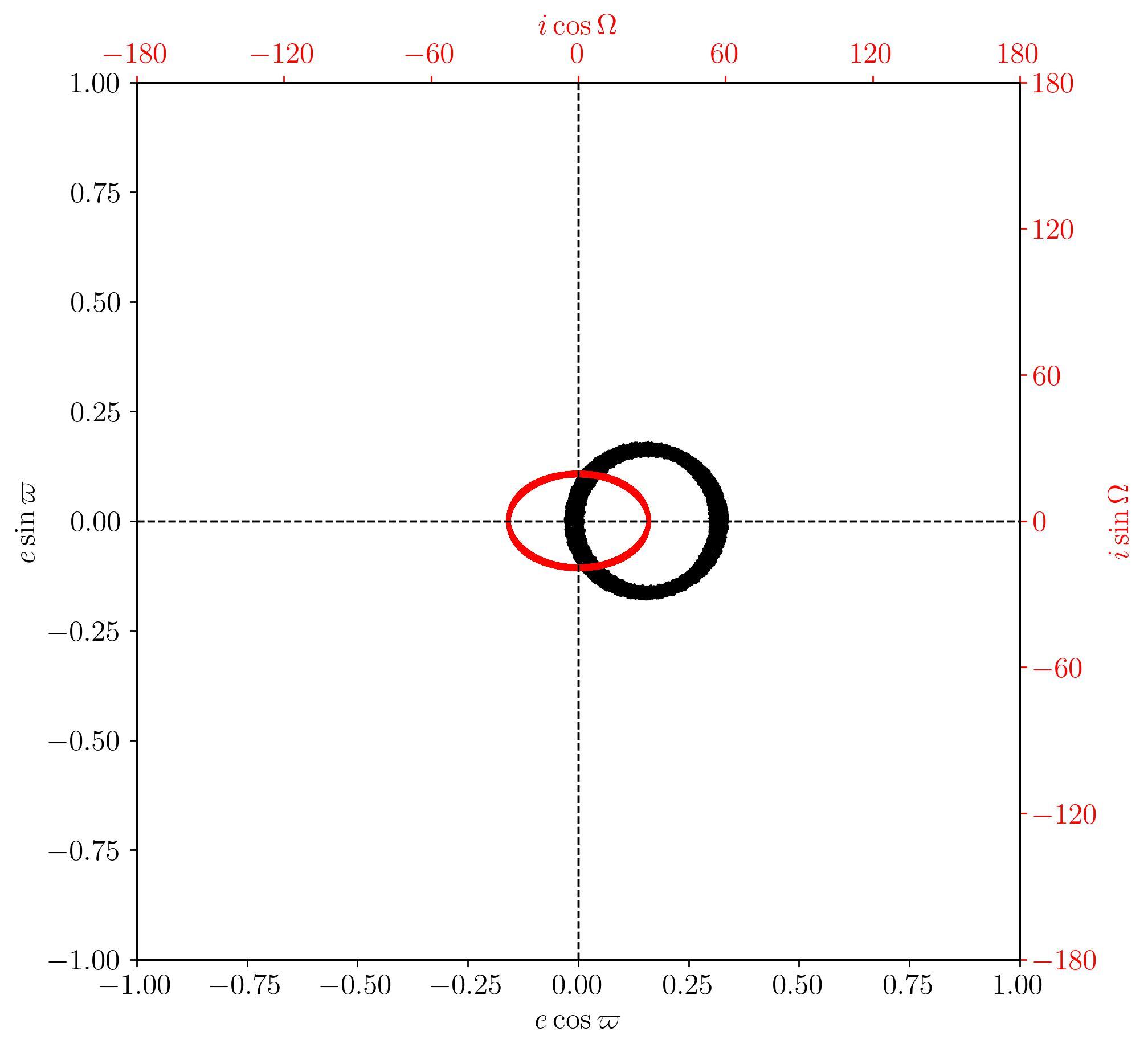}
	\caption{Typical trajectories in the planes $(k,h)$ and $(q,p)$ for a typical classic secular evolution of particle with low $(e,i)$ orbit. The inclination is in degrees.}
	\label{examplehkpg}
\end{figure}

\begin{figure*}[h!]
	\centering
	\begin{tabular}{ccc}
		\includegraphics[width=0.33\textwidth]{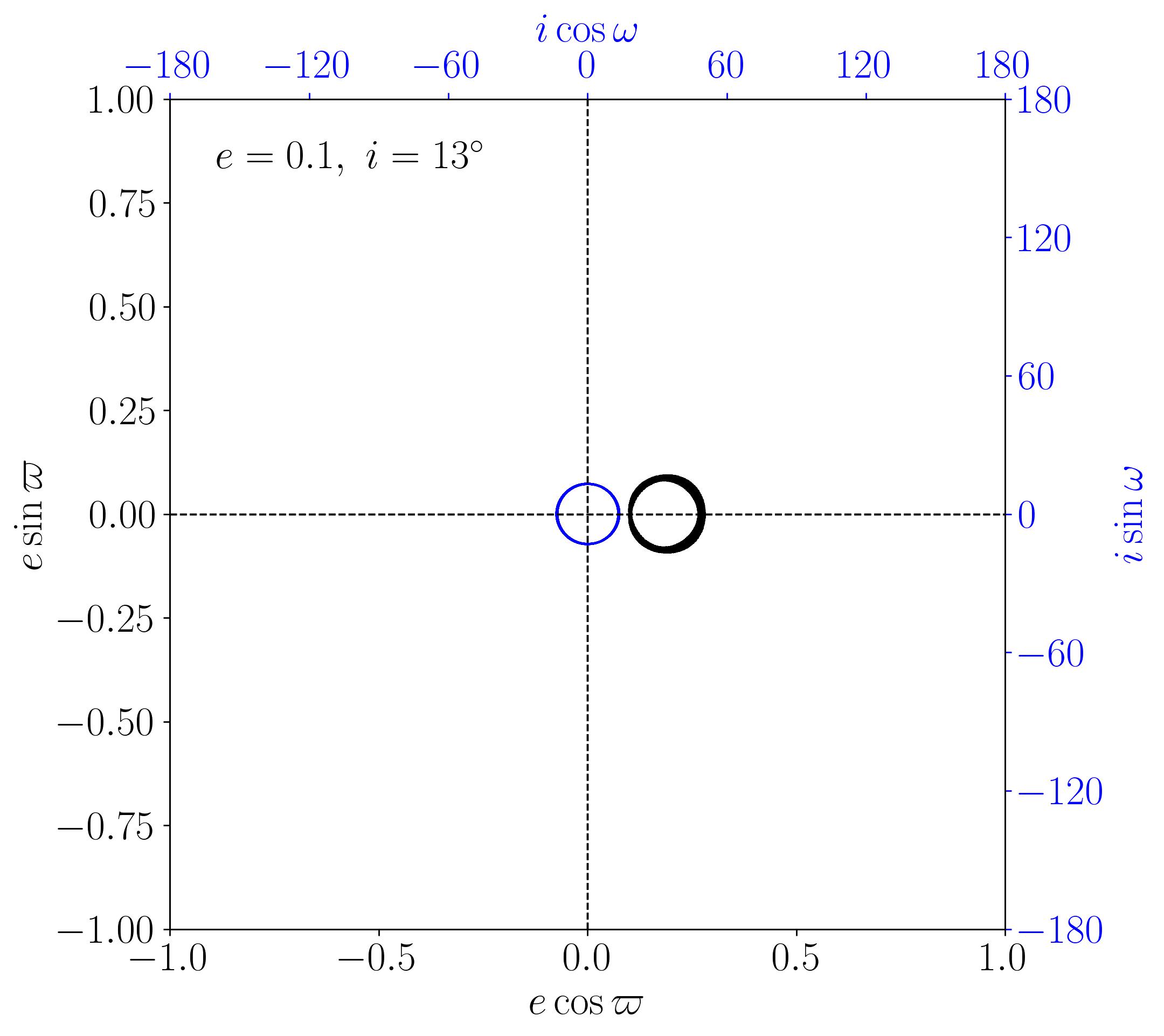}&
		\includegraphics[width=0.33\textwidth]{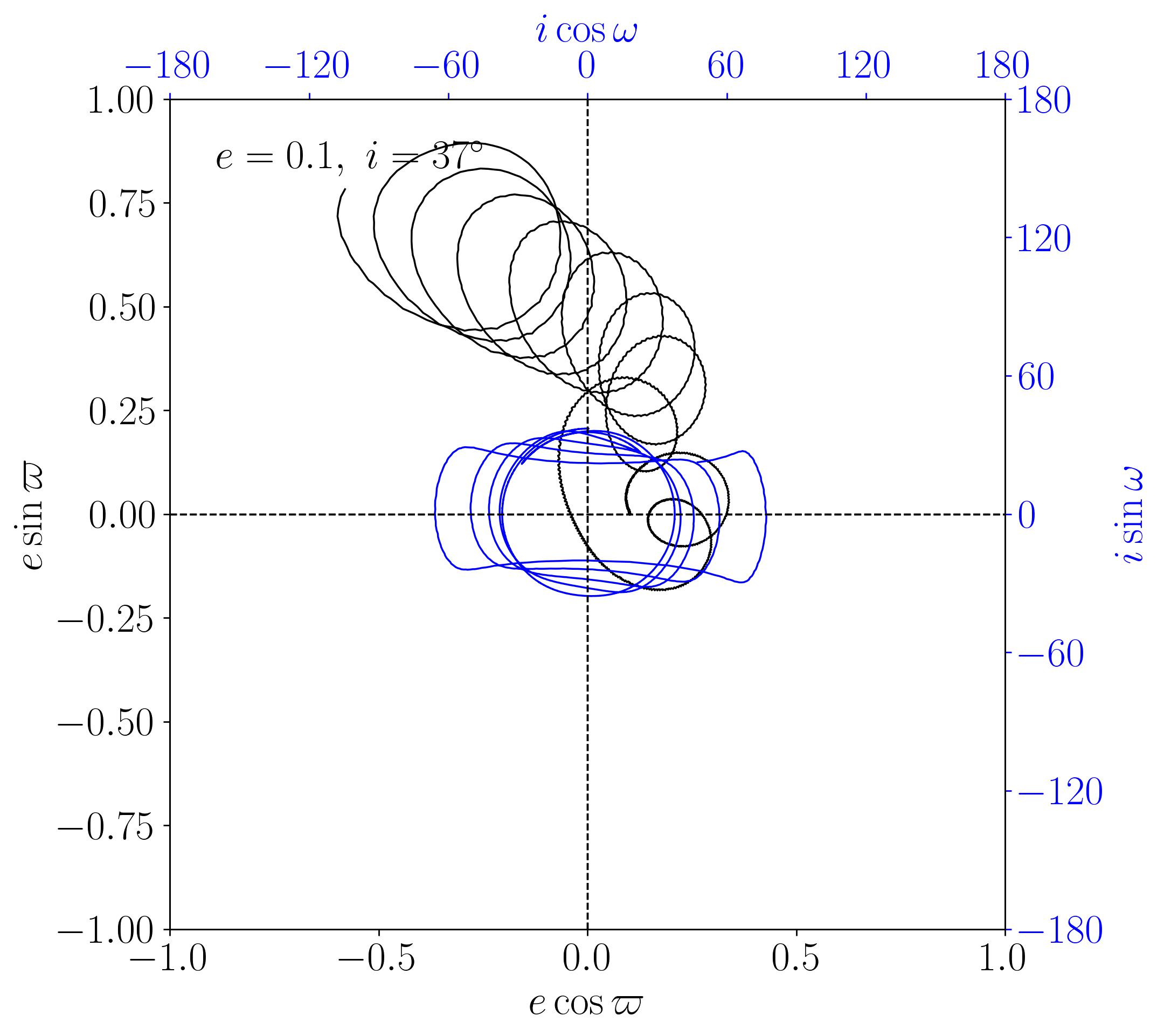}  & \includegraphics[width=0.33\textwidth]{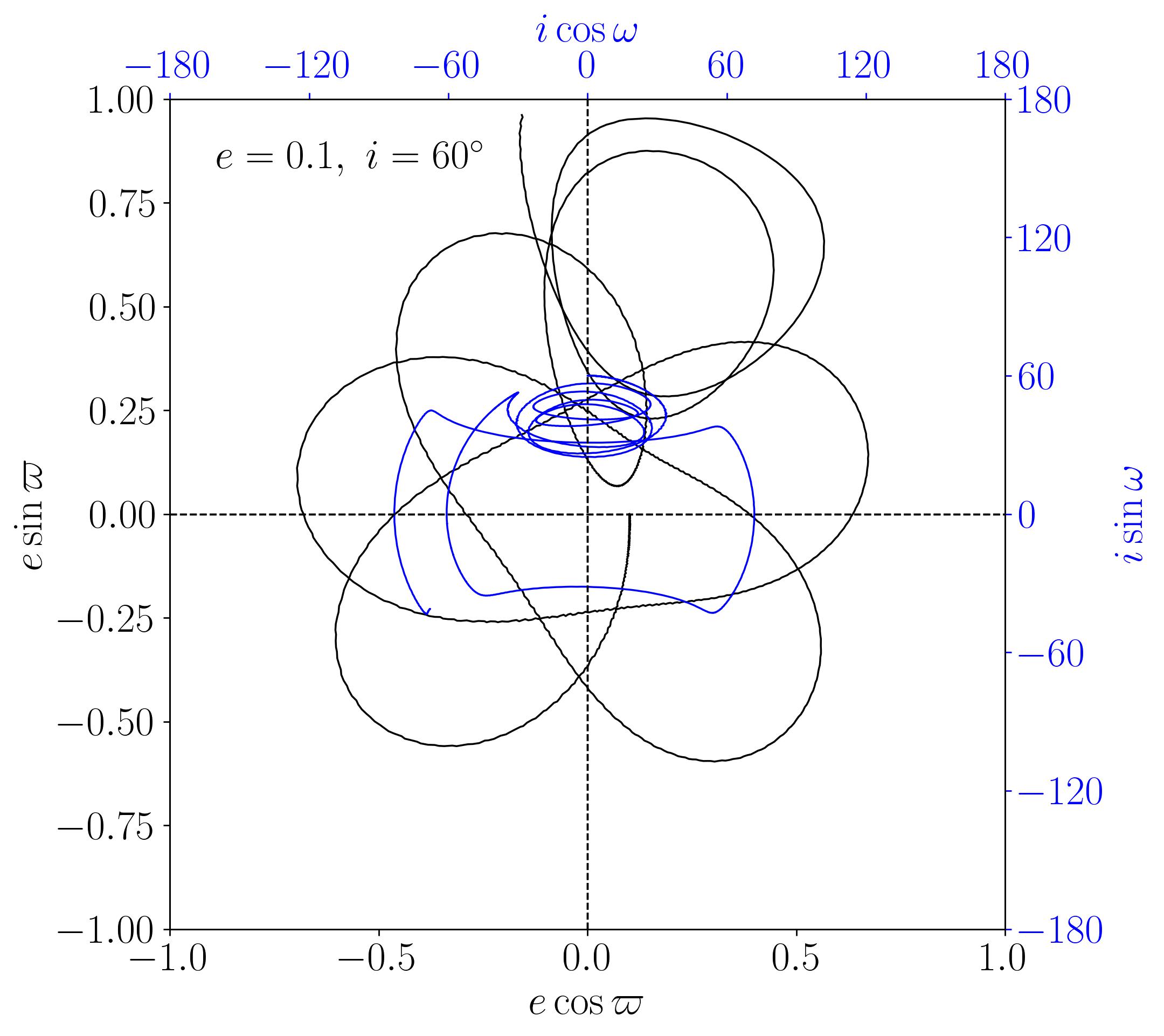}  \\
		\includegraphics[width=0.33\textwidth]{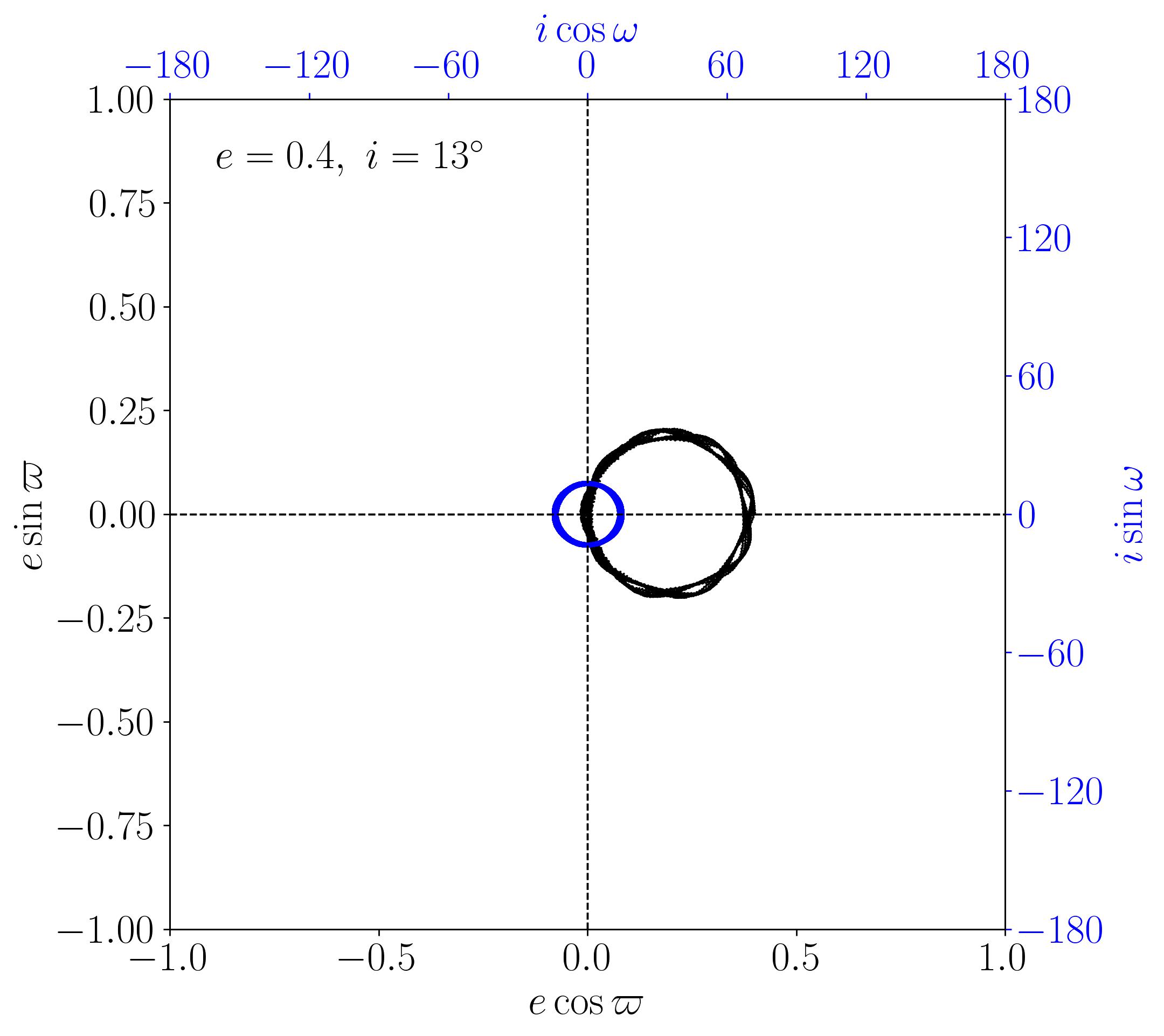}   & \includegraphics[width=0.33\textwidth]{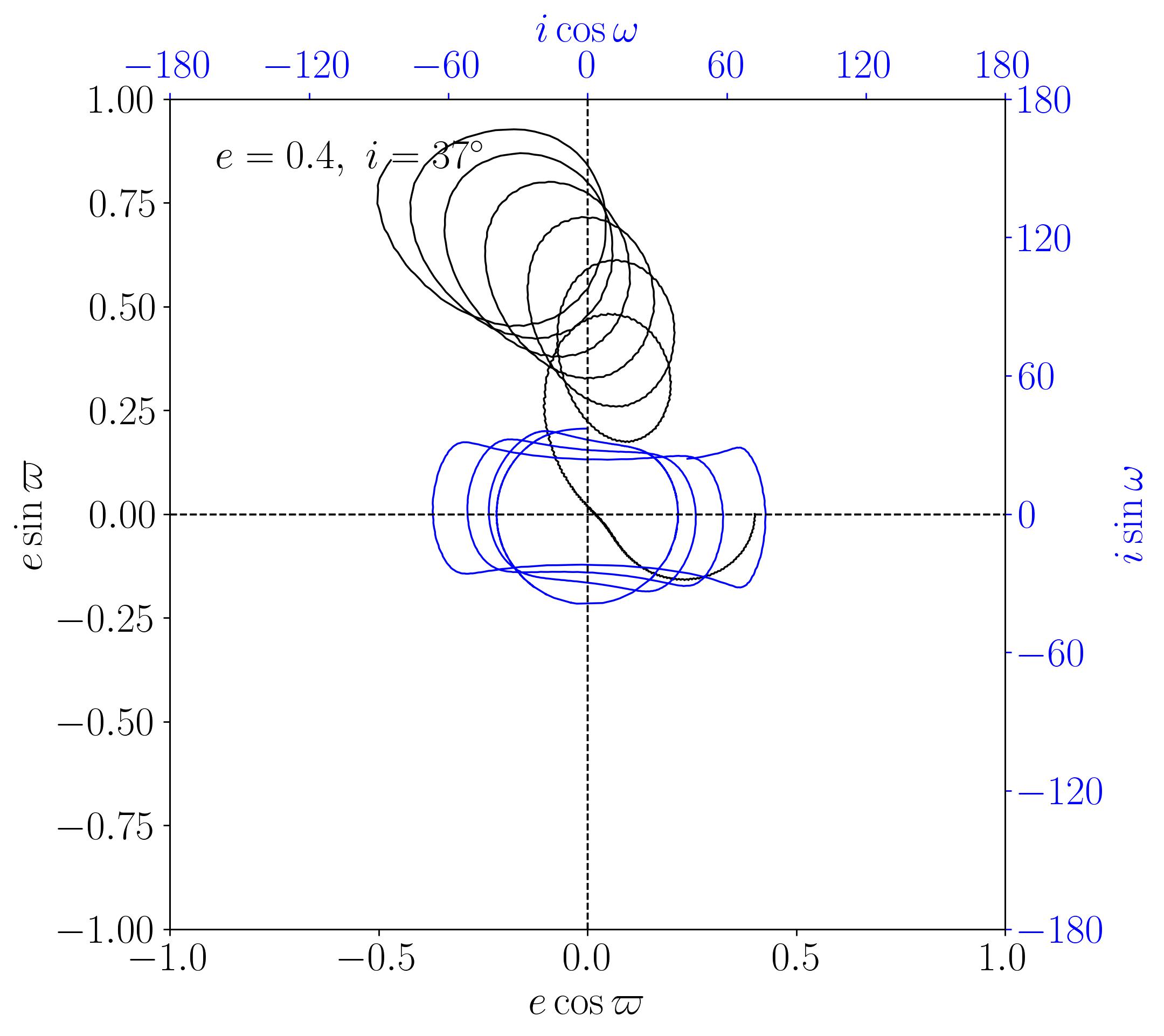}
		& \includegraphics[width=0.33\textwidth]{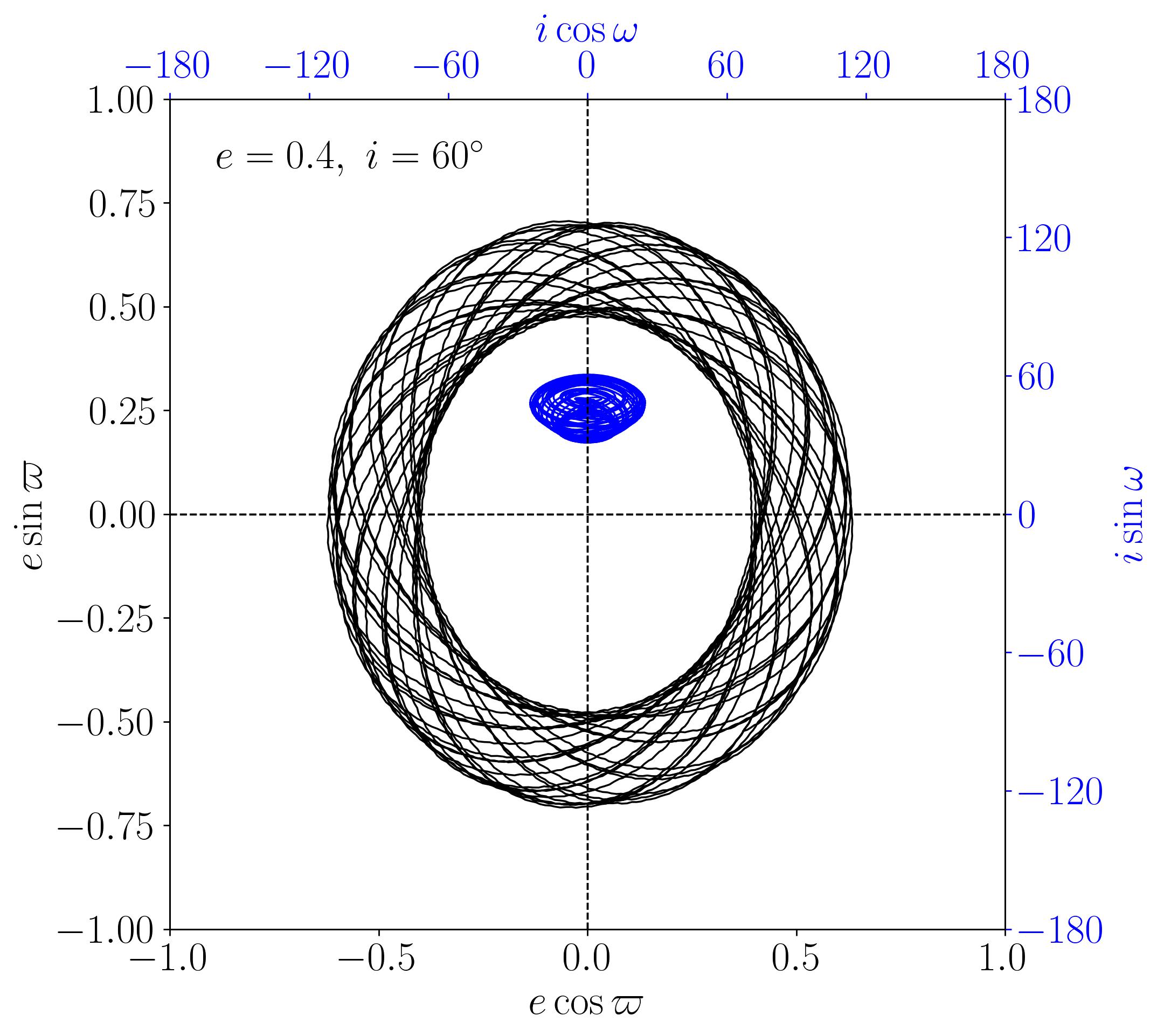}
	\end{tabular}
	\begin{tabular}{cc}
		\\ 
		\includegraphics[width=0.5\textwidth]{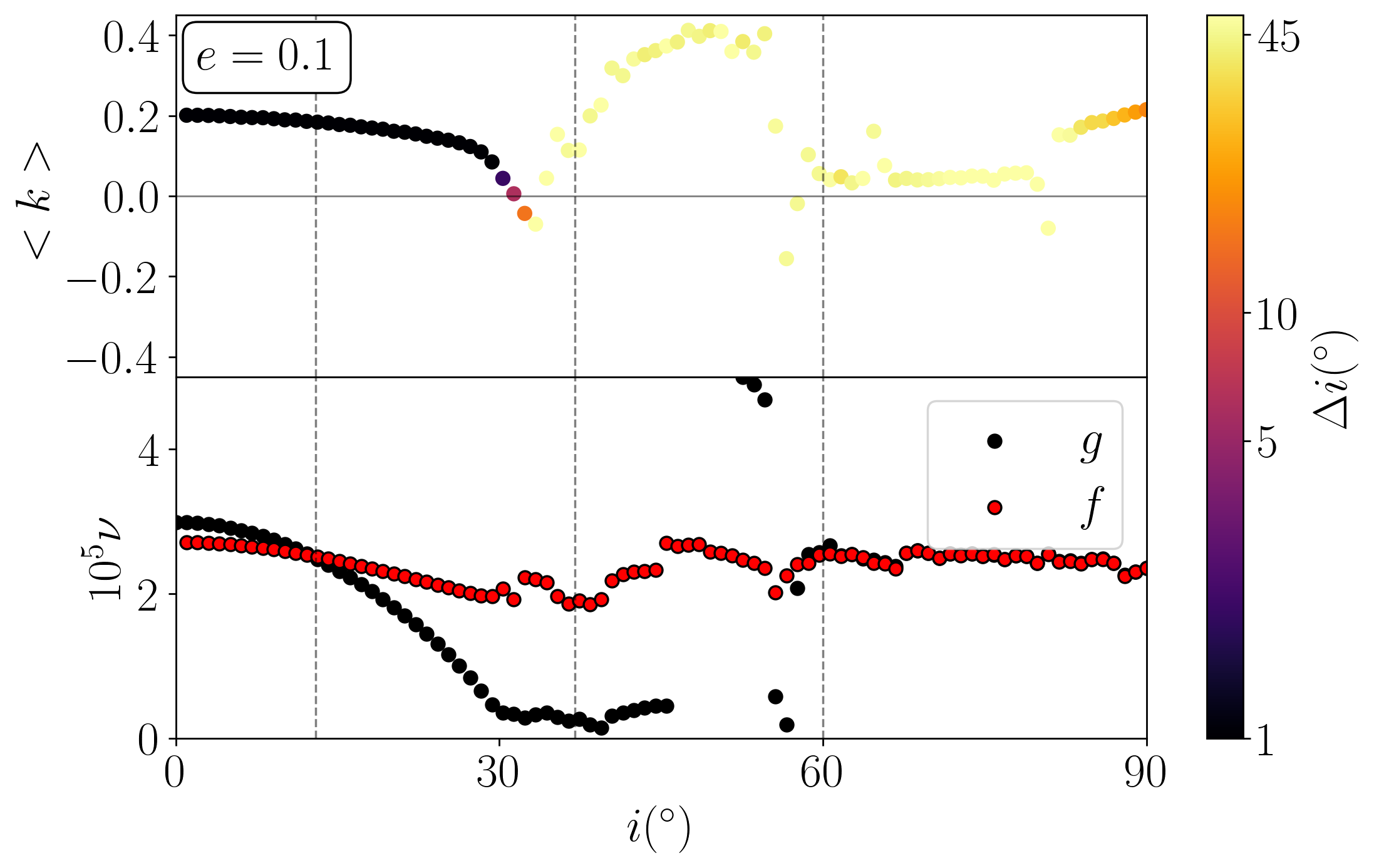}&
		\includegraphics[width=0.5\textwidth]{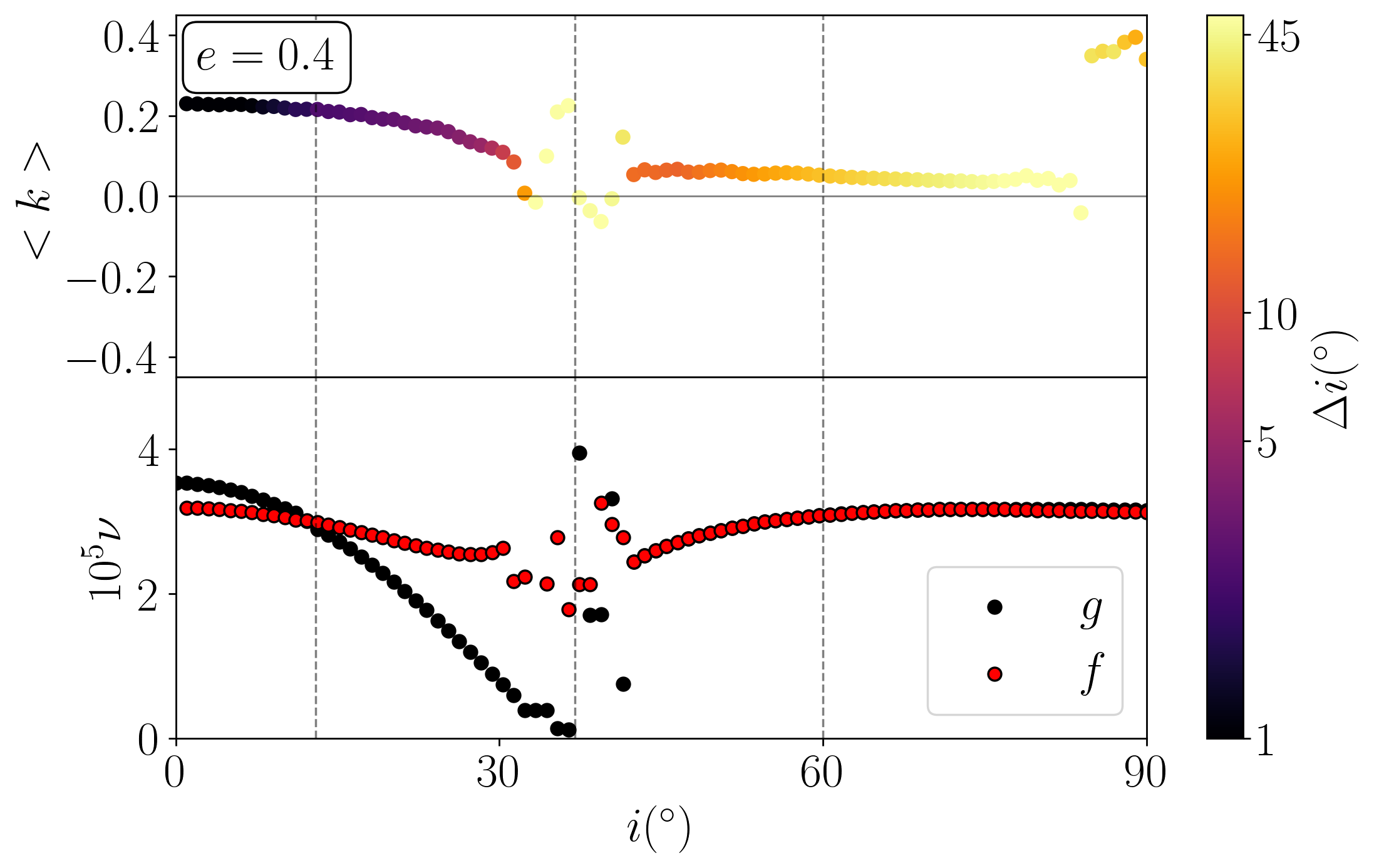} 
	\end{tabular}
	\caption{Numerical integrations for the case of interior particles with $a=2$ au and initial $(\omega,\Omega) = (90\degree,270\degree)$.
		Six upper panels:
		trajectories in $(k,h)$ (black) and $(i\cos\omega,i\sin\omega)$ (blue) 
		for six particles with initial inclinations $(13\degree,37\degree,60\degree)$ for first, second and third column respectively and
		initial eccentricity $(0.1, 0.4)$ for first and second row respectively.
		The two lower panels show the proper frequencies and $<k>$ as function of the initial inclination for initial $e=0.1$ (left) and $e=0.4$ (right). The three vertical lines in each plot correspond to the initial inclinations for the upper panels.
		The plot for $<k>$ also shows in color scale the maximum reached $\Delta i$.}
	
	\label{multi1}
\end{figure*}

\begin{figure*}[h!]
	\centering
	\begin{tabular}{ccc}
		\includegraphics[width=0.33\textwidth]{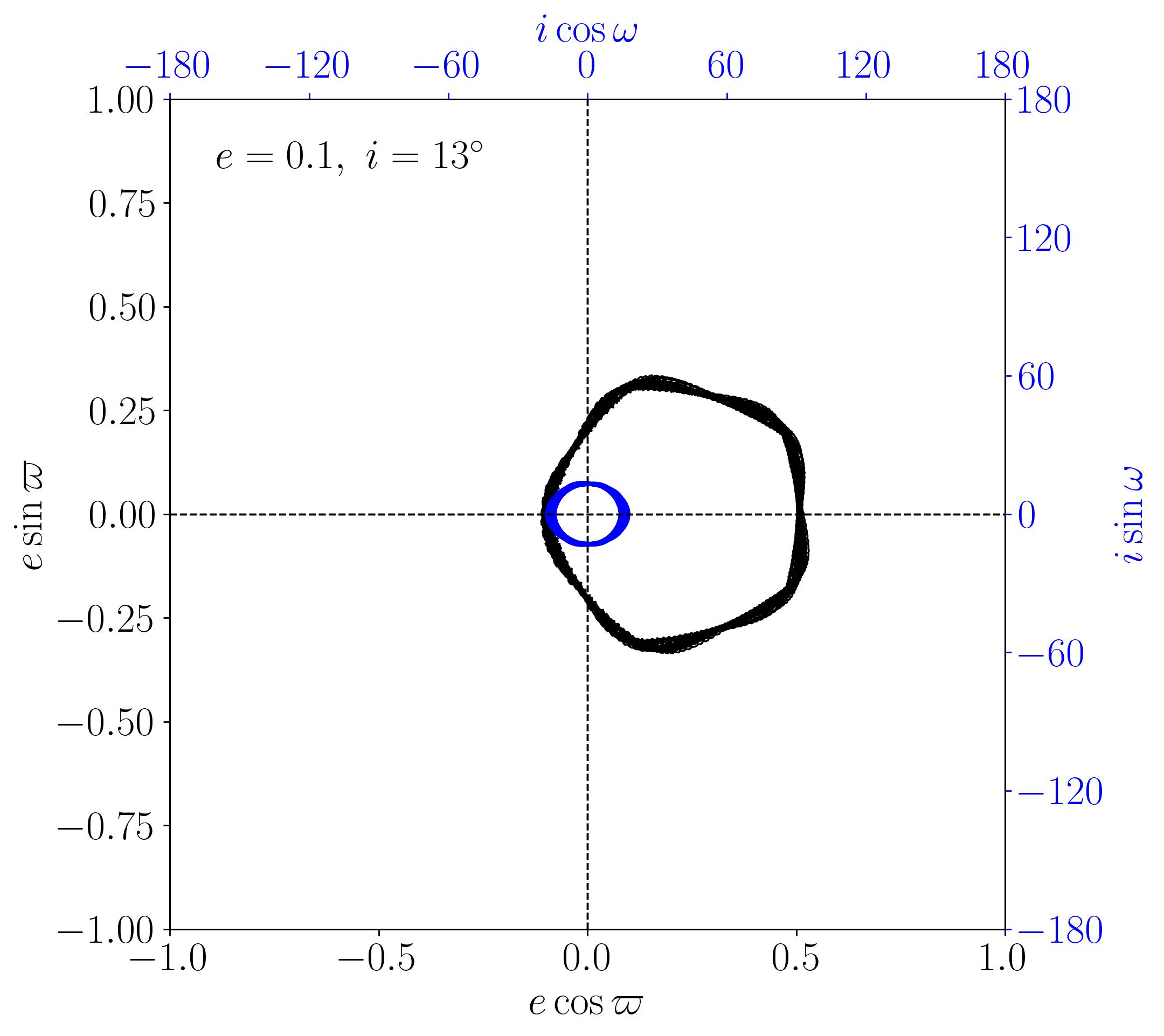}&
		\includegraphics[width=0.33\textwidth]{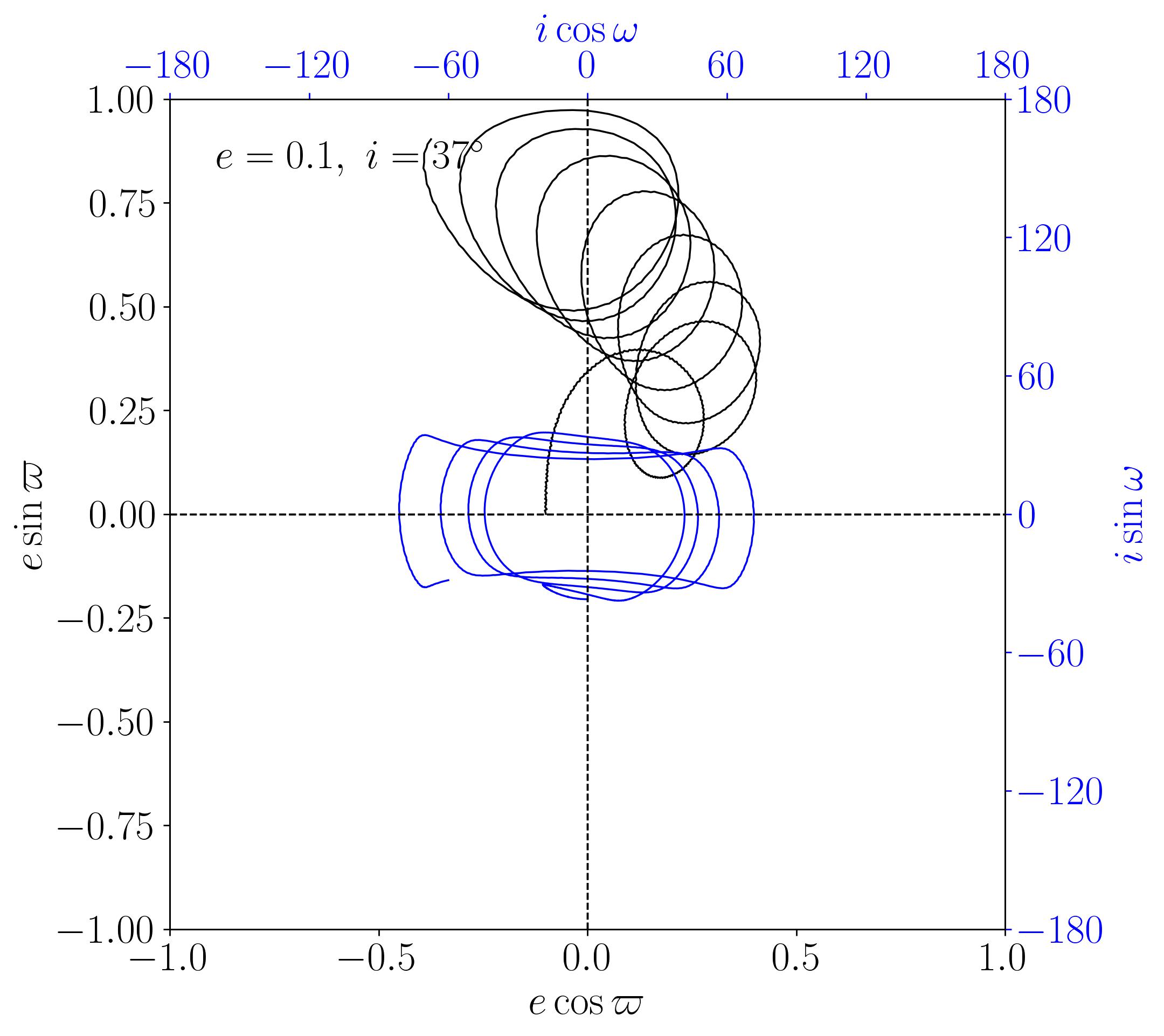}  & \includegraphics[width=0.33\textwidth]{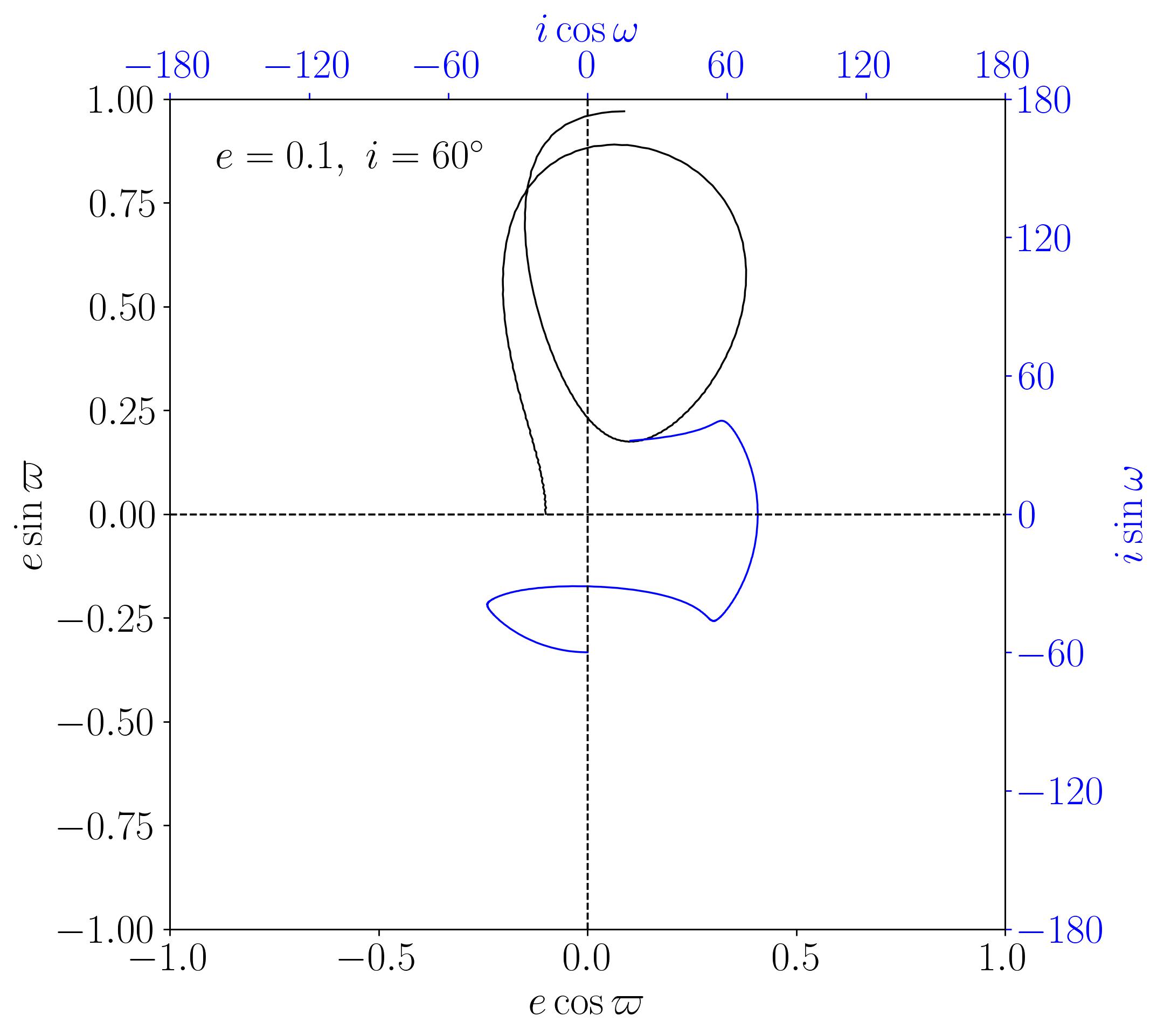}  \\
		\includegraphics[width=0.33\textwidth]{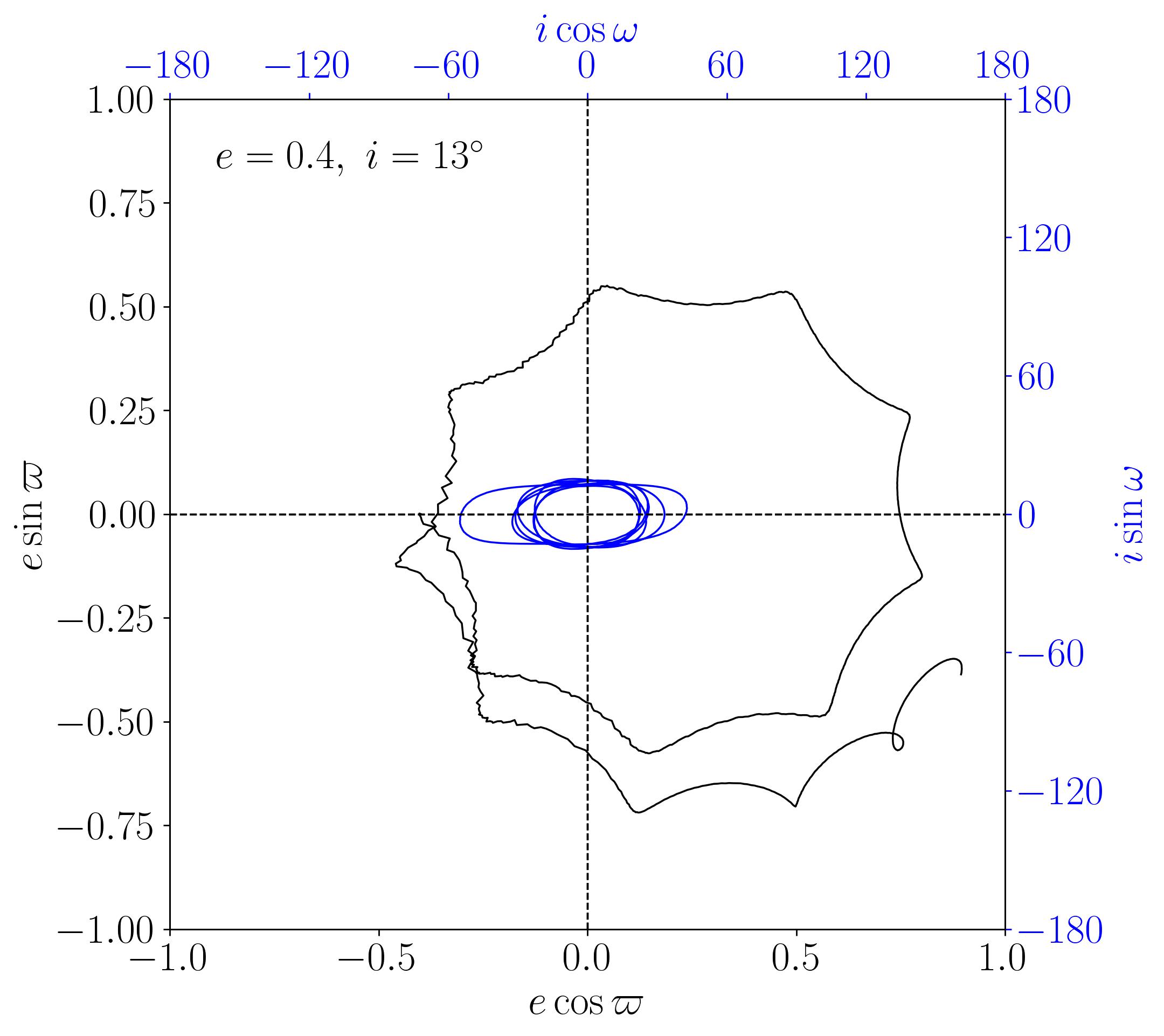}   & \includegraphics[width=0.33\textwidth]{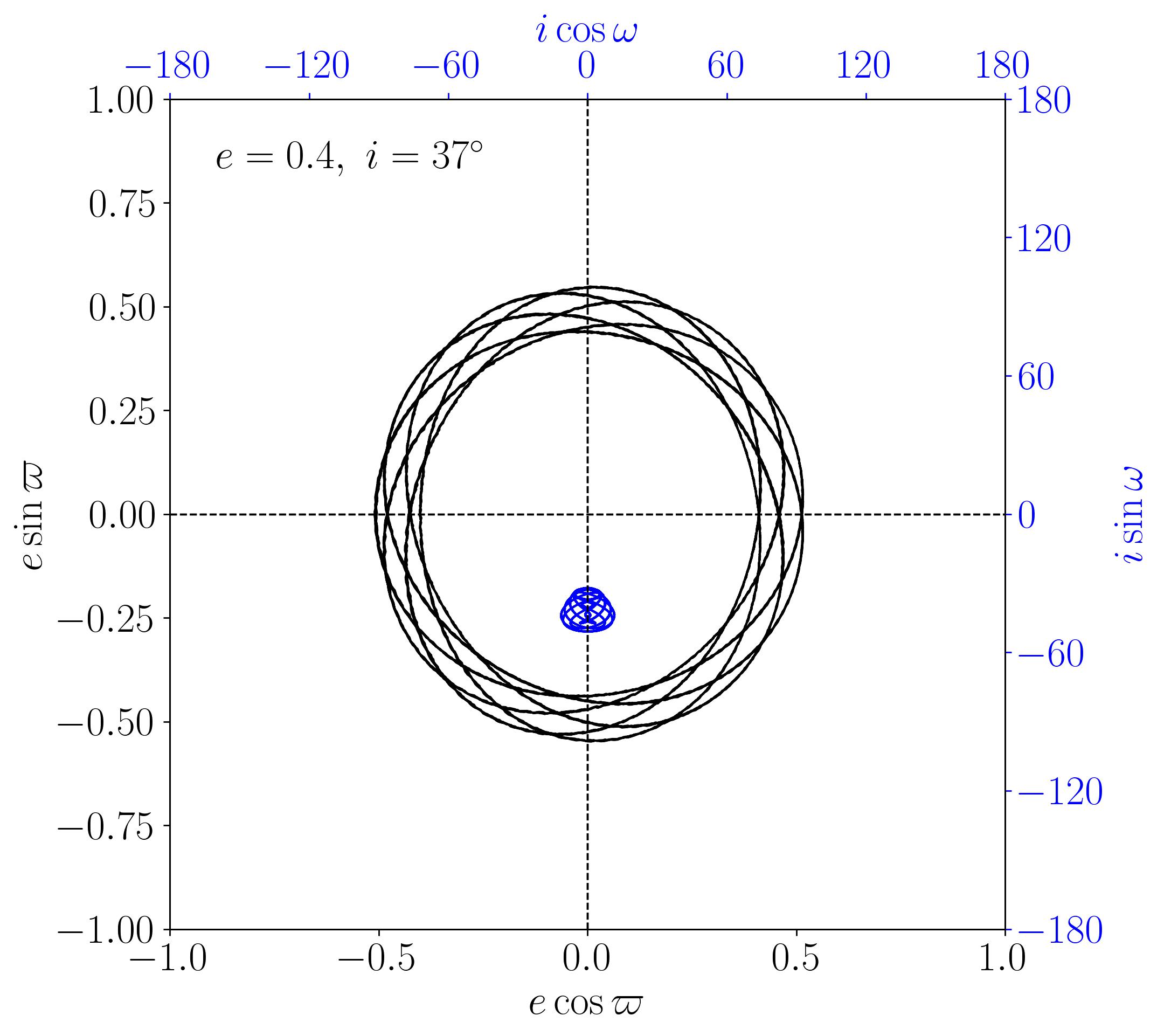}& 
		\includegraphics[width=0.33\textwidth]{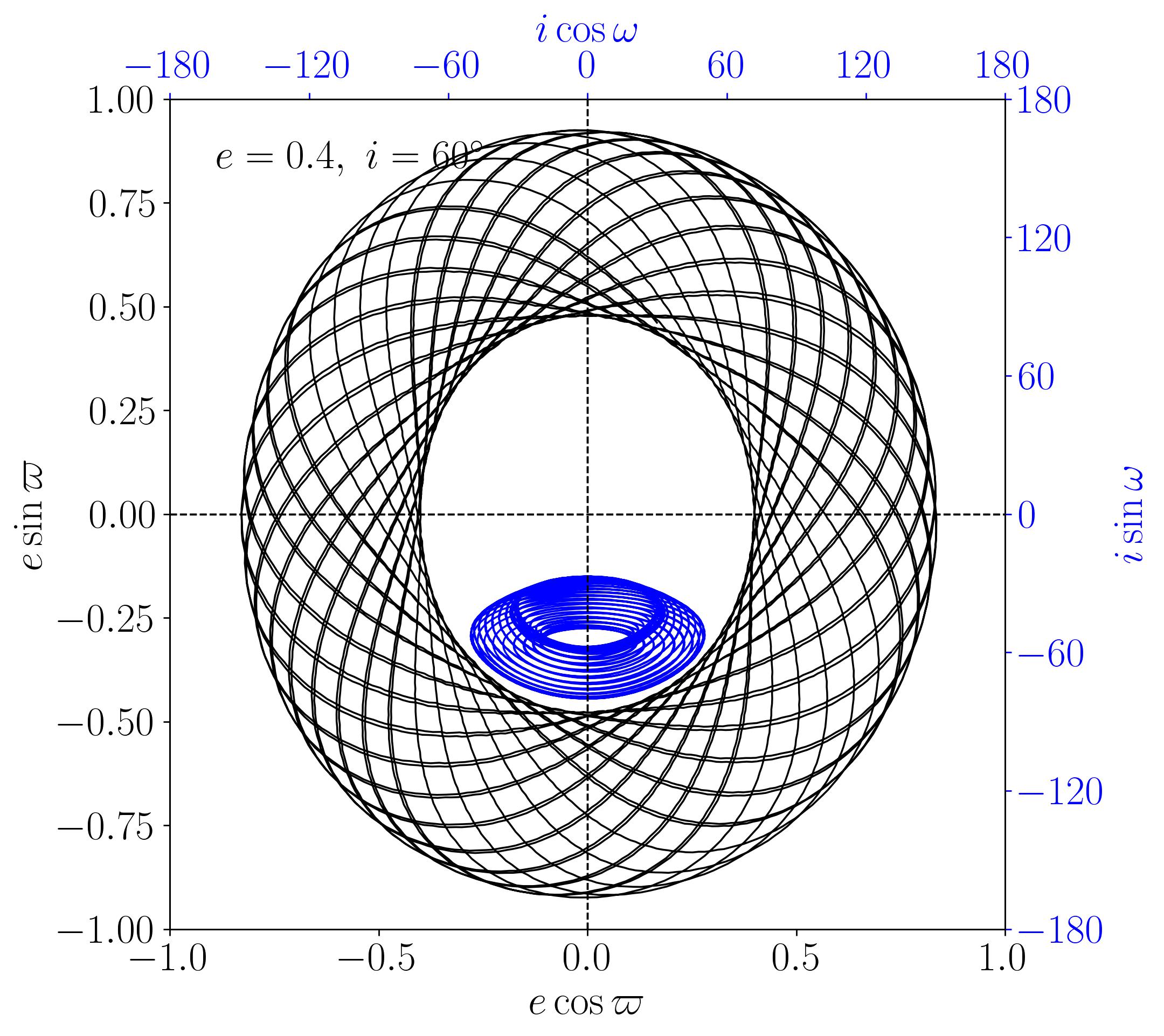}
	\end{tabular}
	
	\begin{tabular}{cc}
		\\ 
		\includegraphics[width=0.5\textwidth]{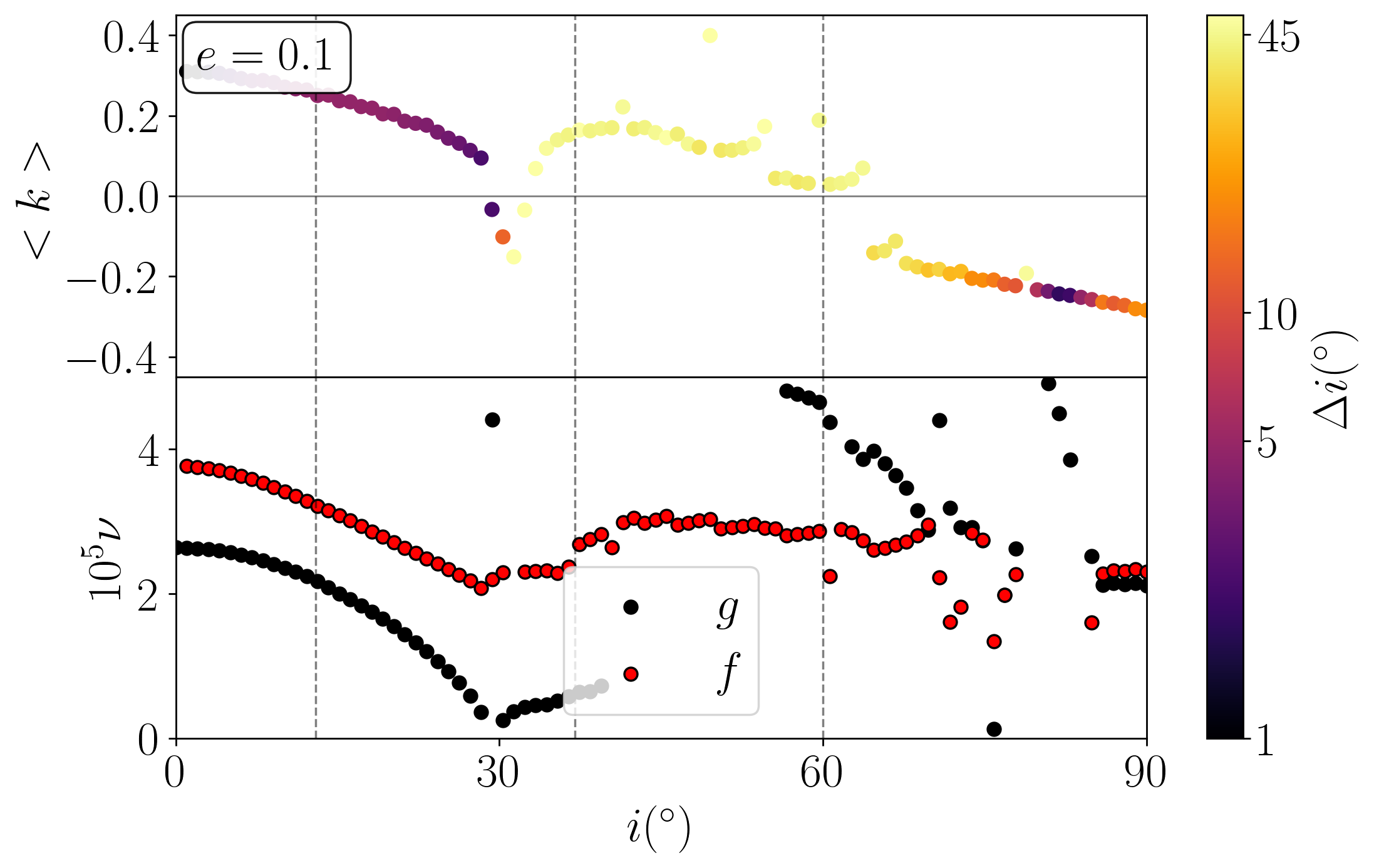}&
		\includegraphics[width=0.5\textwidth]{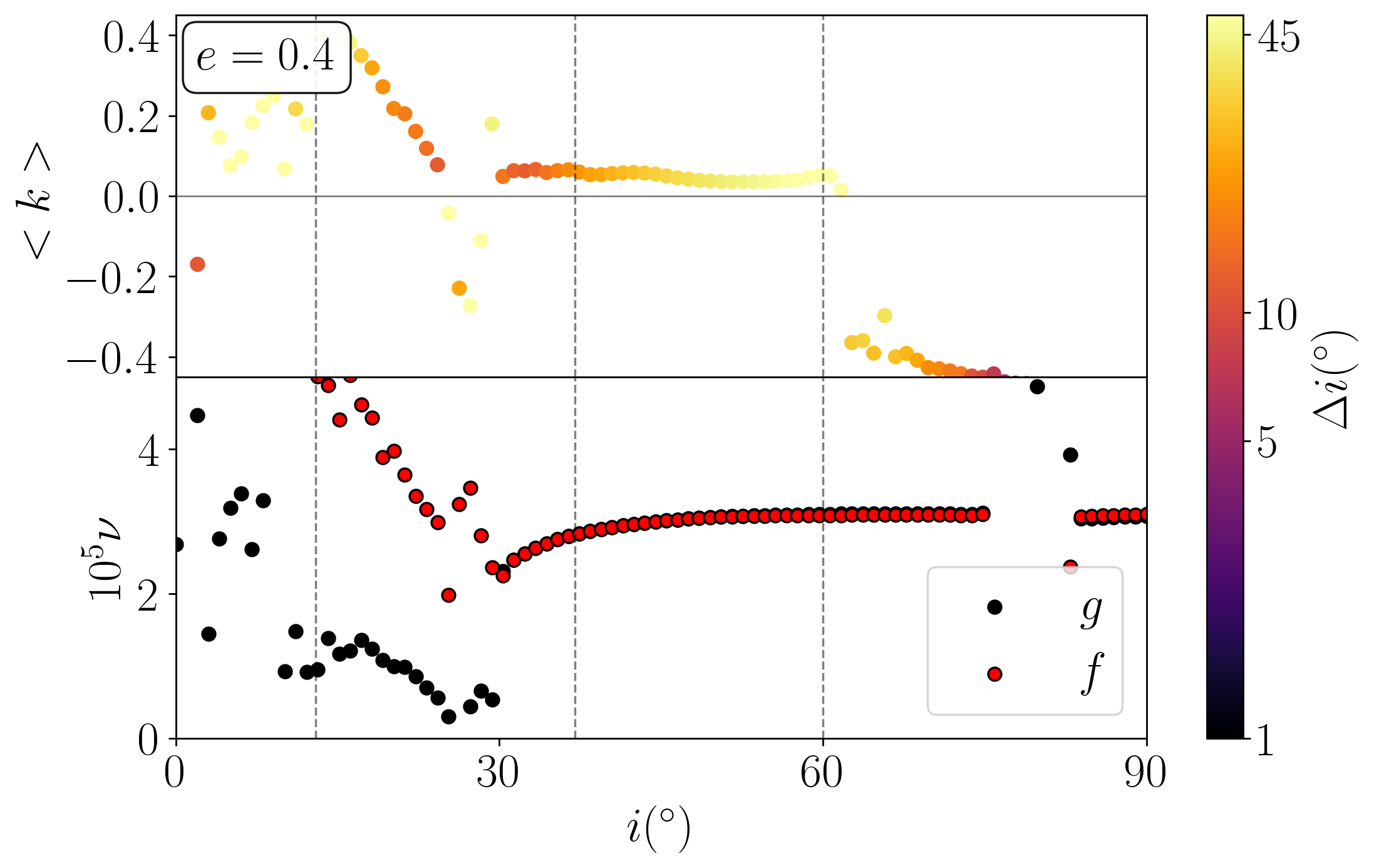} 
	\end{tabular}
	
	\caption{Same as figure \ref{multi1} but for initial 
		$(\omega,\Omega)=(270\degree,270\degree)$.
	}
	\label{multi2}
\end{figure*}

\begin{figure*}[h!]
	\centering
	\begin{tabular}{ccc}
		\includegraphics[width=0.33\textwidth]{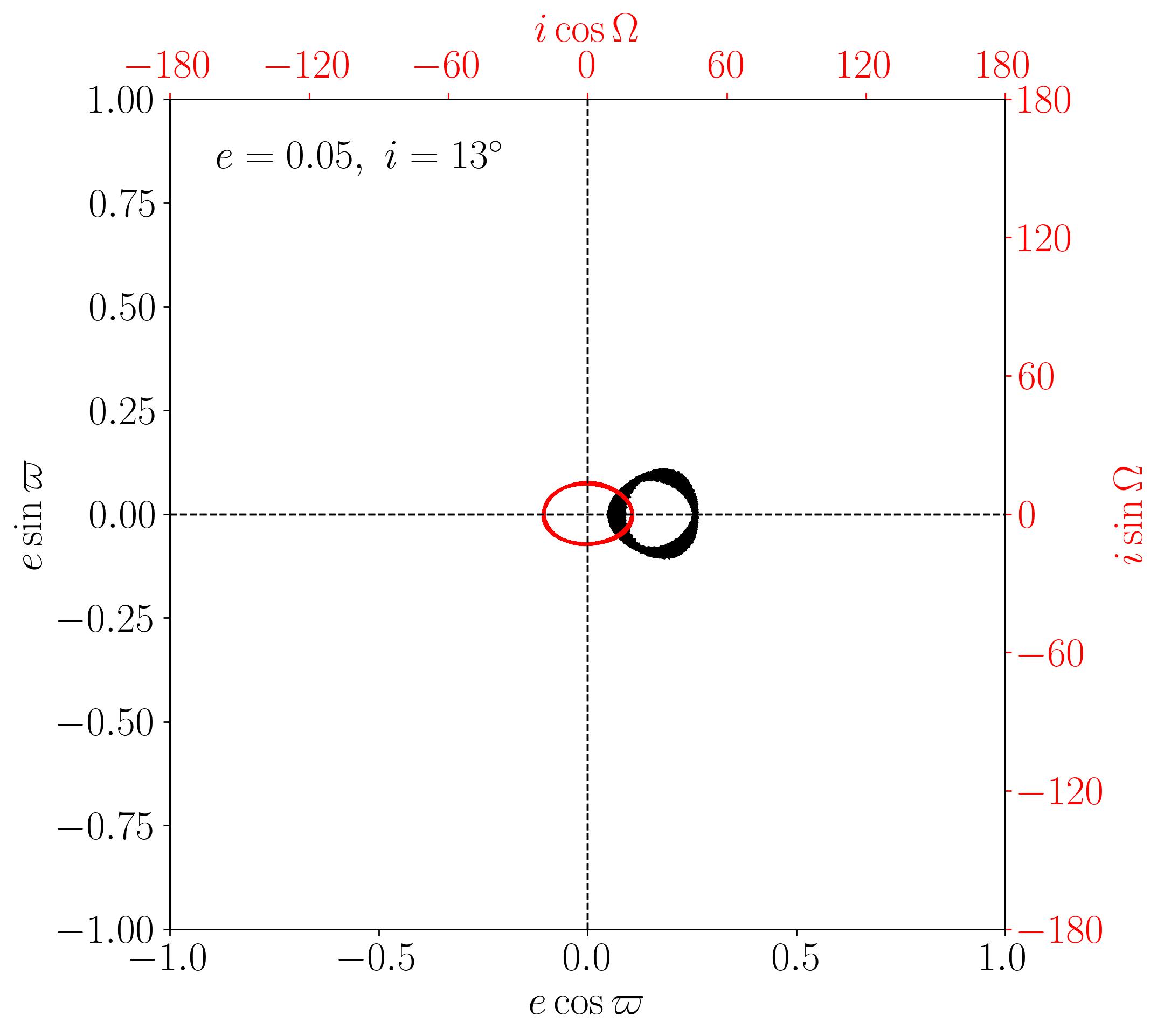}&
		\includegraphics[width=0.33\textwidth]{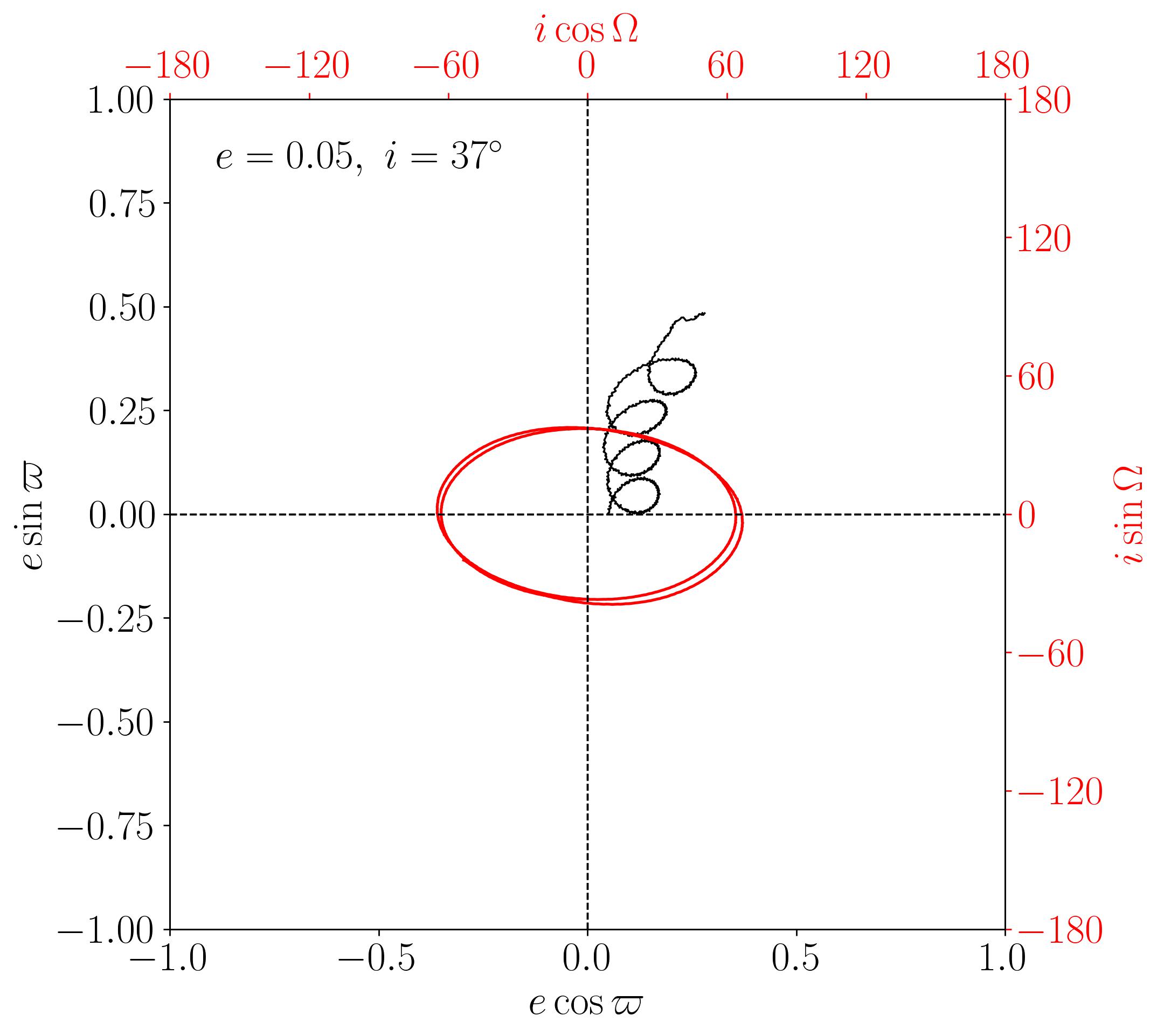}  & \includegraphics[width=0.33\textwidth]{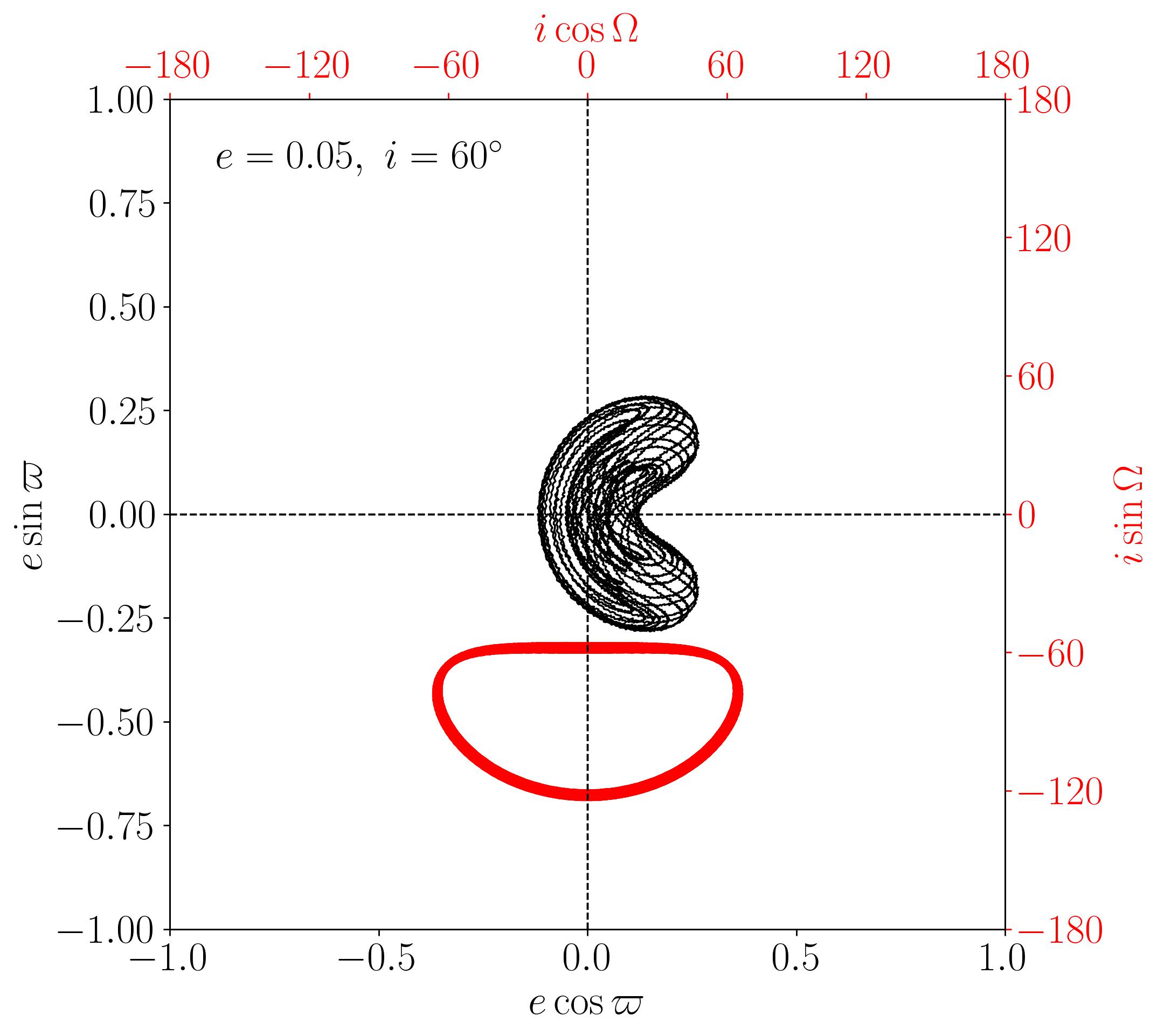}  \\
		\includegraphics[width=0.33\textwidth]{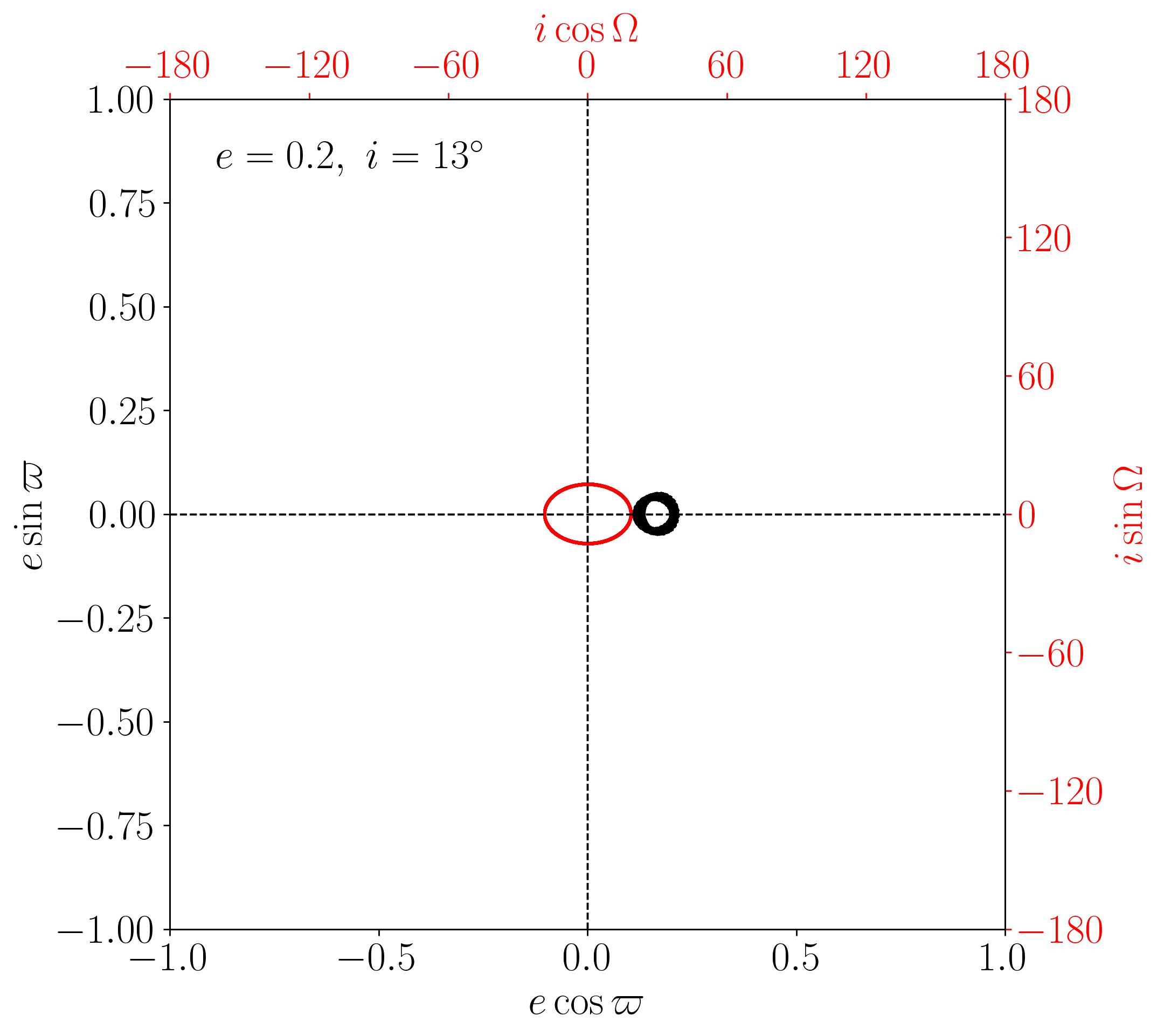}   & \includegraphics[width=0.33\textwidth]{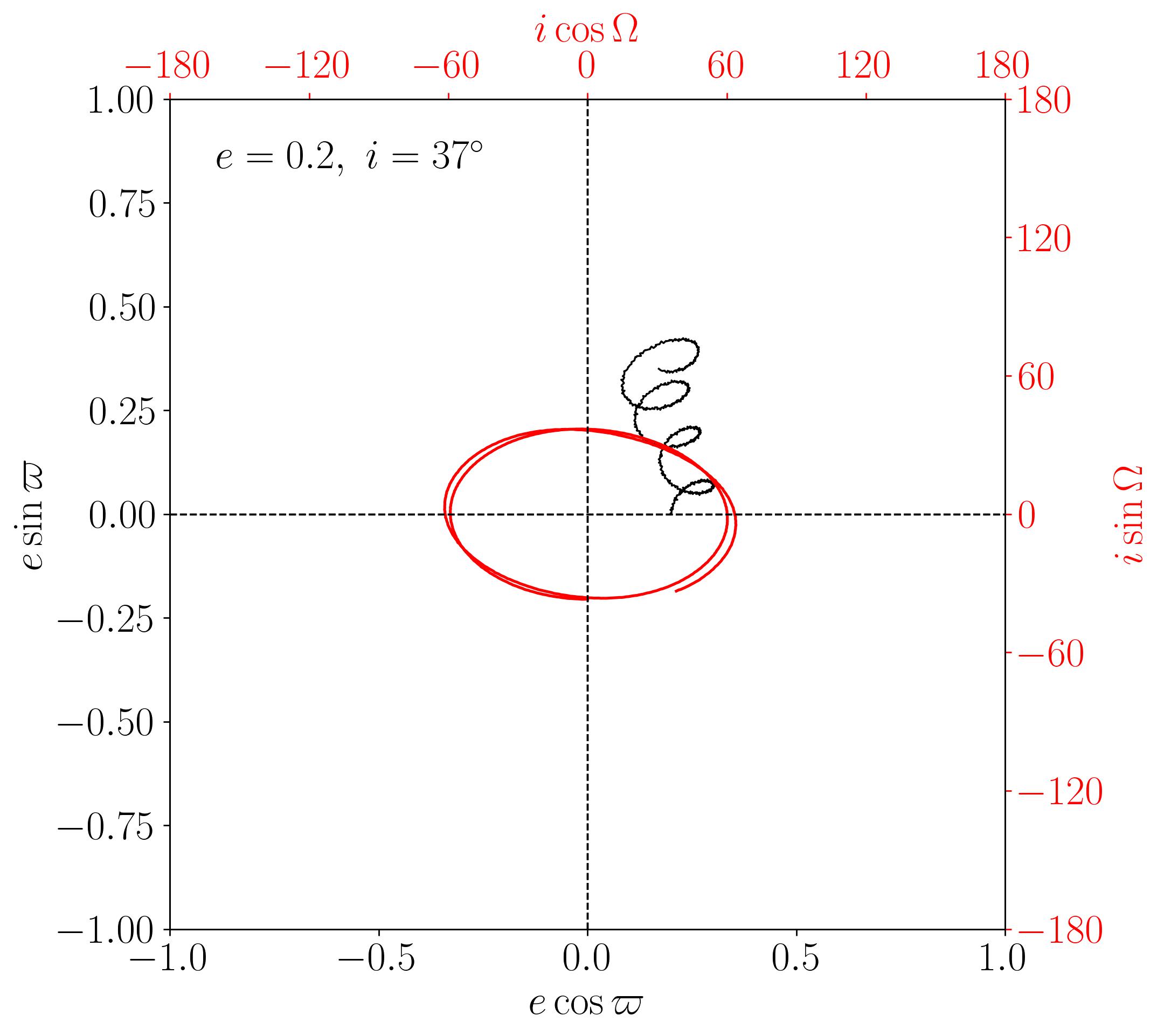}& \includegraphics[width=0.33\textwidth]{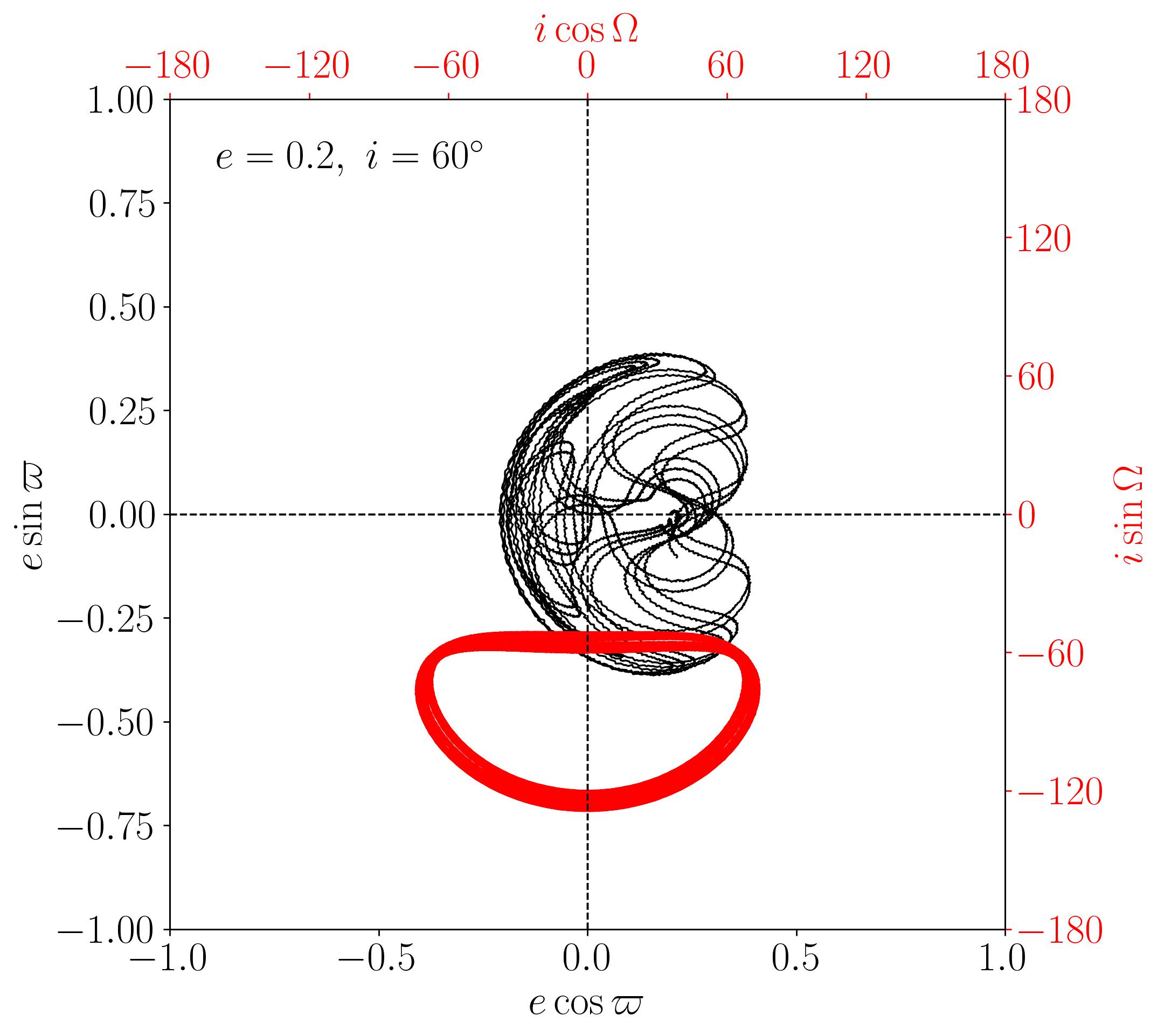}
	\end{tabular}
	\begin{tabular}{cc}
		\\ 
		\includegraphics[width=0.5\textwidth]{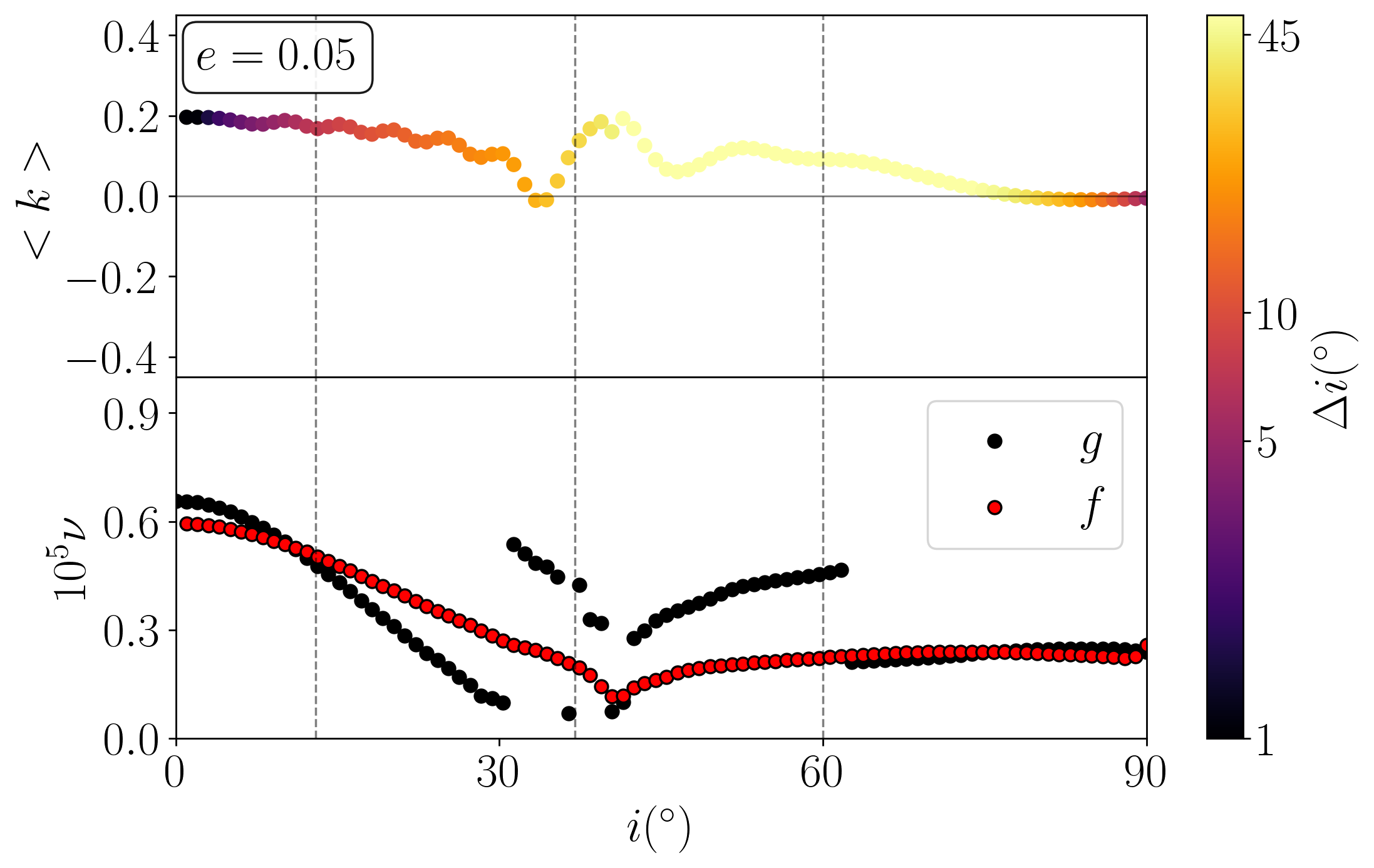}&
		\includegraphics[width=0.5\textwidth]{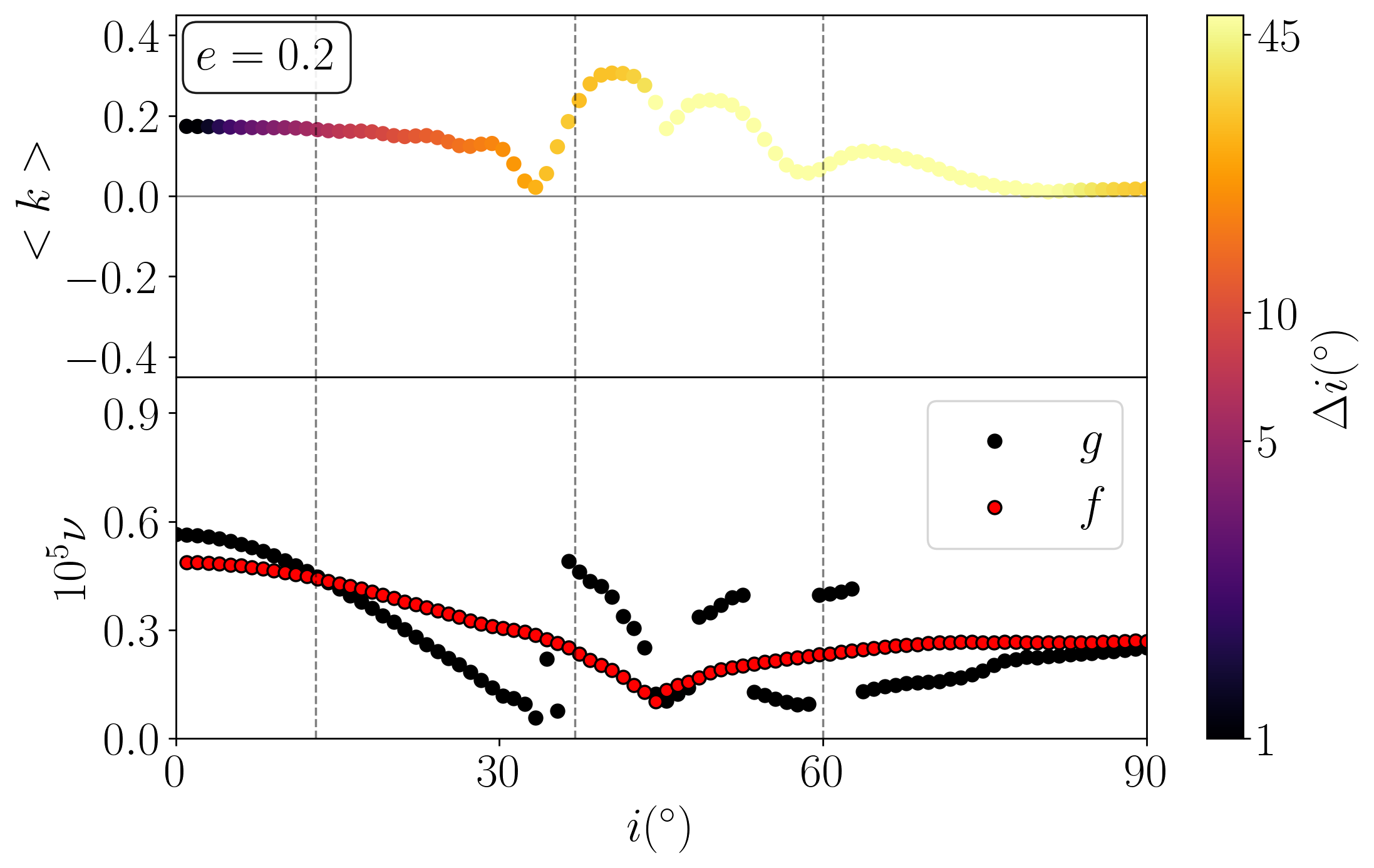} 
	\end{tabular}
	\caption{Same as figure \ref{multi1} but for exterior particles with 
		$a=12.5$ au and $(\omega,\Omega) = (90\degree,270\degree)$ and for $e=0.05$ and $e=0.2$.}
	\label{multi3}
\end{figure*}

\begin{figure*}[h!]
	\centering
	\begin{tabular}{ccc}
      \includegraphics[width=0.33\textwidth]{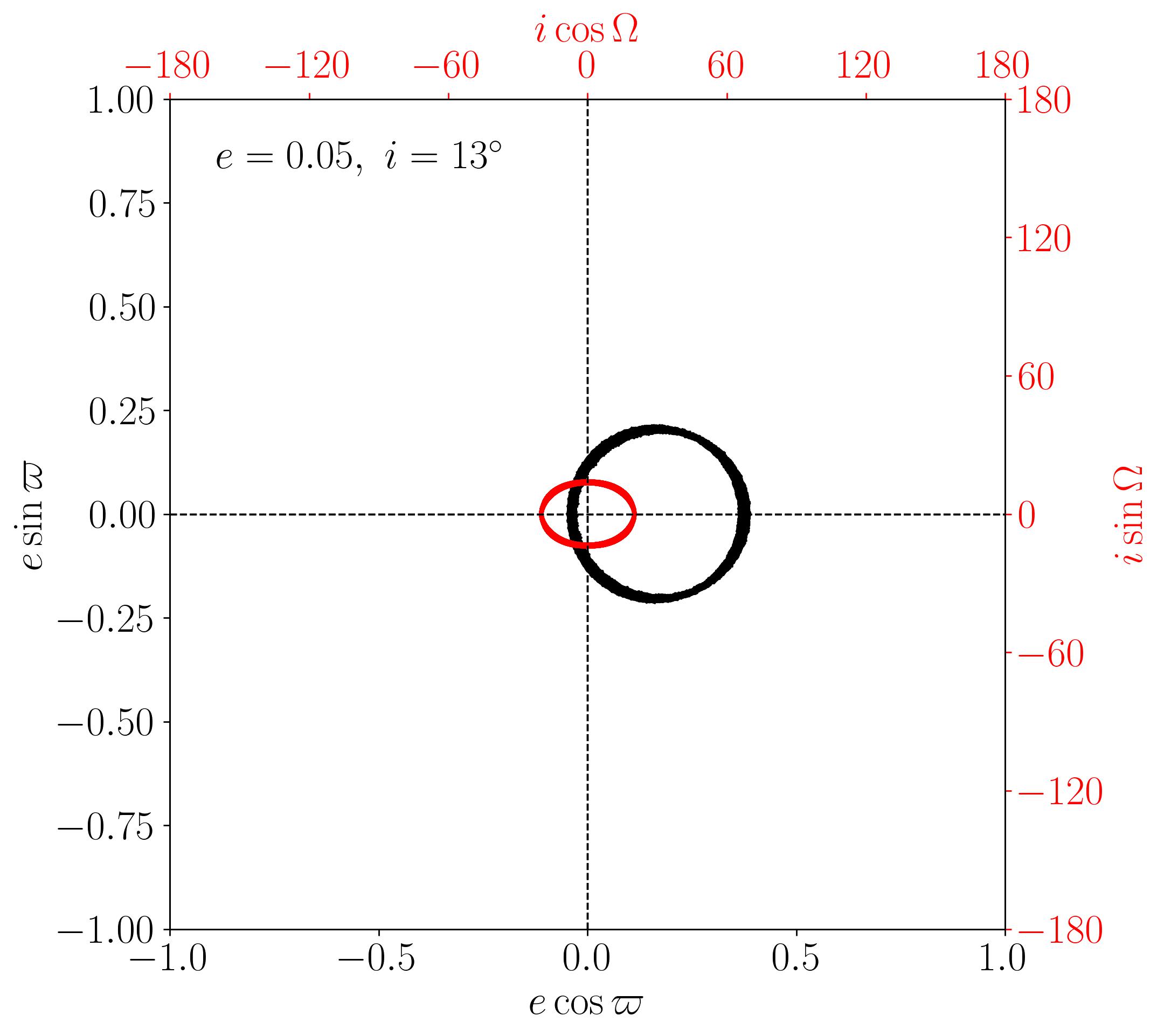}
      &\includegraphics[width=0.33\textwidth]{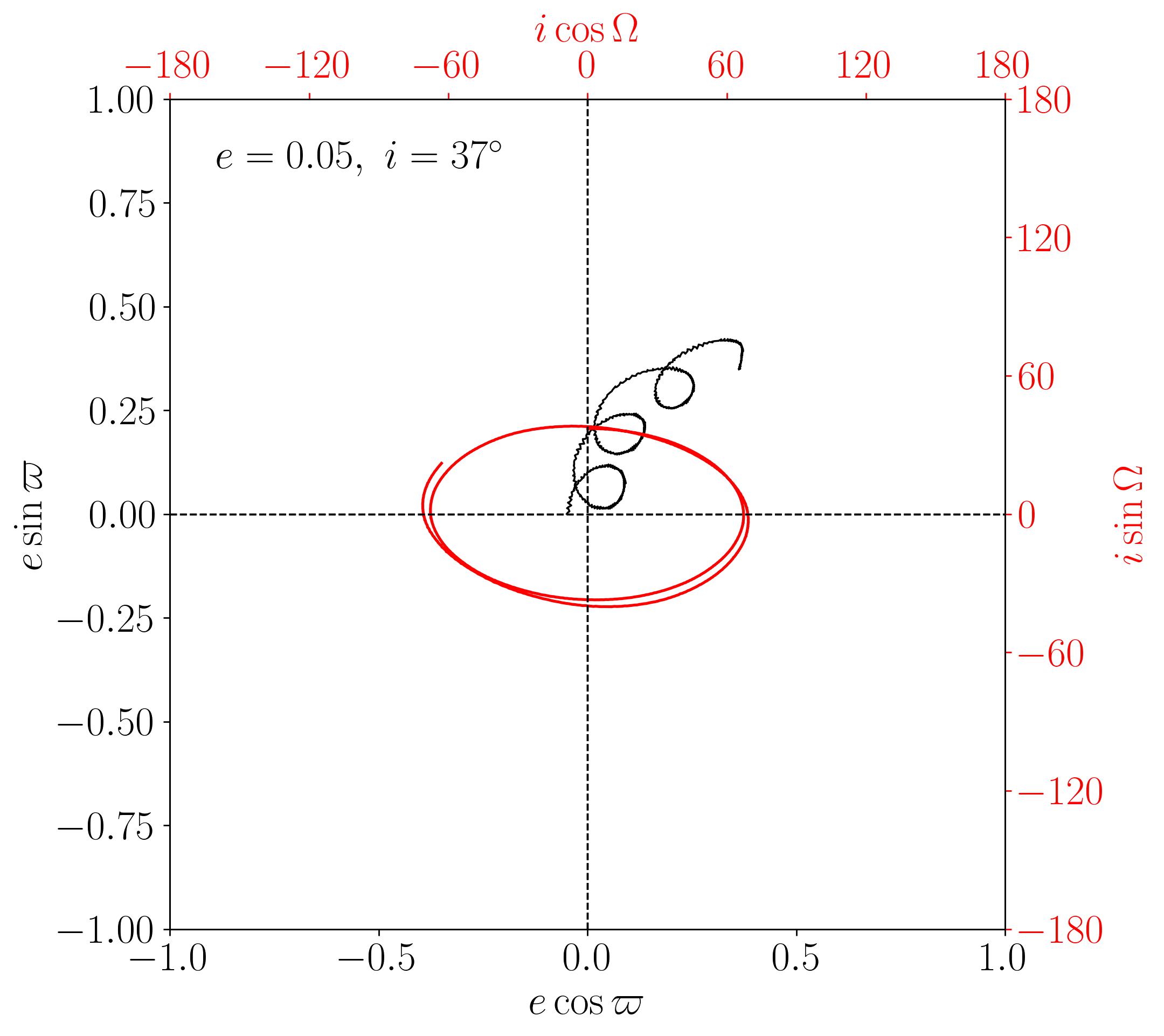}  & \includegraphics[width=0.33\textwidth]{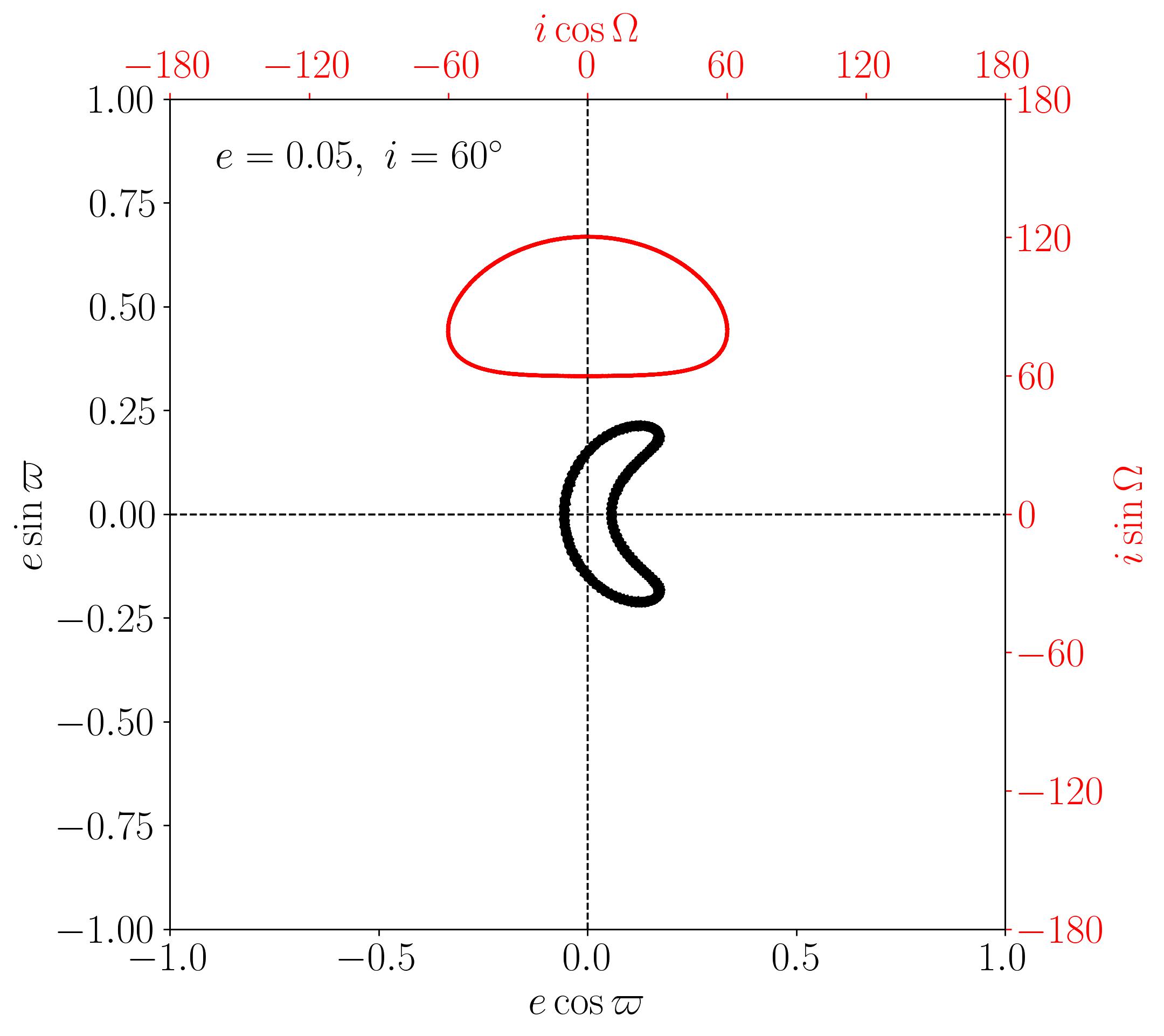}  \\
\includegraphics[width=0.33\textwidth]{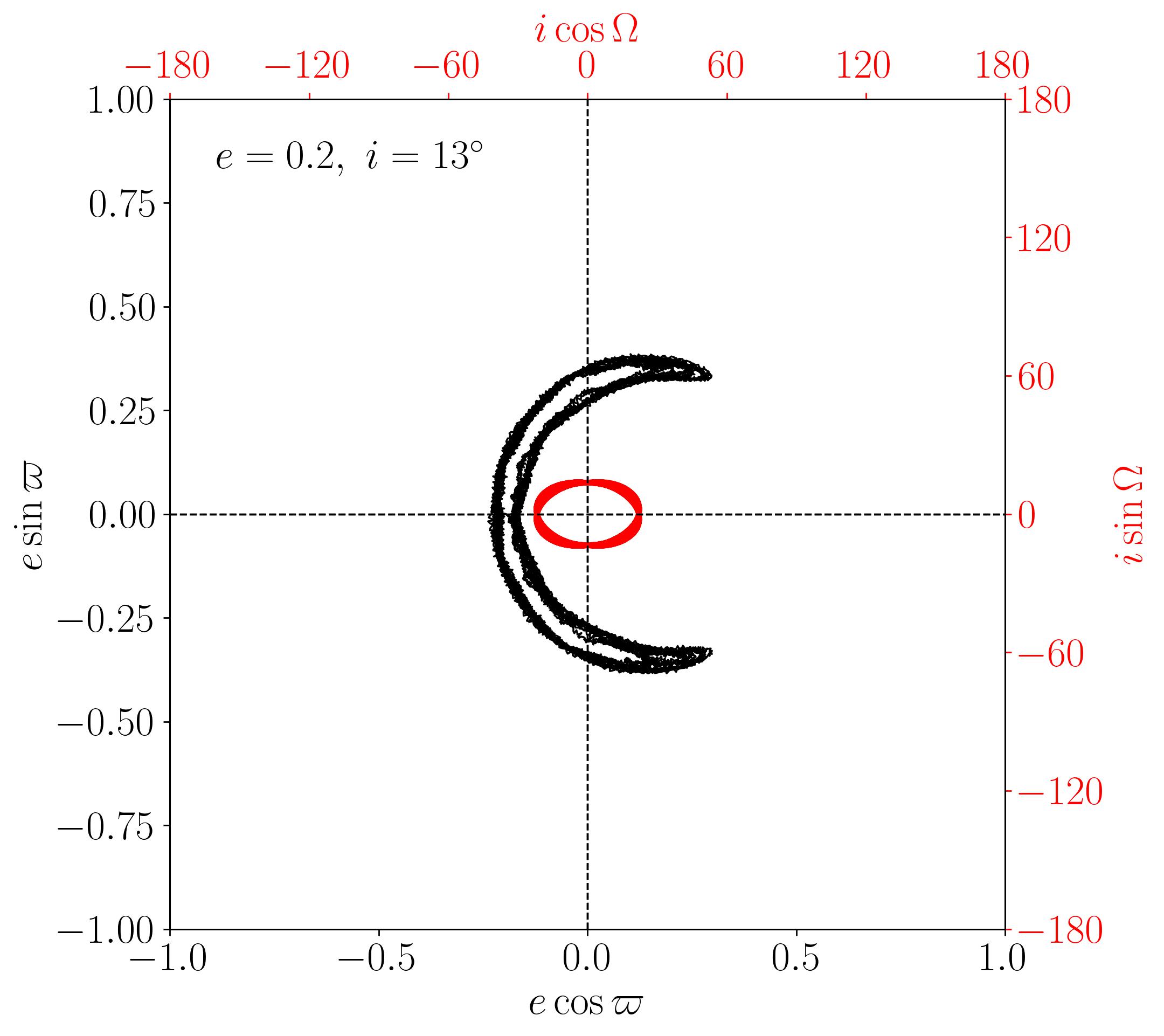}   & \includegraphics[width=0.33\textwidth]{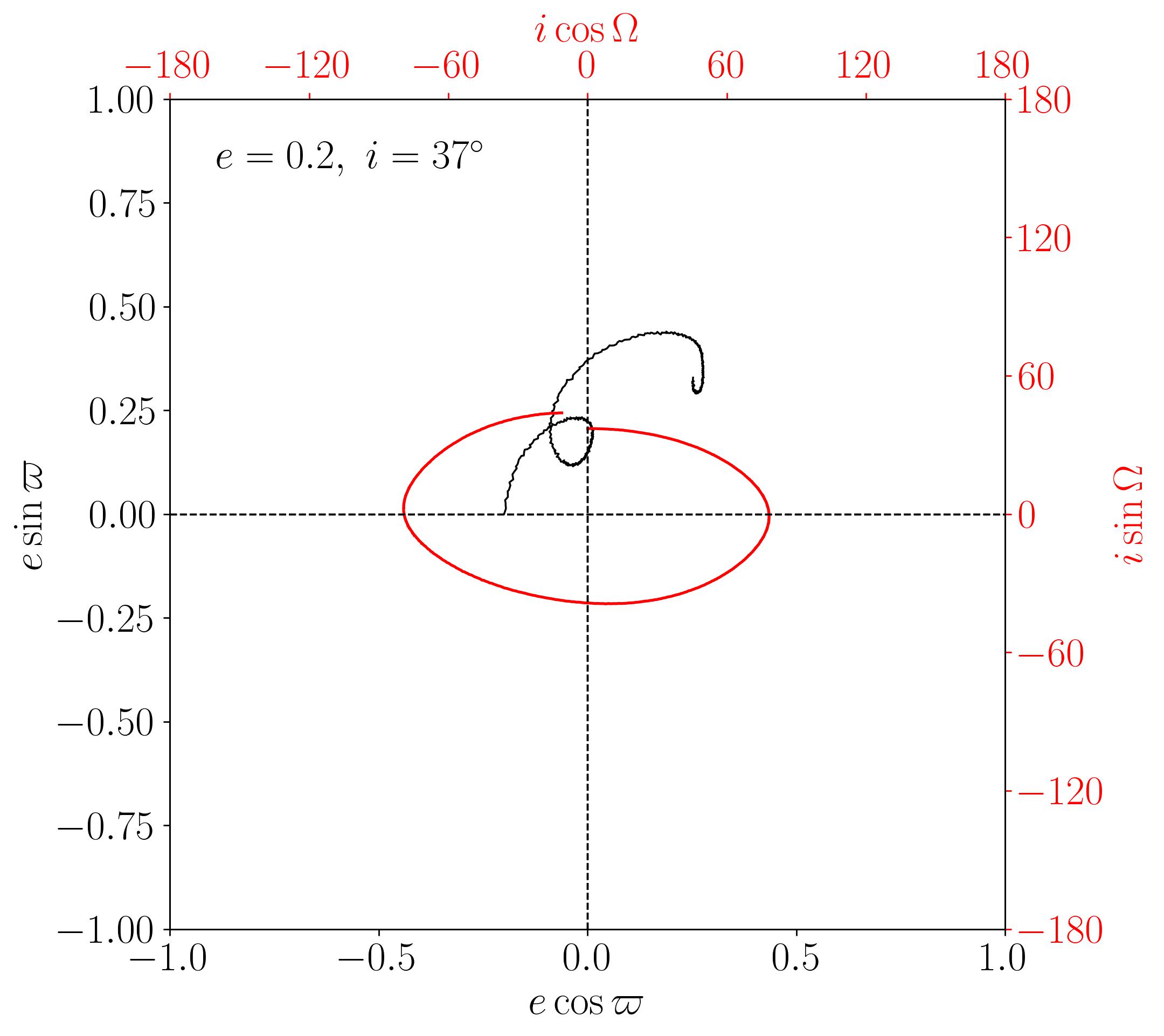}& \includegraphics[width=0.33\textwidth]{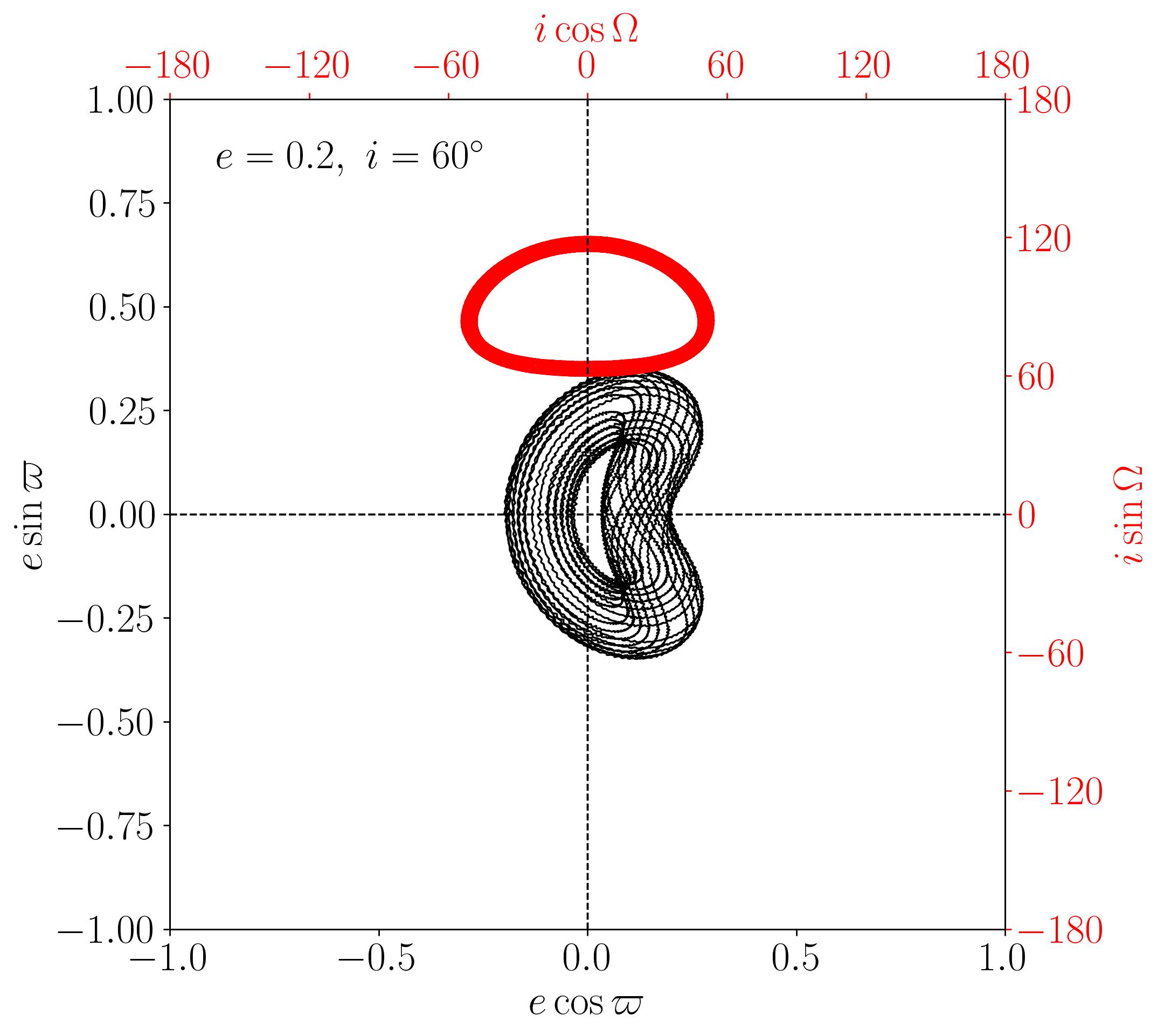}
	\end{tabular}
	\begin{tabular}{cc}
		\\ 
		\includegraphics[width=0.5\textwidth]{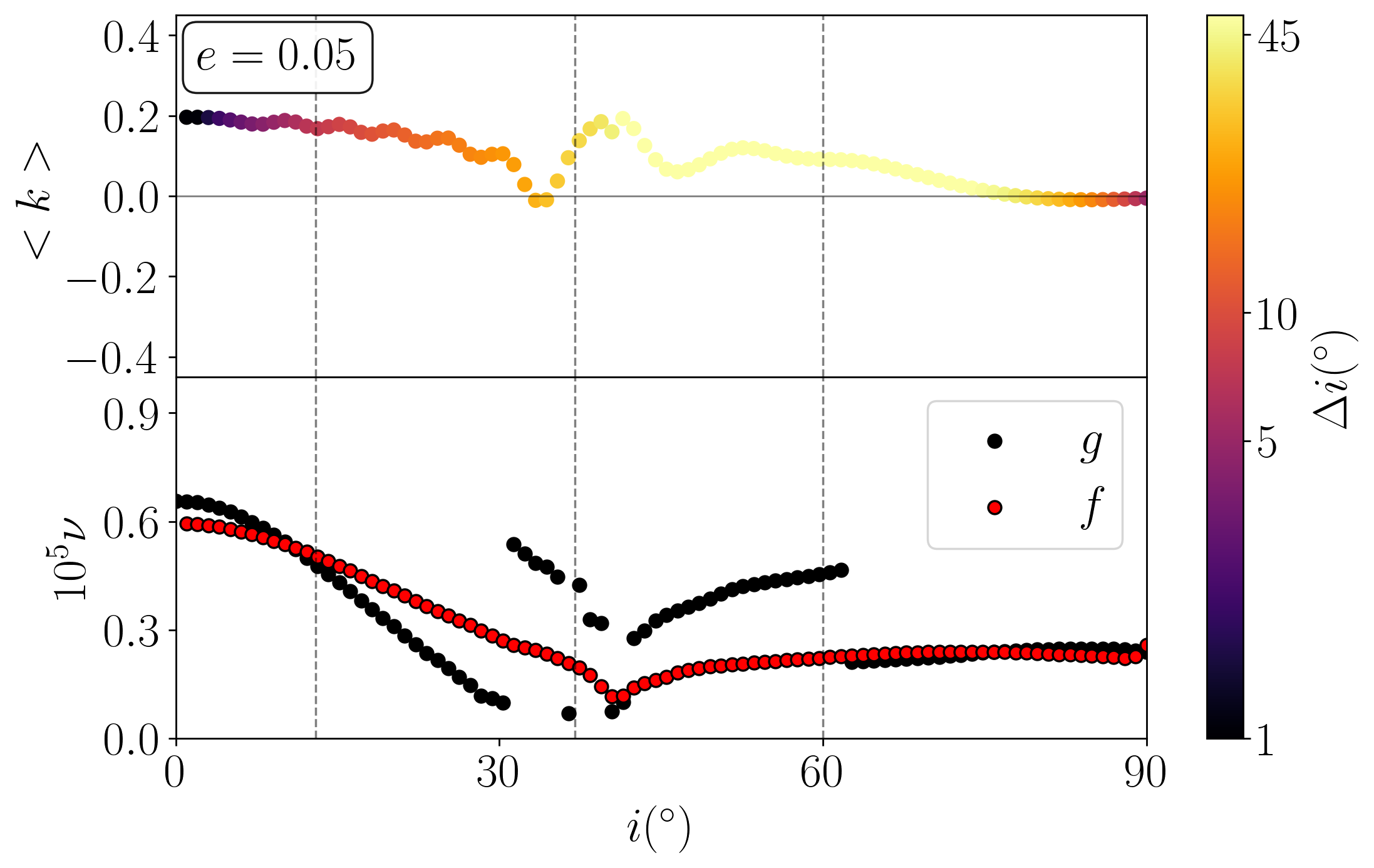}
		&\includegraphics[width=0.5\textwidth]{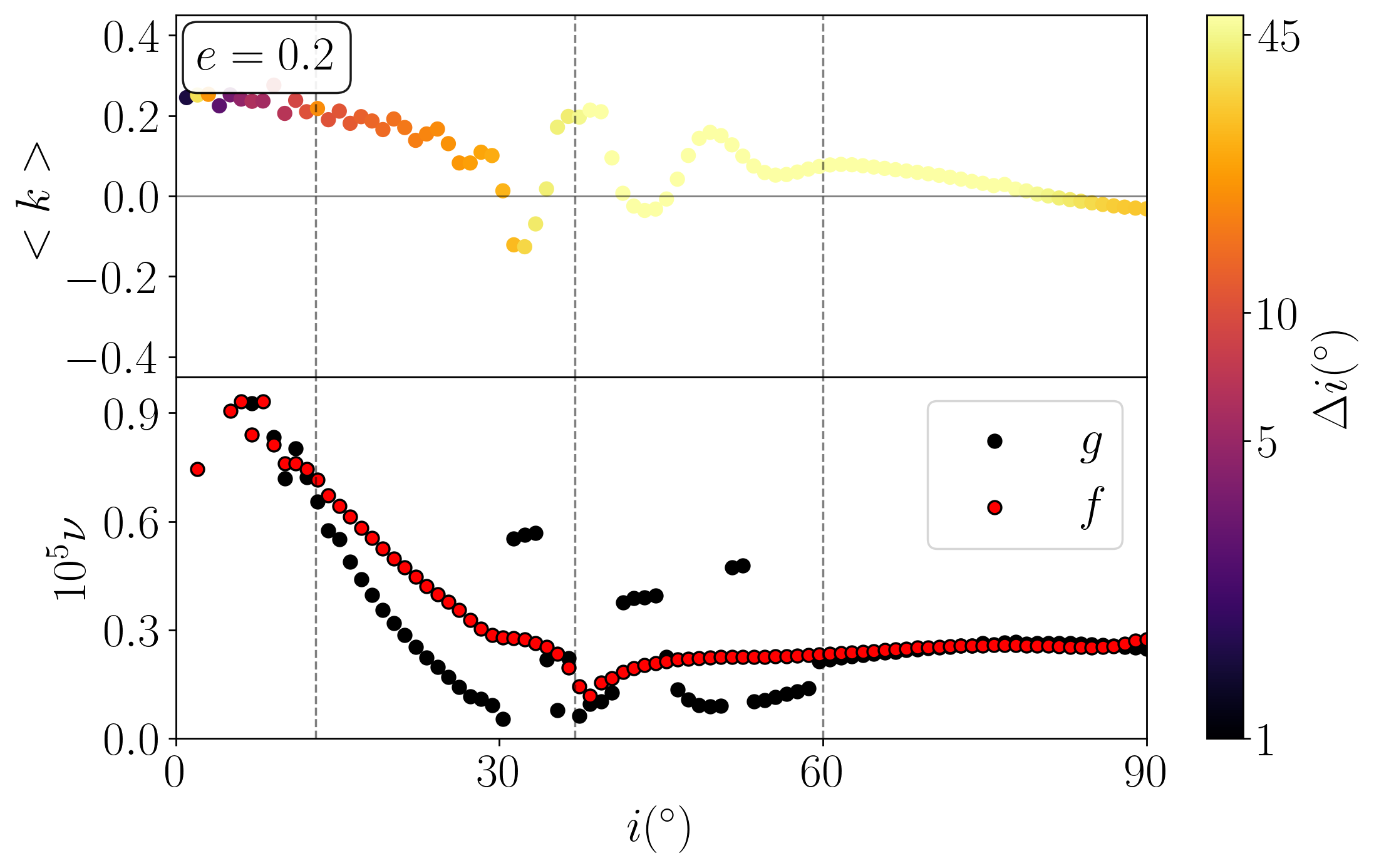} 
	\end{tabular}
	\caption{Same as figure \ref{multi3} for initial
		$(\omega,\Omega)=(90\degree,90\degree)$.}
	\label{multi4}
\end{figure*}

In Fig. \ref{examplehkpg} we show an example of trajectory in the spaces $(k,h)$ in black and $(q,p)$ in red for a particle with $i<i_c$ corresponding to the region in Fig. \ref{frecs} where both proper frequencies diminish with growing inclination. This is a kind of classic secular evolution very similar to the result given by the linear theory for small $(e,i)$ orbits. The $(q,p)$ trajectory is not an exact circle but well centered at $(0,0)$ suggesting there is no forced inclination, due to the adoption of the planet orbital plane as reference. On the other hand, 
the $(k,h)$ trajectory is shifted to a positive $k$ value in the direction given by $\varpi=0\degree$ indicating there is a forced eccentricity due to the planetary eccentricity in the direction of its pericenter which is given by $\varpi_p=0\degree$. Note that calculating numerically the mean value $<k>$ is a simple way to estimate the existence of forced eccentricity and also whether the particle evolves following a quasi classical linear secular dynamics or not.  
With this in mind we have generated some plots like the one in Fig. \ref{frecs} but adding the value $<k>$ in the upper part of the plots, just to give us an idea of how close trajectories are with respect to the neat trajectories of the linear secular theory.

Examples of these plots are shown at Figs. \ref{multi1} and  \ref{multi2} for interior particles with $a=2$ au and at Figs. \ref{multi3} and \ref{multi4} for exterior  particles with $a=12.5$ au.
All the four figures show at the bottom two plots with the sequence of proper frequencies as function of the initial inclination. The left one always corresponds to a particle with initial $e=0.1$ and the right one to an initial $e=0.4$. Both plots include the numerical calculation of $<k>$ for each case in color scale indicating also the registered change in inclination, $\Delta i$. 
At the upper part of the figures the six plots that correspond to the six vertical lines shown in the frequencies plots. The three upper plots correspond to initial $e=0.1$ and the other three correspond to initial $e=0.4$. The three selected inclinations $(13\degree,37\degree,60\degree)$  are the same in all Figs. \ref{multi1} to \ref{multi4}. The first one is for some value less than $i_c$, the second one is for $i>i_c$ close to the unstable zone and the third one is for the region where $g$ and $f$ are linked by a simple relation.
For interior particles (Figs. \ref{multi1} and \ref{multi2}) we plotted the evolution in the planes $(k,h)$  (black) and $(i\cos\omega, i\sin\omega)$ (blue) because in this case the high inclination dynamics is more linked to oscillations of $\omega$. For the exterior particle (Figs. \ref{multi3} and \ref{multi4}) we plotted the evolution in the planes $(k,h)$  (black) and $(q,p)$ (red) because in this case  the high inclination dynamics is linked to oscillations of $\Omega$ associated with flips in inclination.
The initial conditions for Figs. \ref{multi1} and \ref{multi2} are the same except for $(\omega,\Omega)$, so that $\Delta\varpi=0\degree$ for Fig.
\ref{multi1} and $\Delta\varpi=180\degree$  for Fig. 
\ref{multi2}. The same applies for Figs. \ref{multi3} and \ref{multi4}, respectively.

Let us start to analyze the case of Fig. \ref{multi1}. Both low and high eccentricity cases are stable for the low inclination regime (first column, $i=13\degree$). The proper frequencies $g$ and $f$ are well defined and the positive value $<k>$ indicates a forced eccentricity with small $\Delta i$, typical of a classic secular evolution. For growing inclinations $<k>$ goes to zero,  $\Delta i$ increases and then, for greater inclinations,  it follows an erratic behavior, this is because the classic secular model does not apply anymore or just because of unstable dynamics. The second column corresponds to $i=37\degree$  giving rise to unstable evolution. The third column correspond to   $i=60\degree$ which generates unstable evolution for the low eccentricity case and, on the contrary, a stable evolution for the high eccentricity case with an installed ZLK mechanism which is evident because of the oscillations of $\omega$ around $90\degree$.
Note that in the two examples corresponding to $i=37\degree$
the value of $g$ is almost 0, then $\varpi$ is not circulating but oscillating around a value different form $0\degree$ and $180\degree$, and according to Eq. \ref{dedt} the eccentricity will oscillate with large amplitude, which in fact is what is observed generating the unstable evolution.
Figure \ref{multi2} is analogous to the previous figure but the initial $(\omega,\Omega)$ are different and in this case are such that initially $\Delta \varpi=180\degree$. The low eccentricity regime (first row) is stable only for small inclinations and the contrary applies to the high eccentricity regime (second row): only for middle and high inclinations the dynamics is stable by the establishment of the ZLK mechanism, in this case, with oscillations of $\omega$ around $270\degree$.

Figure \ref{multi3} corresponds to exterior particles with initial $\Delta\varpi=0\degree$. For the low inclination case (first column) the behaviors are analogous to the classic secular. For $i=37\degree$ (second column) in both low and high eccentricity cases the evolution are unstable and for  $i=60\degree$ (third column) both are stable with librations of $\Omega$ accompanied by flips ($i$ oscillates around $90\degree$) as is the rule for high inclination orbits in the inverse eccentric ZLK mechanism \citep{DeElia2019}. Note also that in the space $(k,h)$ trajectories appear as banana-like, typical of pericenter secular resonances, situation that we cannot confirm  in this multi dimensional space. Note that both bottom panels show that $f$ decreases substantially for some inclination, situation that does not happen for interior particles.
Figure \ref{multi4} corresponds to exterior particles with initial $\Delta\varpi=180\degree$. For the low inclination cases, the low eccentricity example corresponds to a classic secular dynamics but the high eccentricity case shows a trajectory in  $(k,h)$ typical of secular resonance. With fine tuning the initial conditions in this case it is possible to obtain the complete separatrix or almost.
The second column show the intermediate inclination cases, both unstable. The third column shows the high-inclination cases with stable evolutions with librations of the nodes, orbital flips and, regarding the trajectories in  $(k,h)$, banana-like trajectories typical of pericenter secular resonances.

In all cases we studied, the nullification of the frequency $g$ (and also the indicator $<k>$) delimit the  regions for the  classic secular dynamics. When $g\sim 0$ we systematically observe that $\varpi$ varies very slowly making the eccentricity to grow in agreement with Eq. \ref{dedt}. This situation in general happens for $i_c \sim 30\degree$ for a wide diversity of initial conditions, but our study is limited to some specific semimajor axes in non-hierarchical cases.
For $i>i_c$ an unstable region takes place and for even greater inclinations we have the ZLK-like dynamics with librations of $\omega$ for interior particles and librations of $\Omega$ for exterior particles.
We explored the effect of the planet's eccentricity $(e_p)$ on the critical inclination and find two different behaviors depending on whether the particle is interior or exterior. In the case of the interior particle $i_c$ is almost independent of $e_p$ and always close to $i_c \sim 30\degree$, in agreement with findings by \cite{2011A&A...526A..98F}. On the contrary, for the exterior particle we found
that, as $e_p$ decreases, $i_c$ tends to larger values, up to $i_c\sim 45$, when $e_p=0$. This is consistent with Figure 8 of \cite{2012Icar..220..392G}, where, in the context of circular perturbers, the value of the particle's inclination for which $g=0$ is shown as a function of its semi-major axis.

\begin{figure*}
	\centering
	\begin{tabular}{cc}
		\includegraphics[width=0.5\linewidth]{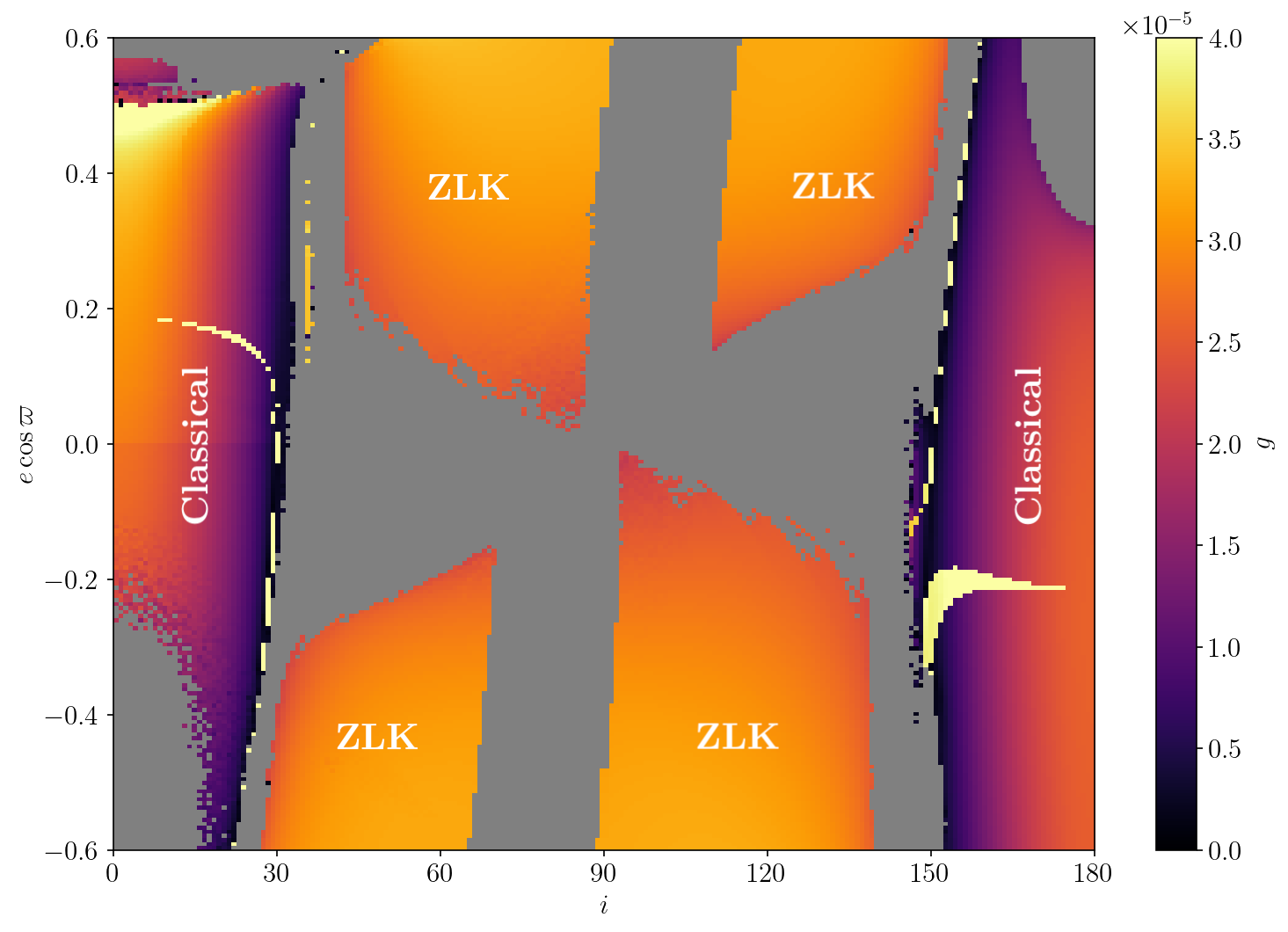} & \includegraphics[width=0.5\linewidth]{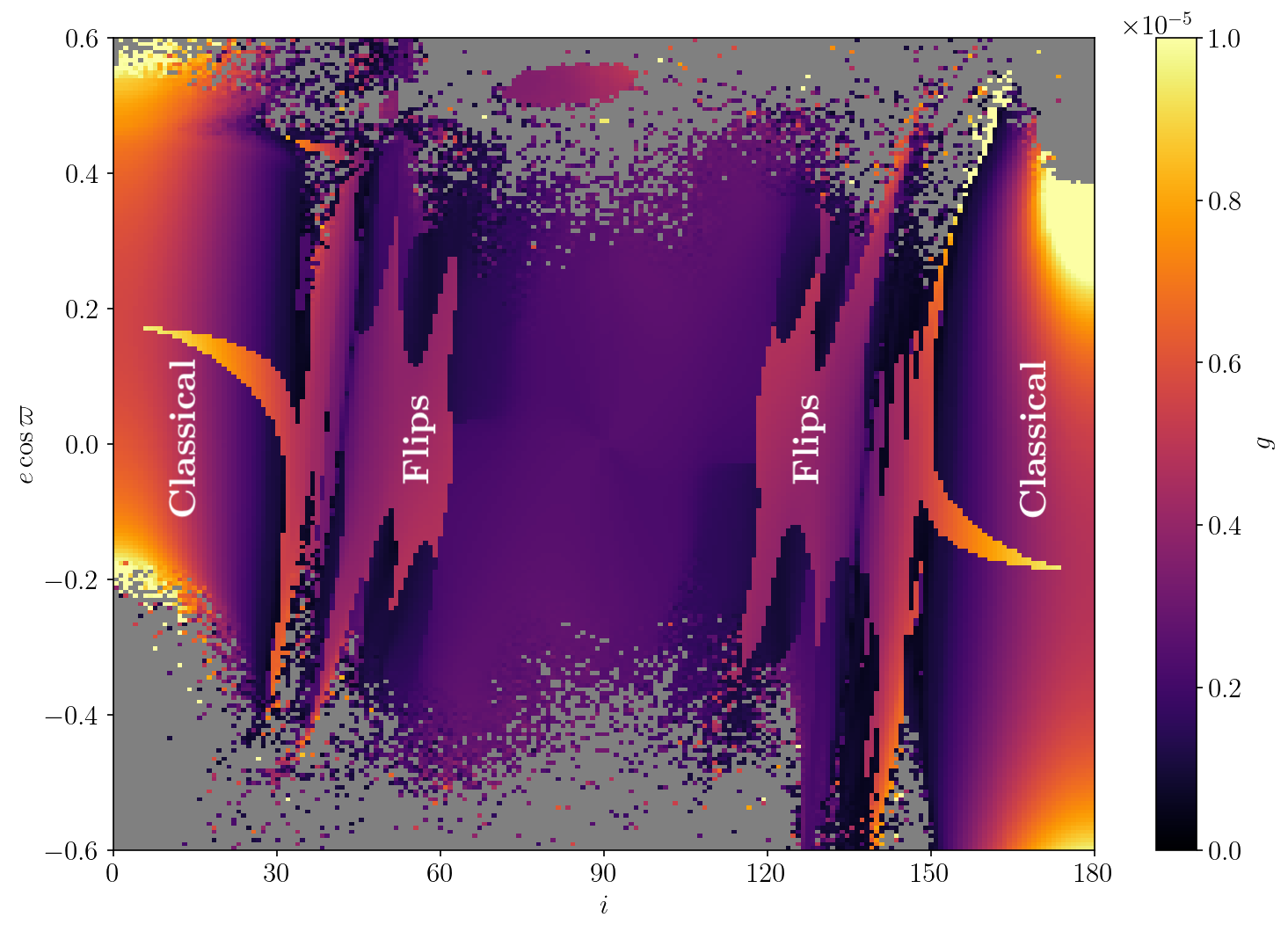}  \\
		\includegraphics[width=0.5\linewidth]{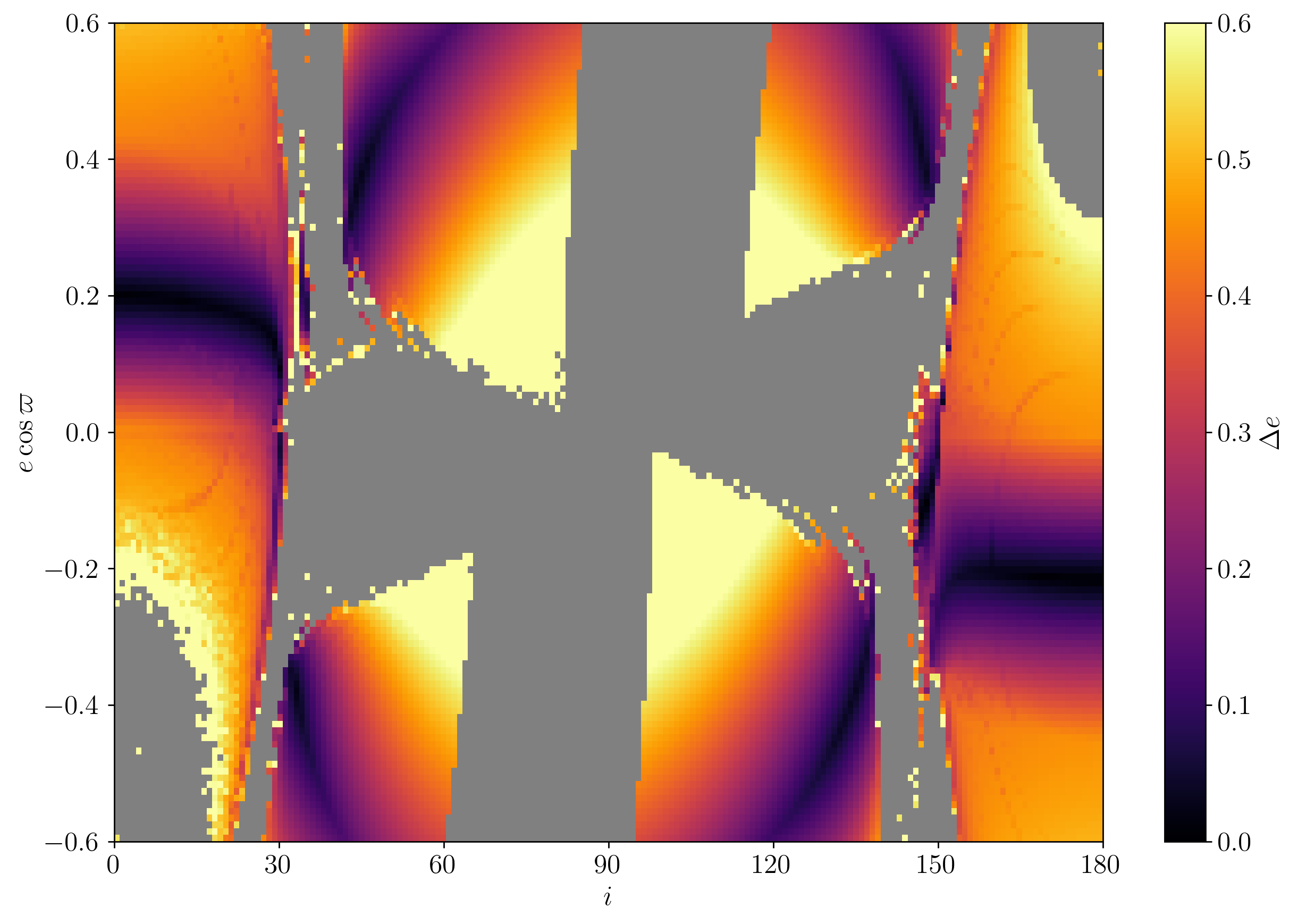} & \includegraphics[width=0.5\linewidth]{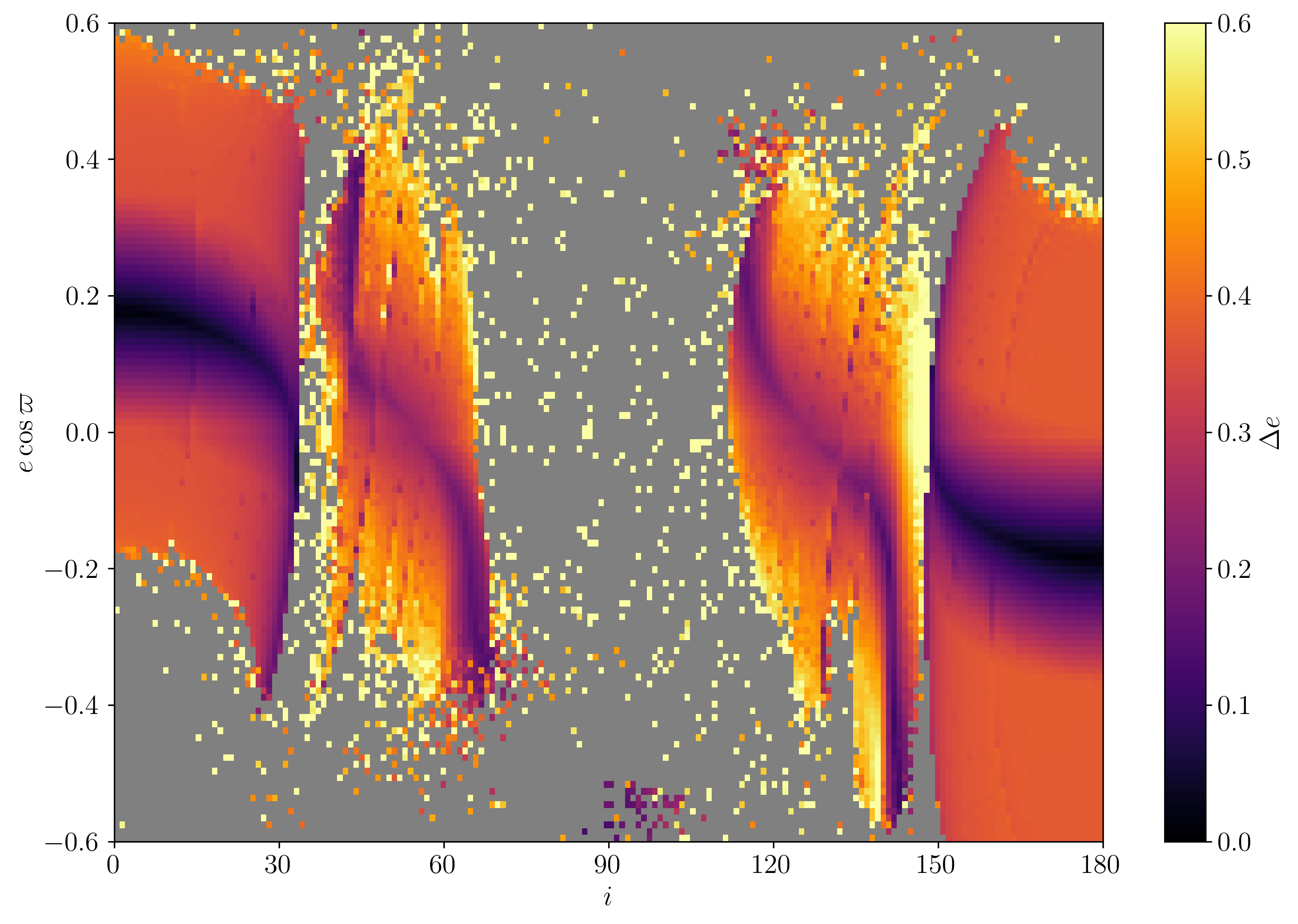} 
		
	\end{tabular}
	\caption{Top: frequency map in the plane $(i,e\cos\varpi)$ for $a=2$ au (left) and  $a=12.5$ au (right) showing frequency $g$ in units of yr$^{-1}$  with color scale. Dark regions correspond to near zero values associated with instabilities. Grey regions indicate unstable evolutions.  Initial  $\Omega=270\degree$. For the positive y-axis we take initial $\omega=90\degree$ and for the negative y-axis we take initial $\omega=270\degree$. Bottom: same as upper panels but showing $\Delta e$ after 100.000 orbital revolutions.}
	\label{freqecc}
\end{figure*}

Considering the relevance of the value of the frequency $g$ for the dynamics we
generate two maps with the calculated main frequency in the time evolution of $(k,h)$ for an interior and an exterior particle  for the total range in $i$, for a wide range in $e$ and for the configurations $\varpi=0\degree$ and $\varpi=180\degree$. Results are at upper panels of Fig.
\ref{freqecc}. 
Dark regions indicate $g \sim 0$, while grey regions represent unstable evolutions leading to the particle colliding with the star or planet or being ejected.
In both interior and exterior cases the nullification of $g$ corresponds very roughly to $i\sim 30\degree\pm 5\degree$ and $i\sim 150\degree\pm 5\degree$, with some dependence with $e\cos\varpi$.
We also calculated two dynamical maps for the same initial conditions shown at bottom panels of Fig. \ref{freqecc}.
In this case we plot the observed changes $\Delta e$ after $10^5$ orbital revolutions which give us an idea of the magnitude of the orbital changes and their relation with $g$. Here, too, grey regions indicate unstable evolutions, where particles either collide with the star or planet or are ejected. Minor discrepancies between the frequency maps and eccentricity maps arise because they were generated using different software (REBOUND and EVORB, respectively) and inconsistent time scales across the plots.
 In both interior and exterior particles, the extreme changes $\Delta e$ at $i\sim 30\degree$ and $i\sim 150\degree$ are clearly related to conditions where $g\sim 0$. Note that for the interior particle quasi circular orbits with $30\degree\lesssim i \lesssim 150\degree$ also experience extreme changes in eccentricity in
agreement with \cite{2011A&A...526A..98F} and \cite{Naoz2016} due to the ZLK mechanism (grey regions). Exterior particles with $60\degree \lesssim i \lesssim 120\degree$ are unstable for all ranges of eccentricities. 

In the Lagrange Laplace secular model (actually valid for low eccentricities) for an asteroid perturbed by an eccentric planet the asteroid's eccentricity is represented by a fixed vector in the direction given by $\varpi_p$, the forced mode, plus a rotating vector, the free or proper mode. If the osculating eccentricity is equal to the forced one and $\varpi=\varpi_p$, the free vector is zero and the variables $(k,h)$ remains fixed in time. In bottom panels of Fig. \ref{freqecc} this situation can be observed in the dark regions corresponding to $\Delta e\sim 0$ for $i \lesssim  30\degree$ and $i \gtrsim  150\degree$
and note also that in the central part of these regions the detected frequency plotted in the upper panels of Fig. \ref{freqecc}  is high because for this particular motion the highest amplitude spectral line is not $g$, which has almost null amplitude, but a higher one. In the region $30\degree\lesssim i \lesssim 150\degree$, when $\Delta e\sim 0$ the reason is that the trajectories in $(k,h)$ are almost circles around the origin. Within this inclination range, the ZLK mechanism is active, causing bidirectional orbit flips (from direct to retrograde and vice-versa) for the exterior particle.

The main conclusion from this subsection is that for non-hierarchical cases and for both interior and exterior particles perturbed by an eccentric planet there is a  \textit{critical inclination} that generates the nullification of $g$ and which is a maximum limit for  classic secular regime with well separated proper frequencies for both pairs of variables $(k,h)$ and $(q,p)$ and for stable evolution.

\subsection{Implications for small body populations}

\begin{figure}[h]
	\centering
	\includegraphics[width=0.6\textwidth]{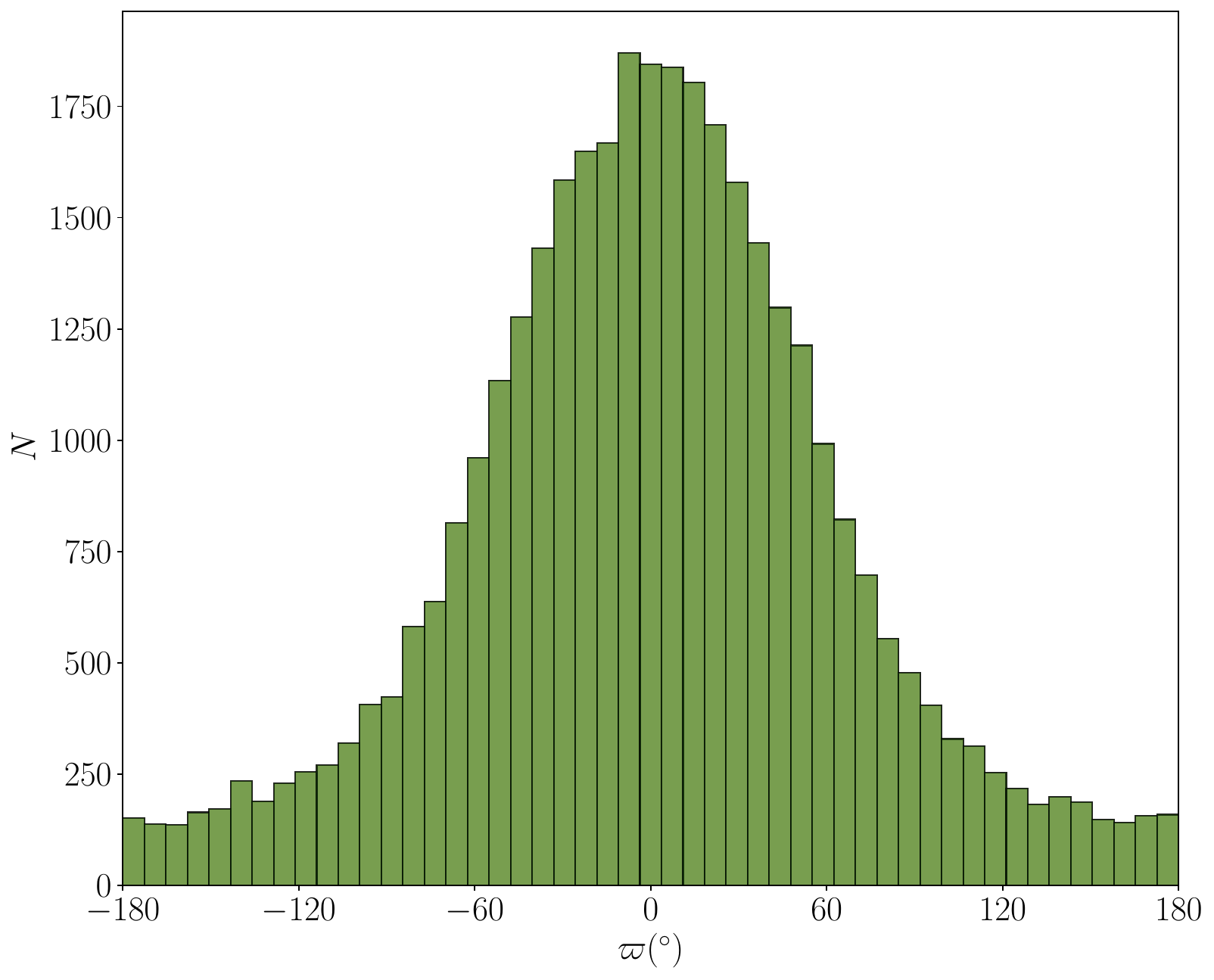}
	\caption{Distribution of final $\varpi$ for all simulated low inclination exterior particles surviving 1 Myr.}
	\label{fig:histovarpi}
\end{figure}

According to our results, small body populations perturbed by an eccentric planet will have different dynamics for interior and exterior cases. But there is a general result that is always present: an instability at $i_c\sim 30\degree$ that prevents to get higher inclination stable orbits and generated by the proper frequency $g$ that falls to almost zero.
For planets with low eccentricity, the secular frequency $g$ for exterior small bodies can approach zero for inclinations up to $\sim 45\degree$, as discussed by \cite{2012Icar..220..392G}; however, this does not necessarily lead to dynamical instability in these specific cases.
The well known pericenter shepherding generated by an eccentric planet that produces clustering of $\varpi$ around $\varpi_p$ applies for both interior and exterior particles and is produced by the forced eccentricity generated by $e_p$ \citep{1999ssd..book.....M}. To illustrate this phenomenon, we generated two synthetic populations of mass-less particles, one interior to the planet and one exterior. Both populations were made of 50000 particles with uniformly distributed inclinations between $0\degree$ and $30\degree$, eccentricities between $0$ and $0.4$, and $\Omega$, $\omega$ and $M$ between $0\degree$ and $360\degree$. Interior particles had semi-major axes between 0.5 and 2, while exterior particles with semi-major axes between 10 and 15. Both populations were integrated for 1 Myr. Figure \ref{fig:histovarpi} is a histogram of the final $\varpi$ for survivor exterior particles showing a clear concentration around $\varpi=\varpi_p=0\degree$. We also observed this concentrations for the population of interior particles, as expected (Solar System's asteroids shows this behavior) \footnote{https://minorplanetcenter.net/iau/plot/OrbEls06.gif}.

\begin{figure}[h]
	\centering    \includegraphics[width=0.6\textwidth]{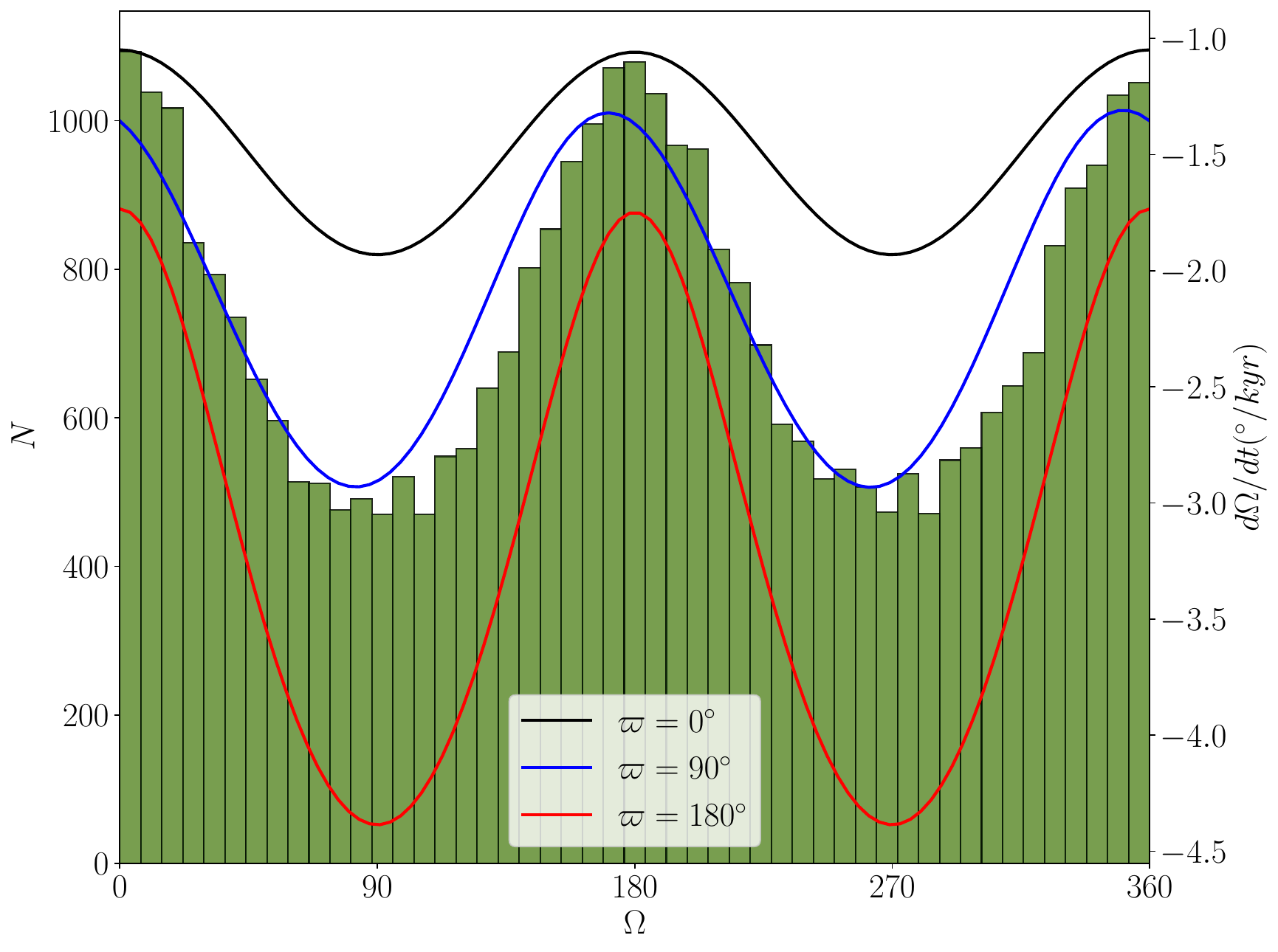}
	\caption{ Left axis: histogram of final $\Omega$ for all simulated exterior surviving particles after 1 Myr. Right axis: Dependence of $d\Omega/dt$ with $\Omega$ computed semianalitically for external particles for three different $\varpi$, computed assuming  $a=13$ au, $e=0.24$ and $i=16.2^\circ$, which correspond to median values across all simulated particles. }
	\label{fig:histoomega}
\end{figure}

\begin{figure}[h]
	\centering    \includegraphics[width=0.6\textwidth]{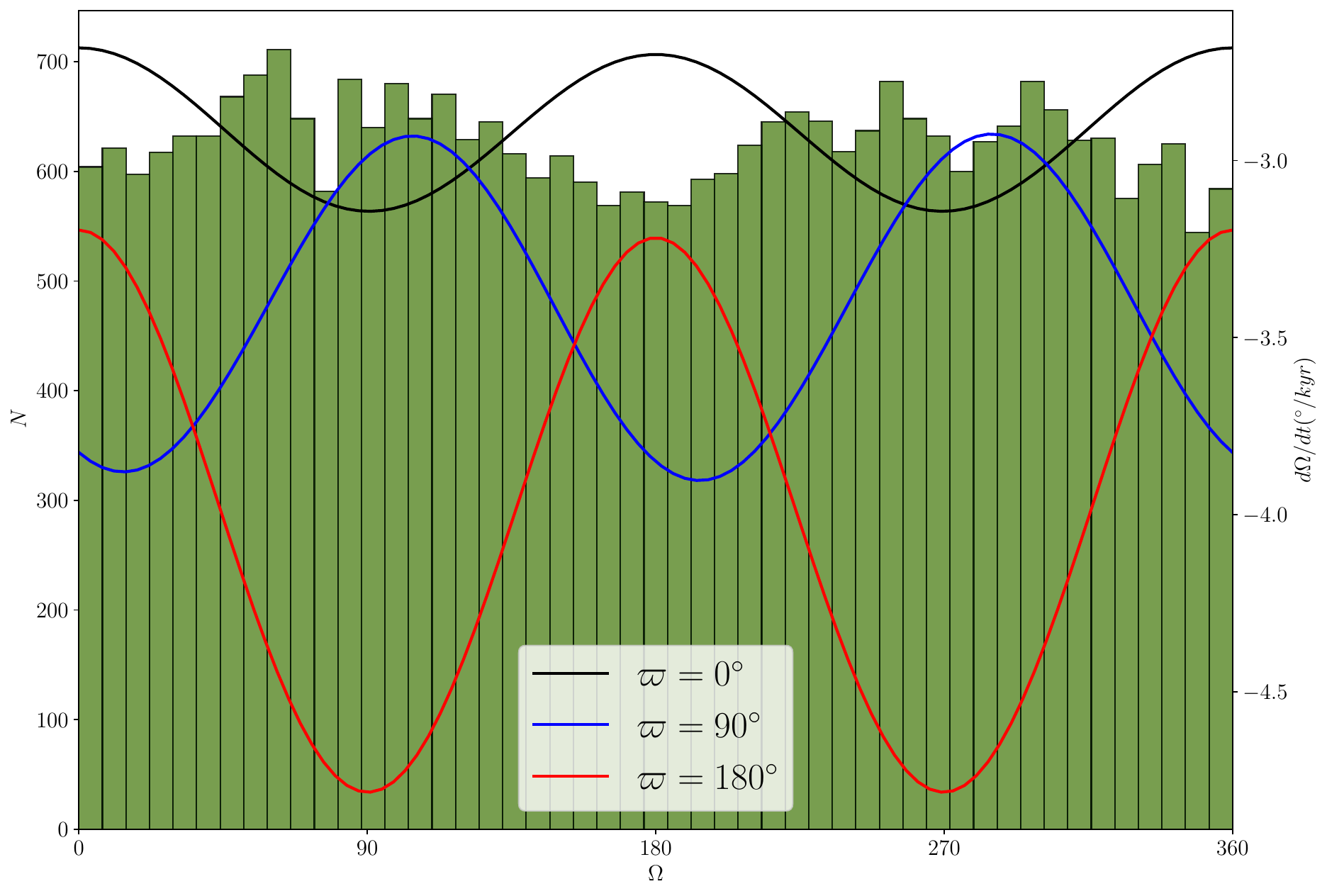}
	\caption{ Left axis: histogram of final $\Omega$ for all simulated interior particles. Right axis: Dependence of $d\Omega/dt$ with $\Omega$ computed semianalitically for interior particles for three different $\varpi$, computed assuming  $a=0.96$, $e=0.23$ and $i=15.9^\circ$, which correspond to median values across all interior particles surviving 1Myr. }
	\label{fig:histoomega2}
\end{figure}

A not so well-known effect is a clustering of $\Omega$ around $\varpi_p$ and $\varpi_p+180\degree$  for external particles only (Fig. \ref{fig:histoomega}). This clustering is not generated by a forced mode, as occurs for systems with planets with some inclination with respect to the reference plane, but by strong changes in the time variation of $\Omega$ which evolves very slowly close to $\Omega=\varpi_p$ and $\Omega=\varpi_p+180\degree$ as is shown in the curves over the histogram of Fig.
\ref{fig:histoomega}. The figure also shows that the location of the maxima
of
$d\Omega/dt$ is almost independent of $\varpi$, so its time variation does not affect the results.
The calculation of $d\Omega/dt$ was done by means of the Lagrange planetary equations calculating numerically $R_s$  and its partial derivatives.  In contrast, as Fig. \ref{fig:histoomega2} shows, an interior population of particles does not show any evident concentration in $\Omega$ because $d\Omega/dt$ has a strong dependence on $\varpi$, which varies over time. Then, due to the strong dependence of $d\Omega/dt$ with $\Omega$ and also its independence of $\varpi$ it is expected that a population of low inclination particles exterior to an eccentric planet will show concentrations of the lines of nodes around  values 
 $\Omega=\varpi_p$ and $\Omega=\varpi_p+180\degree$. Probably the different dynamical behavior of interior and exterior particles came from the fact that for an exterior particle the effect of the planetary orbit can be emulated by a kind of deformation of the central star, while for an interior particle this approximation is impossible.
 This result is valid for low inclination orbits, while for high inclination orbits it is known that $\Omega$ oscillates around $90\degree$ and $270\degree$ as shown by \cite{2017A&A...605A..64Z}.

\section{Discussion and conclusions}

We studied the basic problem of the dynamics of particles perturbed by an eccentric giant planet in both scenarios: interior and exterior orbits. We focused in non-hierarchical cases complementing the well-known results given by analytical models that can be applied when orbits are well separated. Some of our results are analogous to the results obtained by \cite{Gallardo2025} for the case of two massive planets. 
We can summarize our main results as follows:

\begin{itemize}
	
	\item an eccentric giant planet generates a web of strong MMRs and a chaotic region approximately defined in $0.5\lesssim a/a_p \lesssim 1.8$, and covering all inclinations, generated by close encounters an MMRs overlap
	
	\item exterior resonances of the type 1:N dominate over the others
	
	\item an eccentric planet establishes two well separated dynamics according to the inclination of the small bodies
	
	\item hints to the dynamics can be found in the behavior of $R_s(\omega,\Omega)$

	\item for $i<30\degree$  the dynamics is similar to the classic linear secular theory with defined proper frequencies and a forced mode that generates shepherding of pericenter around $\varpi_p$
	
	\item the limit for this regime is imposed by the inclination that makes $g=0$, which for the cases we studied is $i_c\sim 30\degree$, immediately after this inclination the dynamics is  unstable
	
	\item for exterior particles, $i_c$ depends on $e_p$, approaching  $\sim 45\degree$ as the planet's eccentricity nears zero; however, low planetary eccentricity doesn't  lead to dynamical instability.
	
	\item for interior particles with $i>40\degree$ both proper frequencies converge generating the ZLK mechanism characterized by oscillations in $\omega$

	\item we found that only for exterior particles $f$ decreases substantially
	for some inclination $i>i_c$, and this is associated with oscillations of $\Omega$
	
	\item we found exterior particles can exhibit banana like trajectories in $(k,h)$ suggesting secular pericenter resonances

	\item small bodies perturbed by a giant eccentric planet will show clustering in $\varpi$ (shepherding) around $\varpi_p$ and for low inclination  exterior ones also clustering in $\Omega$ around $\varpi_p$ and $\varpi_p+180\degree$.

\end{itemize}

In his seminal paper \cite{1962AJ.....67..591K} provides  a table with the minimum inclination for an interior asteroid, as a function of the semimajor axes ratio $\alpha$, that generates $d\omega/dt=0$ and the surging of oscillations in $\omega$ accompanied by large oscillations in $(e,i)$. He also stated:
\textit{Without the aid of a high-speed computer, it is rather difficult to estimate the effects of Jupiter’s eccentricity}. It took almost 50 years for the first results to appear when considering an eccentric planet.
Our contribution to the study of this case is 
that the key parameter is the pericenter proper frequency $g$ and when it is close to zero the eccentricity grows and that happens for some critical $i_c$ that we can assume it most probably depends on $\alpha$, in analogy to \citet{1962AJ.....67..591K}. We have not explored the dependence of $g\sim 0$  with $\alpha$ but, as Fig. \ref{fig:2fideos} shows, 
we found that the equilibrium at $\Delta\varpi=0\degree$ is broken at different $i$ according to $\alpha$.
We studied some cases with $\alpha$ up to $\sim 0.4$ because for greater values we found always unstable dynamics. 
The stability barrier we found at $i_c\sim 30\degree$ (also at $i_c\sim 150\degree$) seems to be the same  inclination limit for stable dynamics for low eccentricity interior particles found by \cite{2011A&A...526A..98F}. As we have already mentioned,
they also found that the inclination limit is almost independent of the planetary mass and eccentricity and that these parameter mostly have impact on the limits in  $\alpha$ for stable dynamics. Nevertheless, we have found that for exterior particles $i_c$ depends on $e_p$ and goes to $\sim 45\degree$ when $e_p\rightarrow 0$.

Thinking in small body populations as a result of a planetary formation process and perturbed by an eccentric planet, according to our results, it seems very unlikely that due to a secular evolution they can cross the barrier in inclination  at $i_c$  without becoming unstable. This barrier works for both interior and exterior small bodies. So, a stable small body population excited by an eccentric planet will remain with inclinations below $i_c$.
Thinking in the Planet 9 hypothesis (see review by \cite{Batygin2019}), according to our results (see Figs. \ref{multi1} and \ref{multi2}) and taking its orbital plane as reference its perturbations will generate a stable low inclination ($i<30\degree$) interior small bodies population with the well known perihelion shepherding at $\Delta\varpi=0\degree$. 
In our simplified scenario where we have ignored perturbations from the inner giant planets \citep[for a detailed study]{Batygin2019}, small bodies with  higher inclinations could be stable only if they have high eccentricity under the ZLK mechanism, showing no shepherding in $\varpi$ but oscillations in $\omega$ around $90\degree$ and $270\degree$.

\textbf{Acknowledgments.} 
We acknowledge funding for the Project  "Dinamica secular y resonante en sistemas planetarios" from CSIC (Udelar). Support from PEDECIBA is also acknowledged. 
We sincerely thank the two reviewers for their valuable comments and corrections, which significantly helped us improve this work.

\bibliographystyle{elsarticle-harv}

\bibliography{references}

\end{document}